\documentclass[longauth]{aa} % for the long lists of affiliations 

\usepackage{amsmath} % for align
\usepackage{graphicx}
\usepackage{txfonts}
\usepackage{xcolor}

\usepackage{booktabs,siunitx}
\usepackage[breaklinks, colorlinks, citecolor=blue, linkcolor=blue]{hyperref}
\newcommand{\juliet}{\texttt{juliet}}
\newcommand{\serval}{\texttt{serval}}
\newcommand{\exostriker}{\texttt{Exo-Striker}}
\newcommand{\tess}{TESS}
\newcommand{\gaia}{\textit{Gaia}}
\newcommand{\jwst}{JWST}
\newcommand{\cheops}{\textit{CHEOPS}}

\newcommand{\au}{au}
\def\ms{m\,s$^{-1}$}
\def\mearth{$M_\oplus$}
\def\rearth{$R_\oplus$}

\newcommand*\samethanks[1][\value{footnote}]{\footnotemark[#1]}

\makeatletter
\def\instrefs#1{{\def\scsep{\def\scsep{,}}\@for\w:=#1\do{\scsep\ref{inst:\w}}}}
\renewcommand{\inst}[1]{\unskip$^{\instrefs{#1}}$}

\begin{document} 

\title{TOI-1201~b: A mini-Neptune transiting \\ a bright and moderately young M dwarf\thanks{Tables~\ref{tab:rvdata} and \ref{tab:rvdatacomp} and additional data (i.e., stellar activity indicators) are available in electronic form at the CDS via anonymous ftp to \url{cdsarc.u-strasbg.fr} (\url{TBD})}}
\titlerunning{Transiting mini-Neptune around TOI-1201}

\author{
D.\,~Kossakowski\inst{mpia}\thanks{Fellow of the International Max Planck Research School for Astronomy and Cosmic Physics at the University of Heidelberg (IMPRS-HD).}
\and J.~Kemmer\inst{lsw}\samethanks % 0000-0003-3929-1442
\and P.~Bluhm\inst{lsw}\samethanks % 0000-0002-0374-8466
\and S.~Stock\inst{lsw}\samethanks % 0000-0002-1166-9338
\and J.\,A.~Caballero\inst{cabesac} % 0000-0002-7349-1387
\and V.\,J.\,S.~B\'ejar\inst{iac,ull} % age
\and C.~Cardona~Guill\'en\inst{iac,ull} % age
\and N.~Lodieu\inst{iac,ull} % ASAS and age
\and K.\,A.~Collins\inst{cfa} % TESS science office LCO obs and data reduction % 0000-0001-6588-9574
\and M.~Oshagh\inst{iac,ull} % RM effect
\and M.~Schlecker\inst{mpia}\samethanks % multiplanetary systems
\and N.~Espinoza\inst{stsci} % CARMTESS leader
\and E.~Pall\'e\inst{iac,ull} % CARMTESS leader
\and Th.~Henning\inst{mpia} % advisor
\and L.~Kreidberg\inst{mpia} % mpia director/TS
\and M.~K\"urster\inst{mpia} % advisor % 0000-0002-1765-9907
\and P.\,J.~Amado\inst{iaa} % CARMENES Deputy PI
\and D.\,R.~Anderson\inst{warwick} % WASP % 0000-0001-7416-7522
\and J.\,C.~Morales\inst{ice,ieec} % planet detection limits
\and S.~Cartwright\inst{proto} % POC, MAST, and GI 
\and D.~Charbonneau\inst{cfa} % TESS science office
\and P.~Chaturvedi\inst{tls} % carmtess
\and C.~Cifuentes\inst{cabesac} % stellar parameters
\and D.\,M.~Conti\inst{aavso} % SG1 ? % 0000-0003-2239-0567
\and M.~Cort\'es-Contreras\inst{cabesac} % stellar parameters
\and S.~Dreizler\inst{iag} % CARMENES Deputy PI
\and D.~Galad\'i-Enr\'iquez\inst{caha} % CAHA representative % 0000-0003-4946-5653
\and P.~Guerra\inst{oaa} % OAA % 0000-0002-4308-2339
\and R.~Hart\inst{usq} % MKO - passed away
\and C.~Hellier\inst{keele} % WASP leader
\and C.~Henze\inst{ames} % SPOC
\and E.~Herrero\inst{ice,ieec} % carmtess
\and S.\,V.~Jeffers\inst{mig} % CARMENES Deputy PI
\and J.\,M.~Jenkins\inst{ames} % TESS architect
\and E.\,L.\,N.~Jensen\inst{swarthmore} % LCO obs % 0000-0002-4625-7333
\and A.~Kaminski\inst{lsw}
\and J.\,F.~Kielkopf\inst{louisville} % MKO leader % 0000-0003-0497-2651
\and M.~Kunimoto\inst{mit} % TESS science office % 0000-0001-9269-8060
\and M.~Lafarga\inst{ice,ieec} % CCF parameters
\and D.\,W.~Latham\inst{cfa} % TESS architect
\and J.~Lillo-Box\inst{cabesac} % carmtess
\and R.~Luque\inst{iaa} % carmtess
\and K.~Molvaerdikhani\inst{lsw,mpia,lmu} % carmtess / atmosphere % 0000-0002-0502-0428
\and D.~Montes\inst{ucm} % UCM rep
\and G.~Morello\inst{iac,ull} % ldc law  % 0000-0002-4262-5661
\and E.\,.H.~Morgan\inst{mit} % POC, MAST, and GI 
\and G.~Nowak\inst{iac,ull} % 0000-0002-7031-7754
\and A.~Pavlov\inst{mpia}
\and M.~Perger\inst{ice,ieec} % QPC
\and E.~.V.~Quintana\inst{goddard} % POC, MAST, and GI  % 0000-0003-1309-2904
\and A.~Quirrenbach\inst{lsw} % CARMENES PI
\and S.~Reffert\inst{lsw}
\and A.~Reiners\inst{iag} % CARMENES Deputy PI
\and G.~Ricker\inst{mit} % TESS architect % 0000-0003-2058-6662
\and I.~Ribas\inst{ice,ieec} % CARMENES project scientist % 0000-0002-6689-0312
\and C.~Rodr\'iguez~L\'opez\inst{iaa} % CARMENES Deputy PI
\and M.\,R.~Zapatero~Osorio\inst{cabinta} % carmtess
\and S.~Seager\inst{mit,mit_planetarysciences,mit_aeronautics} % TESS architect % 0000-0002-6892-6948
\and P.~Sch\"ofer\inst{iag}
\and A.~Schweitzer\inst{hamburg} % stellar parameters % 0000-0002-1624-0389
\and T.~Trifonov\inst{mpia}
\and S.~Vanaverbeke\inst{leuven,iris} % carmtess % 0000-0003-0231-2676
\and R.~Vanderspek\inst{mit} % TESS architect
\and R.~West\inst{warwick} % WASP
\and J.~Winn\inst{princeton} % TESS architect
\and M.~Zechmeister\inst{iag}
}

\institute{
\label{inst:mpia}Max-Planck-Institut f\"{u}r Astronomie, K\"{o}nigstuhl  17, 69117 Heidelberg, Germany
\and \label{inst:lsw}Landessternwarte, Zentrum f\"ur Astronomie der Universit\"at Heidelberg, K\"onigstuhl 12, 69117 Heidelberg, Germany
\and \label{inst:cabesac}Centro de Astrobiolog\'ia (CSIC-INTA), ESAC, Camino bajo del castillo s/n, 28692 Villanueva de la Ca\~nada, Madrid, Spain
\and \label{inst:iac}Instituto de Astrof\'isica de Canarias (IAC), 38205 La Laguna, Tenerife, Spain
\and \label{inst:ull}Departamento de Astrof\'isica, Universidad de La Laguna, 38206 La Laguna, Tenerife, Spain
\and \label{inst:cfa}Center for Astrophysics \textbar \ Harvard \& Smithsonian, 60 Garden Street, Cambridge, MA 02138, United States of America
\and \label{inst:stsci}Space Telescope Science Institute, Baltimore, MD 21218, United States of America
\and \label{inst:iaa}Instituto de Astrof\'isica de Andaluc\'ia (CSIC), Glorieta de la Astronom\'ia s/n, 18008 Granada, Spain
\and \label{inst:warwick}Department of Physics, University of Warwick, Gibbet Hill Road, Coventry CV4 7AL, UK
\and \label{inst:ice}Institut de Ci\`encies de l'Espai (ICE, CSIC), Campus UAB, C/ de Can Magrans s/n, 08193 Cerdanyola del Vall\`es, Spain
\and \label{inst:ieec}Institut d'Estudis Espacials de Catalunya (IEEC), C/ Gran Capit\`a 2-4, 08034 Barcelona, Spain
\and \label{inst:proto}Proto-Logic LLC, 1718 Euclid Street NW, Washington, DC 20009, United States of America
\and \label{inst:tls}Th\"uringer Landessternwarte Tautenburg, Sternwarte 5, 07778 Tautenburg, Germany
\and \label{inst:aavso}American Association of Variable Star Observers, 49 Bay State Road, Cambridge, MA 02138, United States of America
\and \label{inst:iag}Institut f\"ur Astrophysik, Georg-August-Universit\"at, Friedrich-Hund-Platz 1, 37077 G\"ottingen, Germany
\and \label{inst:caha}Centro Astron\'onomico Hispano Alem\'an, Observatorio de Calar Alto, Sierra de los Filabres, E-04550 G\'ergal, Spain
\and \label{inst:oaa}Observatori Astron\`omic Albany\`a, Cam\'i de Bassegoda s/n, Albany\`a 17733, Girona, Spain
\and \label{inst:usq}Centre for Astrophysics, University of Southern Queensland, Toowoomba, QLD, 4350, Australia
\and \label{inst:keele}Astrophysics Group, Keele University, Staffordshire, ST5 5BG, United Kingdom
\and \label{inst:ames}NASA Ames Research Center, Moffett Field, CA 94035, United States of America
\and \label{inst:mig}Max-Planck-Institut f\"ur Sonnensystemforschung, Justus-von-Liebig Weg 3, 37077 G\"ottingen, Germany
\and \label{inst:swarthmore}Department of Physics \& Astronomy, Swarthmore College, Swarthmore PA 19081, United States of America
\and \label{inst:louisville}Department of Physics and Astronomy, University of Louisville, Louisville, KY 40292, United States of America
\and \label{inst:mit}Department of Physics and Kavli Institute for Astrophysics and Space Research, Massachusetts Institute of Technology, Cambridge, MA 02139, United States of America
\and \label{inst:lmu}Universit\"ats-Sternwarte, Ludwig-Maximilians-Universit\"at M\"unchen, Scheinerstrasse 1, 81679 M\"unchen, Germany
\and \label{inst:ucm} Departamento de F{\'i}sica de la Tierra y Astrof{\'i}sica \& IPARCOS-UCM (Instituto de F\'{i}sica de Part\'{i}culas y del Cosmos de la UCM), Facultad de Ciencias F{\'i}sicas, Universidad Complutense de Madrid, E-28040 Madrid, Spain
\and \label{inst:goddard}NASA Goddard Space Flight Center, 8800 Greenbelt Road, Greenbelt, MD 20771, United States of America
\and \label{inst:cabinta}Centro de Astrobiolog\'ia (CSIC-INTA), Carretera de Ajalvir km 4, 28850 Torrej\'on de Ardoz, Madrid, Spain
\and \label{inst:mit_planetarysciences}Department of Earth, Atmospheric and Planetary Sciences, Massachusetts Institute of Technology, Cambridge, MA 02139, United States of America
\and \label{inst:mit_aeronautics}Department of Aeronautics and Astronautics, Massachusetts Institute of Technology, 77 Massachusetts Avenue, Cambridge, MA 02139, United States of America
\and \label{inst:hamburg}Hamburger Sternwarte, Gojenbergsweg 112, 21029 Hamburg, Germany
\and \label{inst:leuven}Vereniging Voor Sterrenkunde, Brugge, Belgium \& Centre for mathematical Plasma-Astrophysics, Department of Mathematics, KU Leuven, Celestijnenlaan 200B, 3001 Heverlee, Belgium
\and \label{inst:iris}AstroLAB IRIS, Provinciaal Domein ``De Palingbeek'', Verbrandemolenstraat 5, 8902 Zillebeke, Ieper, Belgium
\and \label{inst:princeton}Department of Astrophysical Sciences, Princeton University, 4 Ivy Lane, Princeton, NJ 08544, United States of America
}
% \and \label{inst:cfa} Center for Astrophysics \textbar \ Harvard \& Smithsonian, 60 Garden Street, Cambridge, MA 02138, USA
% \and \label{inst:mit} Department of Physics and Kavli Institute for Astrophysics and Space Research, Massachusetts Institute of Technology, Cambridge, MA 02139, USA
% \and \label{inst:luisville} Department of Physics and Astronomy, University of Louisville, Louisville, KY 40292, USA
% \and \label{inst:mit_planetarysciences} Department of Earth, Atmospheric and Planetary Sciences, Massachusetts Institute of Technology, Cambridge, MA 02139, USA
% \and \label{inst:mit_aeronautics} Department of Aeronautics and Astronautics, MIT, 77 Massachusetts Avenue, Cambridge, MA 02139, USA
% \and \label{inst:oaa} Observatori Astron\'omic Albany\'a, Cam\'i de Bassegoda S/N, Albany\'a 17733, Girona, Spain
% \and \label{inst:swarthmore} Department of Physics \& Astronomy, Swarthmore College, Swarthmore PA 19081, USA
% \and \label{inst:aavso} American Association of Variable Star Observers, 49 Bay State Road, Cambridge, MA 02138, USA
%\and \label{inst:usq} Centre for Astrophysics, University of Southern Queensland, Toowoomba, QLD, 4350, Australia
% \and \label{inst:ames}NASA Ames Research Center, Moffett Field, CA 94035, USA
% \and \label{inst:prince}Department of Astrophysical Sciences, Princeton University, 4 Ivy Lane, Princeton, NJ 08544, USA
   \date{Received 18 June 2021 / Accepted dd September 2021}

\abstract{ 
We present the discovery of a transiting mini-Neptune around TOI-1201, a relatively bright and moderately young early M dwarf ($J \approx$ 9.5\,mag, $\sim$600--800\,Myr) in an equal-mass $\sim$8\,arcsecond-wide binary system, using data from the Transiting Exoplanet Survey Satellite (\tess), along with follow-up transit observations. 
With an orbital period of 2.49\,d, TOI-1201~b is a warm mini-Neptune with a radius of $R_\mathrm{b} = 2.415\pm0.090\,R_\oplus$. This signal is also present in the precise radial velocity measurements from CARMENES, confirming the existence of the planet and providing a planetary mass of $M_\mathrm{b} = 6.28\pm0.88\,M_\oplus$ and, thus, an estimated bulk density of $2.45^{+0.48}_{-0.42}$\,g\,cm$^{-3}$. The spectroscopic observations additionally show evidence of a signal with a period of 19\,d and a long periodic variation of undetermined origin. In combination with ground-based photometric monitoring from WASP-South and ASAS-SN, we attribute the 19\,d signal to the stellar rotation period ($P_\textnormal{rot}=$ 19--23\,d), although we cannot rule out that the variation seen in photometry belongs to the visually close binary companion. We calculate precise stellar parameters for both TOI-1201 and its companion. The transiting planet is an excellent target for atmosphere characterization (the transmission spectroscopy metric is $97^{+21}_{-16}$) with the upcoming James Webb Space Telescope. It is also feasible to measure its spin-orbit alignment via the Rossiter-McLaughlin effect using current state-of-the-art spectrographs with submeter per second radial velocity precision. }


\maketitle
%
%-------------------------------------------------------------------

\section{Introduction} \label{sec:intro}

Results from the \textit{Kepler} \citep{Borucki2010_Kepler,Mathur2017} and K2 \citep{Howell2014_K2} missions have revealed that M dwarfs ($T_\textnormal{eff} \lesssim 4\,000$\,K) host on average $\sim$2.5 planets with radii below 4\,\rearth\ and with periods of less than 200\,d \citep[e.g.,][]{DressingCharbonneau2013,DressingCharbonneau2015}. 
The bimodal distribution of radii of small, close-in planets produces a gap that separates planets with 1--2\,\rearth, the so-called super-Earths, and 2--4\,\rearth, denominated as mini-Neptunes.
For solar-like stars, this radius gap occurs between 1.7\,\rearth\ and 2.0\,\rearth\ \citep{Fulton2017, FultonPetigura2018,VanEylen2018}, and it resides a bit lower for low-mass stars, between 1.4\,\rearth\ and 1.7\,\rearth\ \citep{CloutierMenou2020,VanEylen2021}.
This bimodality suggests that mini-Neptunes are mostly rocky planets that were born with primary atmospheres and were able to retain them, whereas planets below the radius gap lost their atmospheres and were stripped to their cores \citep{Bean2021}.

The mechanism that drives atmospheric loss for these planets remains ambiguous, with the prime candidates being photo-evaporation \citep[e.g.,][]{OwenWu2013,OwenWu2017,LehmerCatling2017,VanEylen2018,Mordasini2020} and core-powered mass loss \citep{Ginzburg2016,Ginzburg2018,Gupta2019,Gupta2021} on timescales of $\sim$100\,Myr and $\sim$1\,Gyr, respectively.
In both models, the heating of the planet's upper atmosphere drives a hydrodynamic outflow. In the case of core-powered mass loss, the planet's heating originates from infrared (IR) radiation coming from the cooling planetary interior, while photo-evaporation is due to extreme ultraviolet photons from the host star.
However, it is not clear which heating mechanism dominates the mass loss observed for mini-Neptune-sized planets. Thus, investigating young planets (100\,Myr--1\,Gyr) and determining accurate masses for these worlds is critical in constraining the mass loss rate to then learn how they evolve over time in hopes of better explaining the origin of the radius gap.

Precise mass and radius measurements alone, however, are not sufficient in establishing the bulk composition, as there are large degeneracies when determining the ratio of rock, water, and hydrogen for the interior structure \citep{RogersSeager2010_degeneracy,LopezFortney2014}. Mini-Neptunes are one of the most commonly detected outcomes of planet formation \citep{Barnes2009,Rogers2011,Marcy2014,Rogers2015}, and understanding the composition of their atmospheres can help reveal the nature, origins, and evolution of these objects \citep{Miller-RicciFortney2010, BennekeSeager2013,CrossfieldKreidberg2017}. Luckily, we find ourselves in an era in which the ongoing Transiting Exoplanet Survey Satellite \citep[\tess;][]{Ricker2015} mission is set to find such targets that will be suitable for transmission spectroscopy using the future James Webb Space Telescope \citep[\jwst;][]{jwst}. 
TOI-1201~b adds to the growing sample of mini-Neptunes around M dwarfs that are promising targets for transmission spectroscopy.

Additionally, the stellar M-dwarf host TOI-1201 is found in a wide binary system.
Observations of planets in multiple-star systems can shed light on stellar and planet formation \citep[see e.g.,][]{Goodwin2007,Goodwin2008,Thebault2015,Monnier2018}. To date, planet discoveries in binary systems where the primary is an M dwarf are scarce despite them being the most abundant stars in our Galaxy \citep{Henry2006,Winters2015,Reyle2021}. An M dwarf is the primary in fewer than 10\,\% of the known binary systems with planet detections\footnote{ \url{https://exoplanetarchive.ipac.caltech.edu/}, accessed on 29 March 2021}. 
Such a low count is not surprising since M-dwarf systems with close visual companions were typically discarded from dedicated detection surveys due to the lack of high-resolution near-IR spectrographs and the high risk of potential light contamination in transits \citep[e.g.,][and references therein]{Cortes-Contreras2017}.

Nearly half of the solar-like stars in our solar neighborhood are members of binaries or multiple systems \citep{DuquennoyMayor1991,Raghavan2010}, and 25\,\% of them are found in wide binaries (projected separation $s>100$\,\au). It has been shown that the planet occurrence rate for these wide binary systems is comparable to that around single stars, presumably because the influence of the stellar companion on the formation of the planet is negligible \citep{Wang2014,Deacon2016}.
Turning our focus back to M dwarfs, the stellar multiplicity rate is believed to be $\sim$16--27\,\% \citep{Janson2012,Janson2014,Cortes-Contreras2017,Winters2019}.
However, the planet occurrence rate in low-mass binaries is still uncertain because these stellar targets have often been neglected in surveys.  

More recently, the CARMENES spectrograph and consortium (introduced in Sect.~\ref{sec:carmenes}) has been responsible for increasing the number of detected planets around M dwarfs in binary systems, particularly around wide separation systems, namely, \object{HD~180617} \citep{Kaminski2018}, \object{Gl~49} \citep{Perger2019}, \object{LTT~3780} \citep{Nowak2020_toi732,Cloutier2020_toi732}, \object{HD~79211} \citep[GJ~338~B,][]{Gonzalez2020_gj338B}, \object{GJ~3473} \citep[TOI-488,][]{Kemmer2020_toi488}, and \object{HD~238090} \citep[GJ~458~A,][]{Stock2020_threesuperearths}.
Precise stellar properties, as well as radial velocities (RVs), are commonly determined solely for the host; the details of the companion are often missed, with some exceptions, such as GJ~338~B \citep{Gonzalez2020_gj338B} or \object{GJ~15}~A and~B \citep{Howard2014_GJ15A,Pinamonti2018_GJ15A,Trifonov2018_carmenespaper}. Studying properties such as metallicity and age can shed light on which environments favor the formation of particular planets \citep{JohnsonLi2012,Hobson2018,Montes2018}. Observing more systems would allow for a better grasp on how stellar multiplicity in M-dwarf systems plays a role in the planet formation process and in the types of planets that can exist. 

In this paper we present the discovery and mass determination of the transiting mini-Neptune TOI-1201~b. The planet orbits one member of the M-dwarf wide binary system \object{PM~J02489--1432}. We calculate precise stellar parameters and present RV measurements for both the host and the companion.
The mini-Neptune, with a period of about 2.5\,d, was initially discovered as a transiting planet candidate in \tess\ data and is confirmed here using ground-based photometry and CARMENES RV measurements.
Future observations with \jwst\ will be able to precisely constrain the atmospheric compositions of this and other similar planets and, thus, provide important constraints on mini-Neptune formation.

The paper is organized as follows. 
The \tess\ photometry is introduced in Sect.~\ref{sec:tessphot}, followed by the various ground-based photometric and spectroscopic observations in Sect.~\ref{sec:data}. The stellar properties of both TOI-1201 and its companion are discussed in Sect.~\ref{sec:stellarprops}. The modeling analysis, which combines all the available data to produce the final planetary parameters, is presented in Sect.~\ref{sec:results}. Finally, in Sect.~\ref{sec:discussion} we unveil the future prospects for TOI-1201~b. We give our final conclusions in Sect.~\ref{sec:conclusions}.

\section{\tess\ photometry} \label{sec:tessphot}

TOI-1201 (TIC-29960110) was observed by \tess\ with the short cadence 2-minute integrations during cycle 1, sector 4 (camera \#1, CCD \#1) between 18 October and 15 November 2018, and also during the first year of the extended mission during cycle 3, sector 31 (camera \#1, CCD \#3) between 21 October and 19 November 2020. 
The planetary signature of TOI-1201~b was detected in the Science Processing Operations Center \citep[SPOC;][]{Jenkins2016} transit search pipeline on 16 January 2019 and was issued as an alert by the TESS Science Office (TSO) on 31 January 2019.
The target was not initially announced as a \tess\ object of interest (TOI) along with the other TOIs from sector 4, but rather its companion (TOI-393, TIC-29960109) that fell on the same pixel was (Sect.~\ref{sec:binary}).

\begin{figure}
    \centering
    \includegraphics[width=\linewidth]{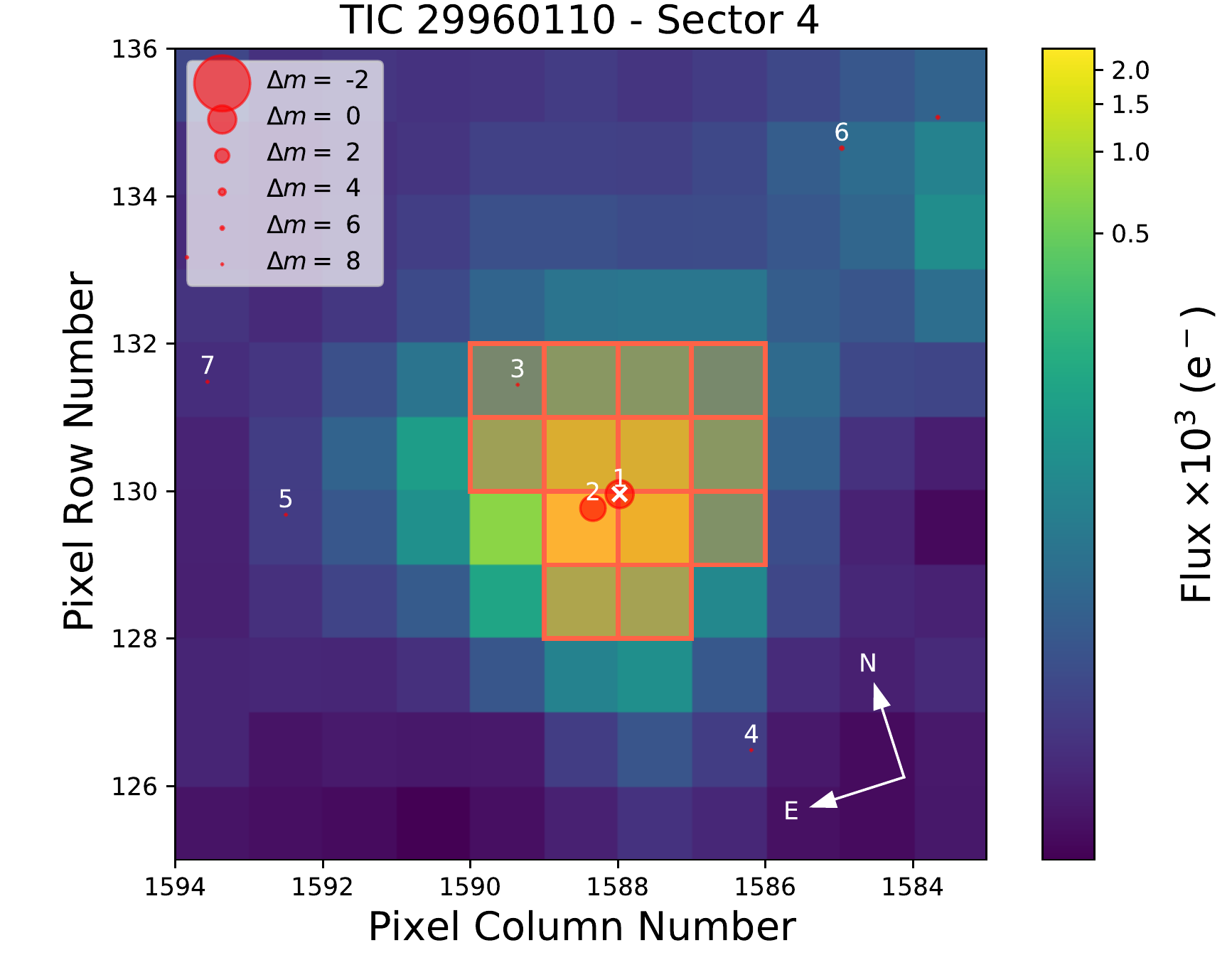}
    \includegraphics[width=\linewidth]{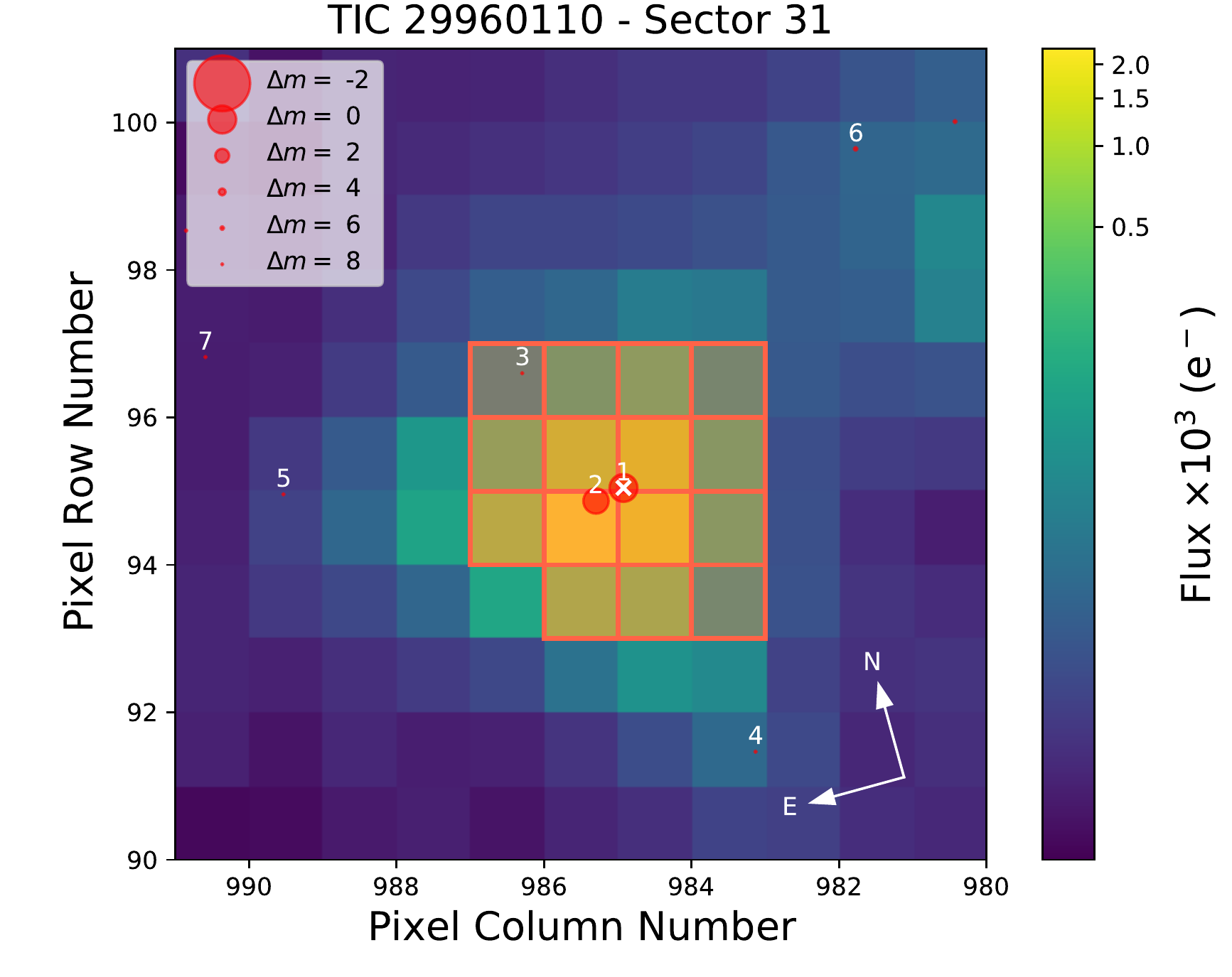}
    \caption[\tess\ TPF plot for TOI-1201 for sectors 4 and 31.]{\tess\ TPF plot for TOI-1201 for sectors 4 and 31. The SAP was computed using the flux counts coming in from the red bordered pixels (mask). The red circles represent neighboring sources listed in \gaia\ DR2, where the size corresponds to the brightness difference with respect to TOI-1201. The close companion to TOI-1201 is indicated as source \#2.} %Credit Eric Jensen
    \label{fig:tpf}
\end{figure}

Fig.~\ref{fig:tpf} displays a plot of the target pixel file (TPF) and the aperture mask used to produce the Simple Aperture Photometry (\texttt{SAP}), created using \texttt{tpfplotter}\footnote{\url{https://github.com/jlillo/tpfplotter}} \citep{Aller2020_tpf}.
Within the Tess Follow-up Observing Program (TFOP) ``Seeing-limited Photometry'' SG1 subgroup\footnote{\url{https://tess.mit.edu/followup/}}, the first follow-up transit photometric data immediately indicated that TOI-1201 was the correct stellar host and not TOI-393 (see Sect.~\ref{sec:followuptransit}). In this work, we only analyzed the 2-minute integrations. In addition, 20-second integrations are available for sector 31.
However, they do not improve the uncertainty on the model parameters and rather increase computation time in comparison to the 2-minute integrations.

% \begin{table}
% \centering
% \small
% \caption{\tess\ observations of TOI-1201.}
% \label{tab:tessphot}
% \begin{tabular}{cccll}
% \hline\hline
% \noalign{\smallskip}
% Sector  &  Camera  & CCD   & Start date & End date\\
% \noalign{\smallskip}
% \hline
% \noalign{\smallskip}
% 4  & 1   &  1  & 18 Oct. 2018 & 15 Nov. 2018\\
% 31  & 1   &  1  & 21 Oct. 2020 & 19 Nov. 2020\\
% \noalign{\smallskip}         
% \hline
% \end{tabular}
% \end{table}

\begin{figure*}
\centering
\begin{minipage}{\textwidth}
  \includegraphics[width=\linewidth]{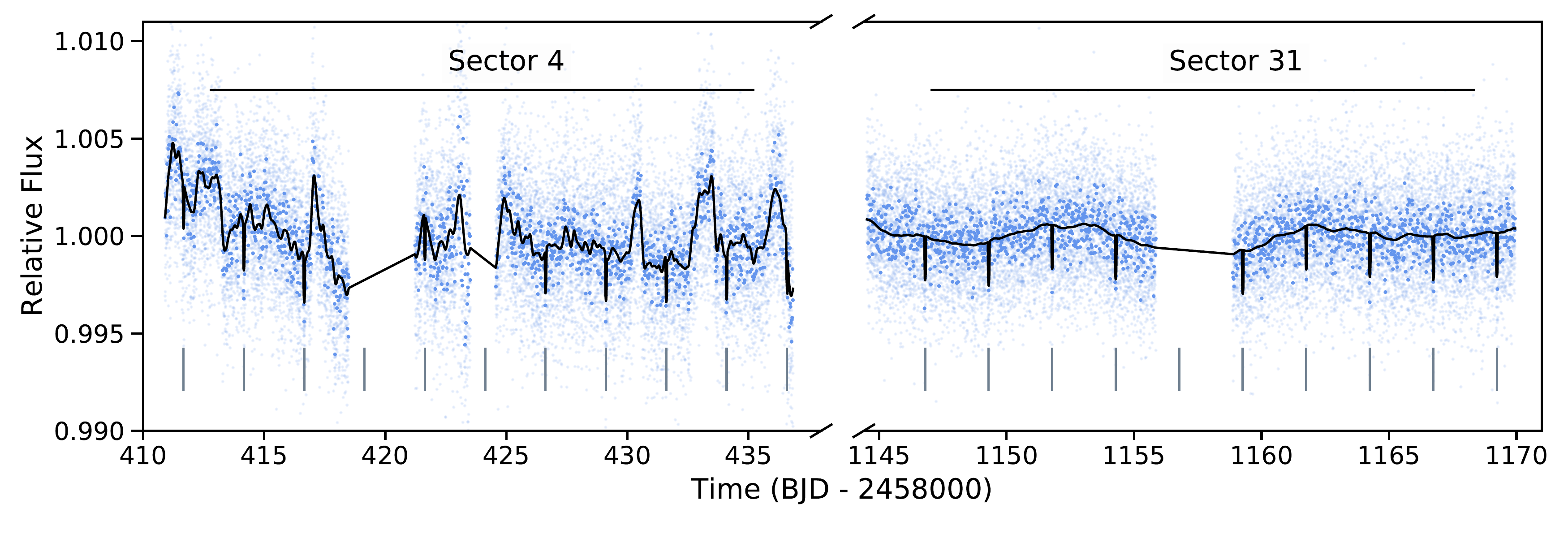}
\end{minipage}
\begin{minipage}{1\textwidth}
  \centering
  \includegraphics[width=0.66\linewidth]{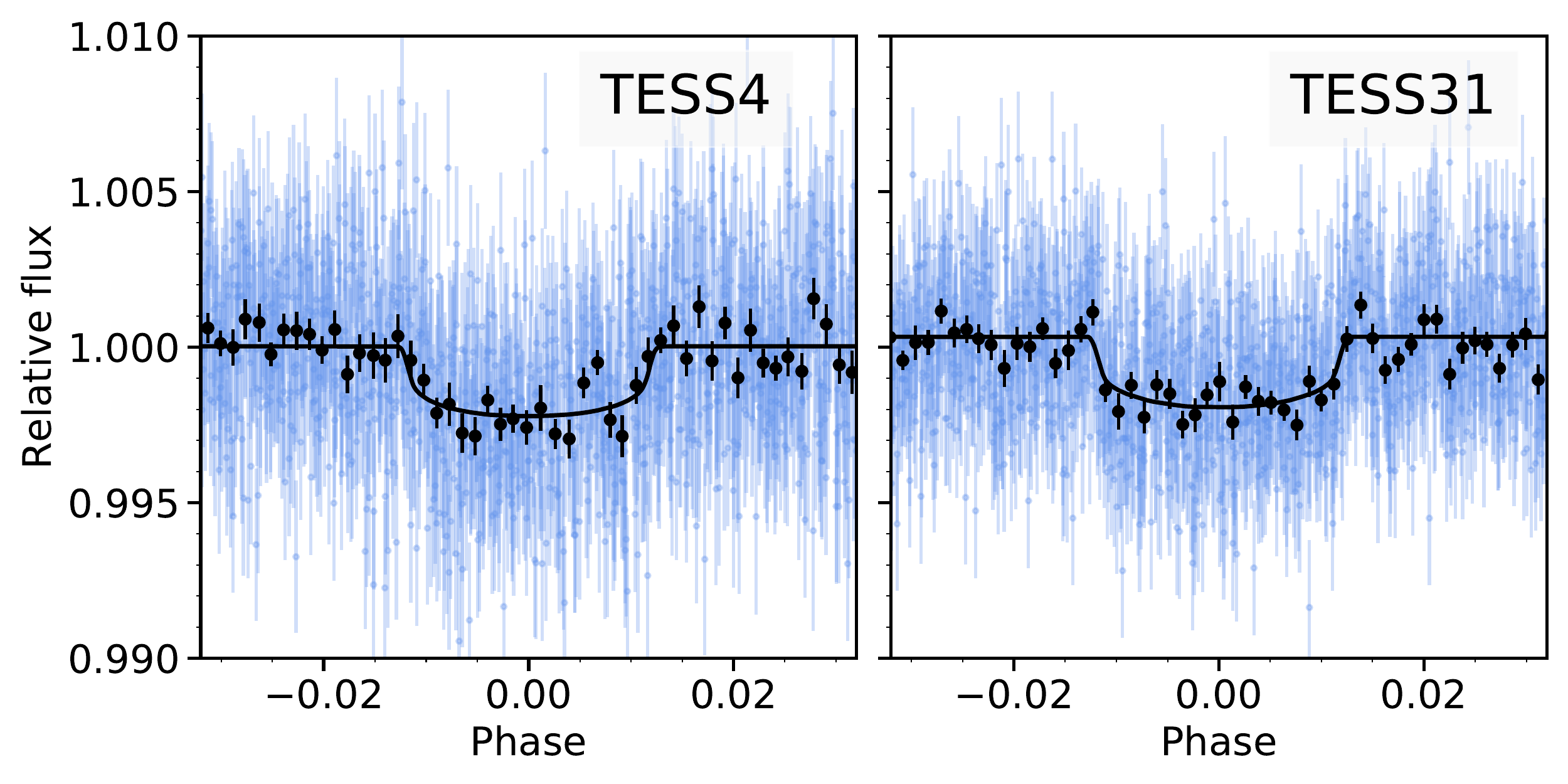}
\end{minipage}
\begin{minipage}{1\textwidth}
  \centering
  \includegraphics[width=1\linewidth]{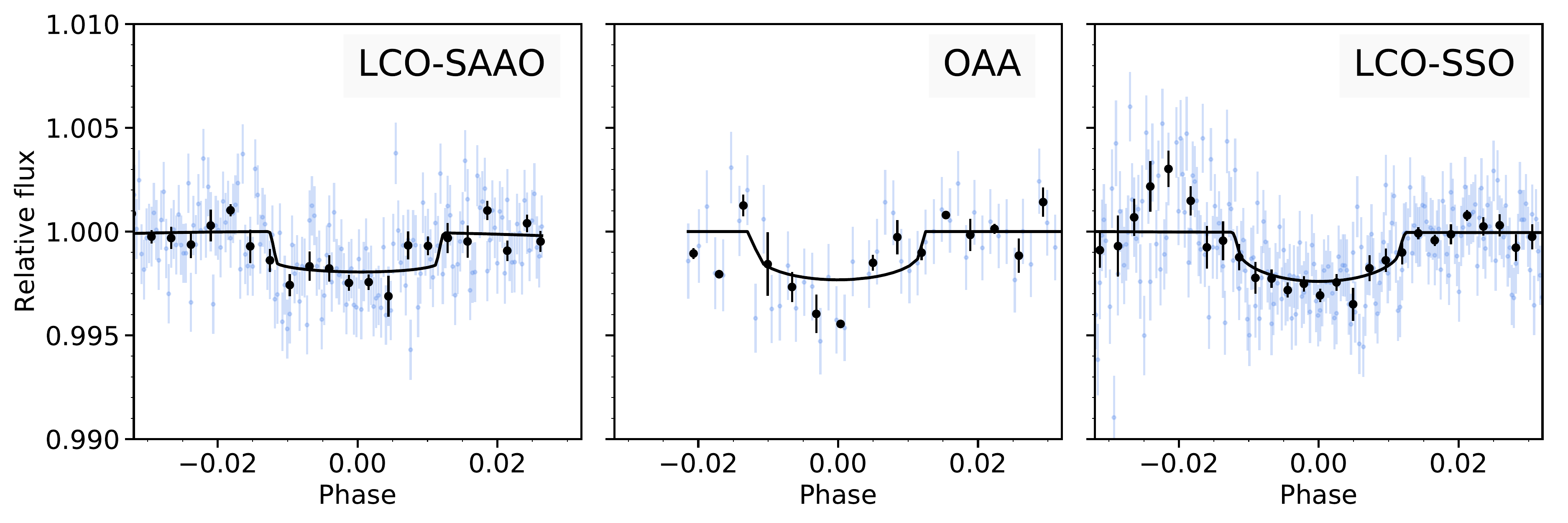}
\end{minipage}
\caption[All transit data and phase-folded transits for TOI-1201~b]{\textit{Top panel}: Full \tess\ photometry for TOI-1201 from sectors 4 and 31. The opaque and black dots are the data binned to 20 minutes in the light curve time series and phase-folded plots, respectively. The vertical gray lines represent the transit times, and the black line is the \juliet\ best-fit model from the joint fit. \textit{Middle and bottom panels}: Phase-folded transits for TOI-1201~b for all photometric instruments: \tess\ sector 4 (\textit{middle left}) and sector 31 (\textit{middle right}), LCO-SAAO $g'$ band (\textit{bottom left}), OAA $V$ band (\textit{bottom middle}), and LCO-SSO $z_s$ band (\textit{bottom right}). Any GP components and linear trends were subtracted out before the curves were phase-folded.}
\label{fig:lc} 
\end{figure*}

% from https://tev.mit.edu/data/collection/193/
% toi393 was alerted 2019 01 16
% toi1201 was alerted 2019 01 31
We downloaded the \tess\ data from the Mikulski Archive for Space Telescopes\footnote{\url{https://mast.stsci.edu}}.
The \tess\ photometric light curve is corrected for systematics as usually carried out \citep[Presearch Data Conditioning, \texttt{PDC\_SAP} flux --][]{Smith2012,Stumpe2012,Stumpe2014} and is presented in Fig.~\ref{fig:lc} with the best-fit model (explained in detail in Sect.~\ref{sec:jointfit}). 

TOI-1201 and its equally bright companion  (0.26\,mag fainter in $G_{BP}$) are just 8\,arcsec away from each other (see Sect.~\ref{sec:binary}) and, therefore, fall on the same \tess\ pixel (21\,arcsec). This raised the issue of ensuring that the radius ratio is correct. This problem was solved since the \texttt{PDC\_SAP} flux is corrected for possible nearby flux contamination. Specifically, the images of the postage stamps for this target have a flat background with only one other bright star in the photometric aperture that is already known with no additional significant crowding, so the sky background bias is expected to be minimal for the sector 4 data. From sector 27 onward, the sky background algorithm was modified and improved such that any sky background bias for sector 31 data is negligible. The transit depth in the sector 4 data is therefore slightly overestimated by 0.3\,\%.
However, this did not raise a serious issue for the analysis. Using the keyword \texttt{CROWDSAP}, the correction for sector 4 and sector 31 data is 0.52 and 0.56, respectively.
These values are in agreement to our calculation for the dilution from converting the \tess\ magnitudes into fluxes and following $f_A/ (f_A + f_B)$ to obtain 0.56, where $f_B$ and $f_A$ are the fluxes for TOI-1201 and its companion, respectively. % I had to use A as toi393
Therefore, the preliminary parameters provided by the SPOC presented in the TOI catalog\footnote{\url{https://tev.mit.edu/data/collection/193/}} are valid. 
The transit signal was initially detected to have a period of 2.49198$\pm$0.00032\,d and a depth of 2128$\pm$160 parts per million (ppm), corresponding to an approximate planetary radius of 2.2$\pm$1.3\,$R_{\oplus}$ and equilibrium temperature of 640\,K. The quoted uncertainty on the radius is rather large, but this is not important because the value is anyhow updated after performing a full analysis (Sect.~\ref{sec:jointfit}).

\section{Ground-based observations} \label{sec:data}

\subsection{Follow-up seeing-limited transit photometry} \label{sec:followuptransit}

\begin{table*}
\caption{Logbook of ground-based transit follow-up observations taken for TOI-1201.}
\label{tab:transitdata}
\centering
\begin{tabular}{llccccccc}
\hline \hline
\noalign{\smallskip}
Instrument & Country & Date & Filter  & Exposure & Duration\tablefootmark{a} & Pixel scale & N$_\textnormal{obs}$ & 10-min rms\tablefootmark{b} \\
& &  &  &  (s) & (min) & (arcsec) &  & (ppt)  \\
\noalign{\smallskip}
\hline 
\noalign{\smallskip}
LCO-SSO & Australia & 2019 Aug. 27 & $z_s$ &  60 & 331 & 0.389 & 306 & 1.27 \\
MKO CDK700 & Australia & 2019 Sep. 01 & $r'$ &  128 & 231 & 0.4 & 74 & 2.73 \\
LCO-SAAO & South Africa & 2019 Sep. 23 & $g'$ &  50 & 215 & 0.389 & 170 & 1.99 \\
OAA & Spain & 2019 Sep. 29 & $V$ &  240 & 234 & 1.44 & 56 & 1.57 \\
\noalign{\smallskip}
\hline 
\end{tabular}
\tablefoot{
\tablefoottext{a}{Time span of the observation.}
\tablefoottext{b}{Root mean square in parts-per-thousand.}
}
\end{table*}

We acquired four sets of full-transit ground-based time-series follow-up photometry of TOI-393 and TOI-1201 as part of the TFOP\footnote{\url{https://tess.mit.edu/followup}} to determine the source of the \tess\ transit-like signal detection, refine the transit ephemeris, and place constraints on transit depth across optical filter bands. We used the \texttt{TESS Transit Finder}, which is a customized version of the \texttt{Tapir} software package \citep{Jensen:2013}, to schedule our transit observations. The photometric data were extracted using the \texttt{AstroImageJ} %(\texttt{AIJ}) 
software package \citep{Collins:2017}. The transit observation logbook is presented in Table~\ref{tab:transitdata}, where technical details, such as facility, filter, exposure, or duration, are included. The following paragraphs describe the various data sets obtained from the ground-based facilities.

\paragraph{MKO CDK700.}
The first follow-up transit observation submitted to TFOP was from the Mount Kent Observatory (MKO) CDK700 telescope near Toowoomba, Queensland, Australia. The CDK700 is a Planewave Corrected Dall-Kirkham 0.7\,m telescope equipped with an Apogee Alta F16 (KAF-16803) detector resulting in a $27\times27$\,arcmin$^2$ field of view. The CDK700 data tentatively ruled out the event on TOI-393 and were suggestive of a shallow event in TOI-1201. However, the TOI-1201 detection was inconclusive due to the observational systematics when separating the two stars. For this reason, the MKO data were not included in our joint system modeling. 

\paragraph{LCOGT.}
% LCO-SSO and LCO-SAAO
We obtained two observations of TOI-393 and its $\sim8$\,arcsec neighbor TOI-1201 using the Las Cumbres Observatory Global Telescope (LCOGT) 1.0\,m network \citep{Brown2013_LCO}. The telescopes are equipped with $4096 \times 4096$\,pixel Sinistro cameras, resulting in a $26\times26$\,arcmin$^2$ field of view. The first LCOGT observation was scheduled according to the initial public TOI-393 ephemeris on 27 August 2019 at the Siding Spring Observatory (SSO) 1.0\,m node in Pan-STARRS $z_s$ band (LCO-SSO hereinafter). The stellar point-spread functions (PSFs) in the images had nominal full width at half maximum (FWHM) of $\sim1.7$\,arcsec. A photometric aperture with radius of 7\,pixels (2.7\,arcsec) was selected to separate most of the flux of TOI-393 and TOI-1201 in two different apertures. A transit-like event consistent with the TESS detection was ruled out in TOI-393, and a transit-like event arriving $\sim40$ minutes early, with depth $\sim2000$\,ppm, was confirmed in TOI-1201. We revised the follow-up ephemeris according to the LCO-SSO observation and observed another predicted full-transit observation on 23 September 2019 from the LCOGT 1.0\,m node at South Africa Astronomical Observatory (SAAO) in Sloan $g'$. A $\sim 2000$\,ppm transit-like event was detected arriving on-time relative to the revised ephemeris (denoted as LCO-SAAO). The combination of the $z_s$ and $g'$ $\sim 2000$\,ppm detections suggests that the transit-like event is achromatic, which reduces of the chances that the \tess\ signal had been caused by a false positive scenario.

\paragraph{OAA.}
We observed another full transit on 29 September 2019 with the 0.4\,m telescope at the Observatori Astron\`omic Albany\`a\footnote{\url{https://www.observatorialbanya.com/}} (OAA) near Girona, Spain, in Johnson $V$ band. The telescope is equipped with a Moravian G4-9000 camera with a field of view of $36\times36$\,arcmin$^2$ . TOI-393 and TOI-1201 could not be cleanly separated, so we selected a photometric aperture with radius 13\,arcsec that included most of the flux from both of the $\sim 8$\,arcsecond-wide neighbors. The transit-like event arrived on-time relative to the revised ephemeris.

\subsection{Long-term photometric monitoring} \label{sec:longtermphot}

We also assembled a list of archival long-time baseline data, namely from the WASP-South and ASAS-SN surveys. 

\paragraph{WASP-South.} 
TOI-1201 was observed with the Wide Angle Search for Planets (WASP) array of eight cameras at the South African Astronomical Observatory in Sutherland \citep[WASP-South;][]{Pollacco2006_WASP} over the course of six years from 2008 to 2014, amounting to about 64\,000 acquired data points. 
From 2008 to 2009, the 200\,mm lenses were used (camera 222, 13.7\,arcsec\,pixel$^{-1}$), and from 2012 to 2014, the data were taken with the 85\,mm lenses (camera 281, 32\,arcsec\,pixel$^{-1}$). 
The root mean square (rms) across all WASP-South cameras is 0.023\,mag.
Since the extraction aperture included both stars, it was not possible to use these data to unequivocally determine which star is producing the sinusoidal modulation found in the data, as examined in Sect.~\ref{sec:rotperiod}.

% 222: 200mm lenses; 2008 and 2009
% 281: 85mm lenses; 2012 to 2014

\paragraph{ASAS-SN.}
The All-Sky Automated Survey for Supernovae
(ASAS-SN) is composed of 24 14\,cm aperture Nikon telephoto lenses, each equipped with a $2048\times2048$ ProLine CCD camera, at locations distributed worldwide \citep{Shappee2014_ASAS, Kochanek2017_ASAS}.
We input the stellar coordinates for TOI-1201 as provided in Table~\ref{tab:stellarparams} to search the ASAS-SN database\footnote{\url{https://asas-sn.osu.edu/}}. Two sources associated with\ these coordinates with a separation of $\sim$3.2\,arcsec were found, so both light curves are surely contaminating each other (assuming that the source APJ024859.45--143214.2 corresponds to TOI-1201 and the other source AP37838964 to the companion). 
Because the ASAS-SN pixel-scale is 8\,arcsec\,pixel$^{-1}$ and the aperture is around 15\,arcsec, blending is an issue.
The extracted data consist of a total of 619 points spanning roughly 1600\,d from March 2012 to January 2019 in four cameras (ba, bb, be, bf), all in the $V$ band, with a combined rms of 0.018\,mag. 

% stellar parameters
\begin{table*}
\centering 
\small
\begin{center}
\caption{Stellar parameters of TOI-1201 and its 8\,arcsecond-wide companion.}
\label{tab:stellarparams}
\centering
\begin{tabular}{lccl}
\hline \hline
\noalign{\smallskip}
Parameter & Primary & Secondary & Reference \\
\noalign{\smallskip}
\hline
\noalign{\smallskip}
\multicolumn{4}{c}{\textit{Names and identifiers}}\\
\noalign{\smallskip}
~~~Name & PM~J02489--1432W & PM~J02489--1432E & LG11  \\
~~~Karmn    & J02489--145W & J02489--145E             & Cab16      \\
~~~\gaia\ EDR3 ID  & 5157183324996790272 &  5157183324996789760  &\gaia\ EDR3          \\
%~~~2MASS ID & J10193634+1952122     & 2MASS                     \\
~~~ TOI & 1201 & 393 & ExoFOP-\tess\ \\
~~~ TIC & 29960110 & 29960109 & Sta18 \\
\noalign{\smallskip}
\multicolumn{4}{c}{\textit{Coordinates and spectral types}}\\
\noalign{\smallskip}
~~~$\alpha$ (J2000) & 02:48:59.27 & 02:48:59.83 & \gaia\ EDR3  \\ %  42.247723502 42.250091340
~~~$\delta$ (J2000) & --14:32:14.9 & --14:32:16.2 & \gaia\ EDR3\\  % -14.537277008 -14.53763994
~~~Sp. type\tablefootmark{a} &  M$2.0\pm0.5$\,V & M$2.5\pm0.5$\,V & This work \\ 
\noalign{\smallskip}
\multicolumn{4}{c}{\textit{Magnitudes}}\\
\noalign{\smallskip}
~~~$NUV$ (mag)      & 21.664 $\pm$ 0.337 & 21.923 $\pm$ 0.491 & \textit{GALEX} \\ 
~~~$B$ (mag)        & \ldots   &  14.111 $\pm$ 0.030 & UCAC4      \\
~~~$g'$ (mag)        & 13.7285 $\pm$ 0.0028   &  13.334 $\pm$ 0.020 & Pan-STARRS1/UCAC4      \\
~~~$G_{BP}$ (mag)   & 13.344 $\pm$ 0.025   & 13.694$\pm$ 0.014 & \gaia\ EDR3     \\
~~~$V$ (mag)        & \ldots   &  12.706 $\pm$ 0.020 & UCAC4      \\
~~~$r'$ (mag)   & 12.5537 $\pm$ 0.0010   &  12.110 $\pm$ 0.080 & Pan-STARRS1/UCAC4      \\ 
~~~$G$ (mag)        & 12.0888 $\pm$ 0.0072  & 12.3710 $\pm$ 0.0052 & \gaia\ EDR3     \\ 
~~~$i'$ (mag)        & 11.4326 $\pm$ 0.038 &  10.900 $\pm$ 0.010 & Pan-STARRS1/UCAC4      \\
~~~$G_{RP}$ (mag)   & 10.9748 $\pm$ 0.0092   & 11.2333 $\pm$ 0.0061  & \gaia\ EDR3     \\
~~~$J$ (mag)        & 9.528 $\pm$ 0.042   &  9.733 $\pm$ 0.024 & 2MASS\\
~~~$H$ (mag)        & 8.876 $\pm$ 0.057   &  9.125 $\pm$ 0.027 & 2MASS\\
~~~$K_s$ (mag)      & 8.646 $\pm$ 0.029   &  8.875 $\pm$ 0.021 & 2MASS\\
~~~$W1$ (mag)       & 7.721 $\pm$ 0.014  &  7.731 $\pm$ 0.011 & AllWISE/WISE     \\
~~~$W2$ (mag)       & 7.693 $\pm$ 0.013   &  7.670 $\pm$ 0.012 & AllWISE/WISE      \\
~~~$W3$ (mag)       & 7.918 $\pm$ 0.020   &  8.063 $\pm$ 0.019 & AllWISE/WISE      \\
~~~$W4$ (mag)       & 7.709 $\pm$ 0.129   &  7.794 $\pm$ 0.120 & AllWISE/WISE      \\
%~~~$r'$ (mag)& $7.8500\pm0.0010$ & APASS?\\
%~~~$i'$ (mag)& $7.7130\pm0.0020$ & APASS?\\
%~~~$z'$ (mag)& $7.4690\pm0.0020$ & APASS?\\

\noalign{\smallskip}
\multicolumn{4}{c}{\textit{Parallax and kinematics}}\\
\noalign{\smallskip}
~~~$\pi$ (mas)      & 26.571 $\pm$ 0.022  & 26.539 $\pm$ 0.023 & \gaia\ EDR3     \\
~~~$d$ (pc)         & 37.636 $\pm$ 0.032   & 37.680 $\pm$ 0.033  & \gaia\ EDR3     \\
~~~$\mu_\alpha \cos \delta$ (mas\,yr$^{-1}$)     & +164.069 $\pm$ 0.025   & +174.433 $\pm$ 0.033 &  \gaia\ EDR3   \\
~~~$\mu_\delta$ (mas\,yr$^{-1}$)                 & +46.549 $\pm$ 0.027   & +45.465 $\pm$ 0.029 & \gaia\ EDR3     \\
~~~$V_r$\tablefootmark{b} (k\ms) & +31.771 $\pm$ 0.018 & +31.868 $\pm$ 0.018 & This work \\
~~~$U$ (k\ms) & --38.495 $\pm$ 0.018 & --39.577 $\pm$ 0.020 & This work \\
~~~$V$ (k\ms) & --17.125 $\pm$ 0.018 & --18.521 $\pm$ 0.019 & This work \\
~~~$W$ (k\ms) & --12.652 $\pm$ 0.019 & --11.956 $\pm$ 0.020 & This work \\
~~~Galactic population & \multicolumn{2}{c}{Young disk} & This work \\
~~~Stellar kinematic group & \multicolumn{2}{c}{Hyades Supercluster} & This work \\
\noalign{\smallskip}
\multicolumn{4}{c}{\textit{Photospheric parameters}}\\
\noalign{\smallskip}
~~~$T_\textnormal{eff}$ (K)         & $3476 \pm 51$ & $3437 \pm 51$  & This work \\ 
~~~$\log g_\star$ (cgs)                 & $4.80\pm 0.04$ & $4.80\pm 0.04$ & This work \\ 
~~~[Fe/H] (dex)                     & $0.05\pm 0.16$ & $0.05\pm 0.16$ & This work \\ 
~~~$v \sin{i}$ (k\ms)              & $<$2 & $<$2 & This work \\
~~~H$\alpha$ index                        & 0.931$\pm$0.025 & 0.913$\pm$0.023 & This work \\
~~~pEW(H$\alpha$)                        & $-$0.437$\pm$0.053 & $-$0.419$\pm$0.050 & This work \\
~~~$P_\textnormal{rot}$ (d) & 19--23 & $\dots$ & This work \\
\noalign{\smallskip}
\multicolumn{4}{c}{\textit{Physical parameters}}\\
\noalign{\smallskip}
~~~$L_\star$ ($10^{-5}\ L_\odot$)            & 3400 $\pm$ 57 & 2683 $\pm$ 25 & Cif20 \\  
~~~$R_\star$ ($R_\odot$)            & 0.508 $\pm$ 0.016  & 0.462 $\pm$ 0.014 &  This work \\ 
~~~$M_\star$ ($M_\odot$)            & 0.512 $\pm$ 0.020 & 0.463 $\pm$ 0.018 & This work \\ 
~~~$\rho_\star$ (g\,cm$^{-3}$)       & $5.50^{+0.58}_{-0.49}$   & $6.40^{+0.75}_{-0.68}$ & This work \\
\noalign{\smallskip}
\hline
\end{tabular}
\tablefoot{\tablefoottext{a}{Photometrically derived spectral types.}
\tablefoottext{b}{RAVE DR4 \citet{rave} tabulated $\gamma$ = +6.6$\pm$6.1\,k\ms\ and $T_\textnormal{eff}$ = 3600$\pm$310\,K for the  primary and $\gamma$ = +31.7$\pm$1.4\,k\ms\ and $T_\textnormal{eff}$ = 3840$\pm$70\,K for the secondary. The RAVE spectrum of the primary had a very poor signal-to-noise ratio (S/N = 4.2), which led to wrong $V_r$ and stellar parameter determination (e.g., $\log{g}$ = 1.5$\pm$1.0, inconsistent with the star's main-sequence nature).}}
\tablebib{
    Cab16: \cite{Caballero2016}; 
    \gaia\ EDR3: \cite{GaiaEDR3}; 
    \textit{GALEX}: \cite{Bianchi2011};
    Sta18: \cite{Stassun2018};
    UCAC4: \cite{Zacharias2013};
    2MASS: \cite{Skrutskie2006_2MASS}; 
    Cif20: \cite{Cifuentes2020}; 
    LG11: \cite{LepineGaidos2011};
    Pan-STARRS1: \cite{Kaiser2010};
    WISE/AllWISE: \cite{Cutri2012,Cutri2014}.
} 
\end{center}
\end{table*}

\subsection{High-resolution spectroscopy with CARMENES} \label{sec:carmenes}
\begin{table}
\caption{CARMENES spectroscopic observations for TOI-1201 and its companion.}
\label{tab:rvobservations}
\centering
\scriptsize
\begin{tabular}{llcccc}
\hline \hline
\noalign{\smallskip}
Target & Start date & End date & $N_\textnormal{obs}$ & $\sigma_\textnormal{RV}$\tablefootmark{(a)} & rms\tablefootmark{(b)}  \\
  & &  &  & (\ms) & (\ms) \\
\noalign{\smallskip}
\hline 
\noalign{\smallskip}
TOI-1201 & Nov. 2019 & Feb. 2020 & 33 & 2.22 & 7.95  \\
PM~J02489–-1432E & Nov. 2019 & Jan. 2020 & 23 & 2.18 & 3.33  \\
\noalign{\smallskip}
\hline 
\end{tabular}
\tablefoot{
\tablefoottext{a}{Median value for the uncertainty of the RV measurement.}
\tablefoottext{b}{Root mean square.}
}
\end{table}
We obtained 34 high-resolution spectra for the transit host TOI-1201 with the CARMENES\footnote{Calar Alto high-Resolution search for M dwarfs with Exo-earths with Near-infrared and optical \'Echelle Spectrographs, \url{http://carmenes.caha.es}} instrument located at the 3.5\,m telescope at the Calar Alto Observatory in Almer\'ia, Spain. 
We used only the data collected with the VIS channel, which covers the spectral range 520--960\,nm with a spectral resolution of $\mathcal{R}$ = 94\,600  \citep{CARMENES, CARMENES18}. 
One of the measurements was missing a drift correction and was therefore discarded, which resulted in 33 observations that were used for the analysis.
The companion at about 8\,arcsec (see Sect.~\ref{sec:binary}) was also observed by CARMENES 23 times from November 2019 to January 2020. Details on the quality of the CARMENES spectroscopic data are given in Table~\ref{tab:rvobservations}.
The simultaneously collected spectra from the near-IR channel (960--1710\,nm, $\mathcal{R}$ = 80\,400) had an rms of 25\,\ms\ and do not show any significant signals, and were, thus, not considered in the analysis as in other CARMENES works \citep[e.g.,][]{Bauer2020}.

% rv = np.loadtxt(fil,unpack=True,usecols=(1))
% val,valup,valdown = juliet.utils.get_quantiles(rv)
% print(valup-val, val-valdown) # the sigmas
% 8.521450874559438 10.1041863263425
% wrms for VIS: 6.8823
% wrms for VIS comp: 3.1196
% wrms for NIR: 25.2853
The data are pipelined through the standard guaranteed time observations data flow \citep{Caballero2016_SPIE}, and the resulting are measured using \serval\footnote{\url{https://github.com/mzechmeister/serval}} \citep{serval}. They are corrected for barycentric motion, secular acceleration, instrumental drift, and nightly zero-points using our standard approach \citep{serval,Trifonov2020}. 
The RVs with their uncertainties for TOI-1201 and its companion are tabulated in Tables~\ref{tab:rvdata} and~\ref{tab:rvdatacomp}, respectively. 
Additionally, \serval\ provides a list of stellar activity indicators, namely the chromatic index (CRX), differential line width (dLW), H$\alpha$ index, and the Ca~\textsc{ii} IR triplet (IRT).
Following \cite{Lafarga2020}, we applied binary masks to the spectra and computed the cross-correlation function (CCF) and its FWHM, contrast (CTR), and bisector velocity span (BVS) values. 
The pseudo-equivalent width (pEW), as defined in and provided by \cite{Schoefer2019}, of the photospheric lines TiO~$\lambda\ 7050\,\AA$, TiO~$\lambda\ 8430\,\AA$, and TiO~$\lambda\ 8860\,\AA$ were also derived from the CARMENES spectra.

\section{Stellar properties} \label{sec:stellarprops}
\begin{figure*}%[hbt!]
    \centering
    \includegraphics[width=0.95\linewidth]{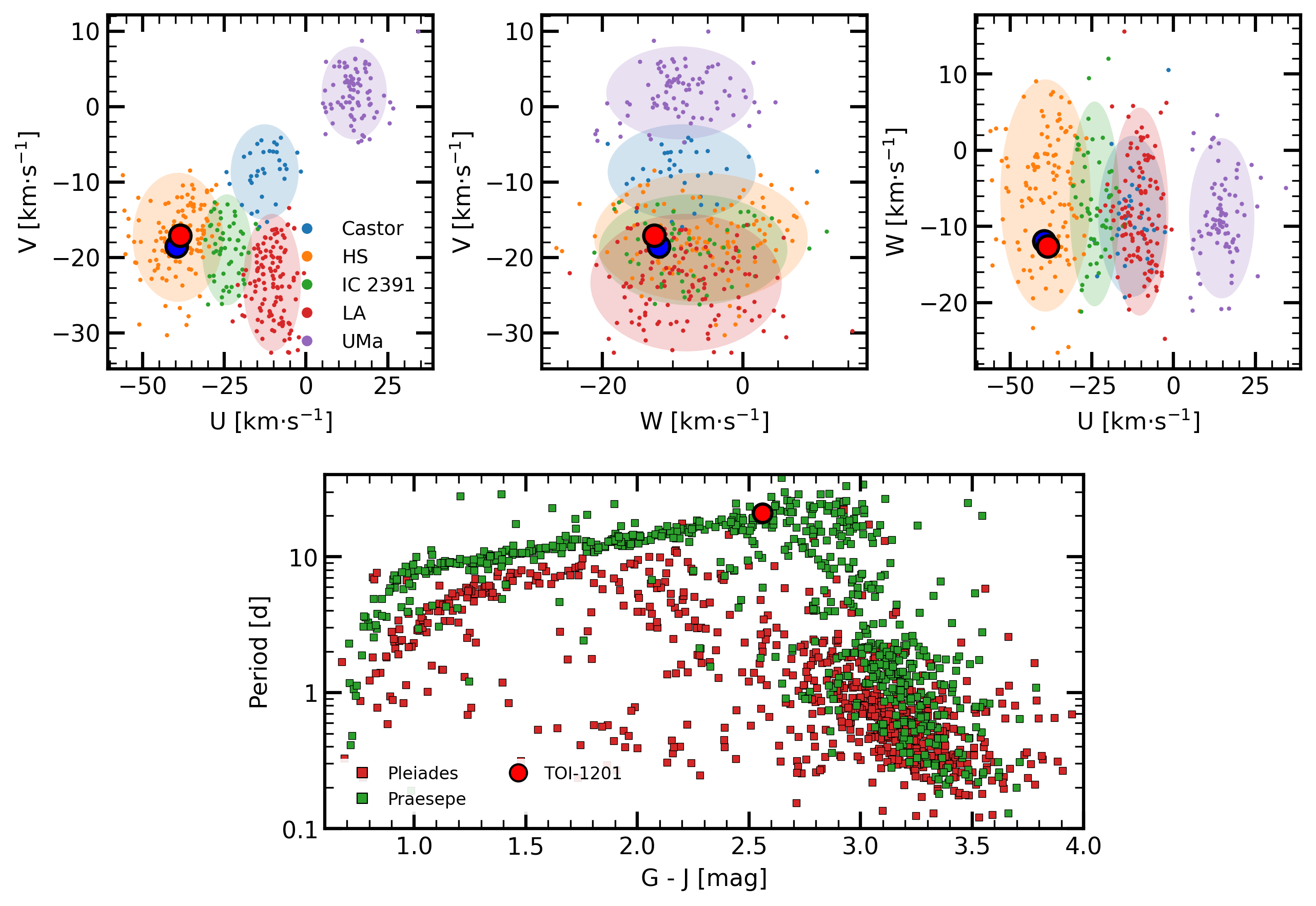}
    \caption{\textit{Top panel:} $ UVW $ velocities of TOI-1201 (red dot) and PM~J02489--1432E (blue dot) compared against members of young moving groups from \citet{Montes2001}, namely Castor (blue), the Hyades Supercluster (HS, orange), the IC~2391 Supercluster (green), the Local Association (LA, red), and Ursa Major (UMa, purple). The ellipses represent the 2$ \sigma $ value of the $UVW$ for each group. The two targets agree with the Hyades Supercluster. \textit{Bottom panel:} Rotational period of TOI-1201 alongside members of the Pleiades (in red) and Praesepe (in green) open clusters from \citet{Curtis2019} and references therein.}
    \label{fig:uvw}
\end{figure*}

\subsection{Basic astrophysical parameters}

Table~\ref{tab:stellarparams} summarizes the stellar parameters of TOI-1201 and its companion.
Both stars are poorly investigated late-type dwarfs \citep[e.g.,][]{LepineGaidos2011,Frith2013}.
We first got the equatorial coordinates, parallaxes, and proper motions of the two stars from \gaia\ Early Data Release 3 \citep[EDR3;][]{GaiaEDR3} and the blue-optical to mid-IR photometry already compiled by \citet{Cifuentes2020}.
Next, we integrated the spectral energy distributions as the latter authors for computing the bolometric luminosities, $L_\star$.
Following the $\chi^2$ stellar synthesis of \citet{Passegger2019} and using only the CARMENES VIS channel spectra, we derived the photospheric parameters $T_\textnormal{eff}$, $\log{g}$, and [Fe/H].
From them, with the Stefan-Boltzman law and the mass-radius relation of \cite{Schweitzer2019}, we obtained the stellar radii and masses, from which we determined the stellar bulk density. The mass-radius relation of \cite{Schweitzer2019} is valid only for objects older than a few hundred megayears, though it is applicable for the age of this system as we currently infer it (see below). Another systematic uncertainty might have originated from the accuracy of the models used for the determination of the photospheric parameters \citep{Passegger2020}. The presented relative uncertainties of 3--4\% in the radius and mass of the two M dwarfs stem from the measurement uncertainties of 1--2\% for the effective temperature and photometry of the stars, and, thus, were carried over accordingly.

We also used the same CCF methodology with the weighted binary masks of \citet{Lafarga2020} for computing the stellar RVs, $V_r$, on the CARMENES spectra.
Using this value and the \gaia\ astrometry, we calculate the $UVW$ galactocentric space velocities as in \citet{JohnsonSoderblom1987}. The $UVW$ values (see Fig.~\ref{fig:uvw}) for the two stars indicate that this system belongs to the young disk (Cort\'es-Contreras 2021, in prep.) and suggest a membership to the Hyades Supercluster \citep{Eggen1958,Eggen1984} when compared to the kinematics of other members of this young moving group from \citet{Montes2001}.

The rotational period of 19--23\,d for TOI-1201 (see Sect.~\ref{sec:rotperiod}) matches other members of the Praesepe open cluster \citep[600--750~Myr;][and references therein]{Douglas2019} with similar effective temperature as TOI-1201 \citep[][Fig.~\ref{fig:uvw} in this paper]{Curtis2019}. Furthermore, the H$\alpha$ feature in faint emission \citep{Jeffers2018,Schoefer2019} is also compatible with members of the Praesepe and the Hyades open cluster \citep[$ \sim$600--800\,Myr --][and references therein]{Martin2018,Lodieu2018b,Douglas2019,Lodieu2019b,Lodieu2019a} with $T_\textnormal{eff}$ $\sim$3500\,K \citep{Fang2018}, which supports the membership of this system to the Hyades Supercluster and, thus, implies its age of $\sim$600--800\,Myr.

We also computed projected rotation velocities, $v \sin i$, on the CARMENES spectra as \citet{Reiners2018b} and estimated spectral types M2.0\,V and M2.5\,V for the primary and secondary, respectively, from absolute magnitudes and colors as \citet{Cifuentes2020}, consistent with the effective temperatures.

\subsection{The stellar host and its companion} \label{sec:binary}

The earliest name in the literature of TOI-1201 is \object{PM~J02489--1432W}.
At only about 8\,arcsec to the east, \citet{LepineGaidos2011} tabulated a second star, \object{PM~J02489--1432E}, slightly fainter (by 0.28\,mag and  0.20\,mag in the $G$ and $J$ bands, respectively) and redder (by 0.08\,mag in $G-J$  color). 
While both stars are located at the same \gaia\ parallactic distance ($\Delta \pi = 0.032\pm0.032$\,mas) and have the same RVs ($\Delta V_r = 0.097\pm0.025$\,km\,s$^{-1}$), their proper motions are similar but not identical ($\Delta \mu_\alpha \cos{\delta} = 10.340\pm0.041$\,mas\,yr$^{-1}$, $\Delta \mu_\delta = 1.084\pm0.040$\,mas\,yr$^{-1}$).
Despite the total proper motion difference of only 6\,\%, the pair would pass any kinematic and astrometric binarity criterion \citep[e.g.,][]{Montes2018}, but such a difference may be indicative of a relative orbital motion as shown by, for example, \citet{MakarovKaplan2005}.

The Washington Double Star catalog \citep[WDS;][]{Mason2001} tabulates the pair as WDS~J02490--1432 (KPP~2871).
Although first reported by \citet{LepineGaidos2011} and next tabulated by \citet{El-BradyRix2018}, it was \cite{KnappNanson2019} who performed the first multi-epoch astrometric analysis and gave the WDS discovery designation (KPP~2871).
However, as of 19 April 2021, WDS listed only five independent epochs between 1998.6 and 2015.5.
For complementing those data, we retrieved SuperCOSMOS \citep{Hambly2001} digitizations of First Palomar Observatory Sky Survey (POSS-I) and United Kingdom Schmidt Telescope (UKST) photographic plates taken  between December 1955 and September 1989, and measured angular  separation $\rho$ and position angle $\theta$ between the primary and  secondary as in \citet{Caballero2010}.
Besides, we recomputed the angular separation $\rho$ and position angle $\theta$ for the 2MASS \citep{Skrutskie2006_2MASS}, DENIS \citep{denis}, and \gaia\ DR1, DR2, and EDR3 \citep{gaiadr1,gaiadr2summary,GaiaEDR3} observations.
The 11 resulting astrometric epochs, which cover 61.97\,yr, are displayed in Table~\ref{tab:binary}.
The pair was not resolved, though, by a number of all-sky surveys: GSC2.3/USNO-A2/USNO-B1, SDSS, UKIDSS LAS, WISE \citep[e.g.,][]{Monet2003,Lawrence2007,Alam2015, Marocco2021}.

In the six decades of astrometric observations of the binary, there has been an appreciable increase in $\rho$ from 7.2\,arcsec to  8.4\,arcsec.
Unfortunately, any variation in $\theta$ has been masked by the  uncertainty of the individual observations, typically larger than 1\,deg.
% s = rho * d > 8.391 * 1000/26.3653
% a ~ s > 318.3 +/- 1.3 au
% (M1+M2) * P^2 = a^3
% M1 = 0.509 +/- 0.020 Msol, M2 = 0.467 +/- 0.020 Msol
% (M1+M2) = 0.976 +/- 0.028 Msol
Taking the \gaia\ EDR3 value as the minimum angular separation of the pair, and at the stars heliocentric distance, we obtain $s$ = 316.4$\pm$1.3\,\au, which is also the minimum semimajor axis, $a$, of the binary.
Assuming a circular orbit, together with the individual masses of the primary and secondary in Table~\ref{tab:stellarparams} and Kepler's third law, the minimum orbital period $P_\textnormal{orb}$ of the binary would be 5709$\pm$86\,yr.
The astrometric monitoring of 61.97\,yr, thus, represents only $\sim$1\,\% of the orbit in time.
Certainly, a prohibitively long astrometric and RV follow-up will be needed to dynamically characterize the system in detail, but a physical separation of about 320\,\au\ between the two nearly identical stars (mass ratio $0.904\pm0.027$) may impose restrictions on the original protoplanetary disk size and planet stability in the system.

\subsection{Rotation period} \label{sec:rotperiod}

To determine the stellar rotation period, we considered the available photometric data in the archive from the WASP and ASAS-SN catalogs, as well as various stellar activity indicators provided by the spectra.

\paragraph{Long-term photometry.}
Solely focusing on the WASP data first, we found a signal of 21 days in the periodogram.
We confirmed this periodicity by implementing in \juliet\ (described in Sect.~\ref{sec:transitonlymodeling}) a rotation term analog to the one in \texttt{celerite2}\footnote{\url{https://celerite2.readthedocs.io/en/latest/}} \citep{celerite2}, which is the sum of two stochastically driven, damped harmonic oscillator (SHO) terms. The power spectrum of each SHO term was given by \cite{Anderson1990}, 

\begin{subequations} \label{eq:dsho}
\begin{align}
    \textnormal{SHO}_1(\omega_\textnormal{GP}) = \sqrt{\frac{2}{\pi}} \frac{S_0\omega_1^4}{(\omega^2_\textnormal{GP}-\omega^2_1)^2 + \omega^2_1\omega_\textnormal{GP}^2/Q^2_1}
\end{align}\\
\textnormal{and}\\
\begin{align}
    \textnormal{SHO}_2(\omega_\textnormal{GP}) = \sqrt{\frac{2}{\pi}} \frac{S_0\omega_2^4}{(\omega^2_\textnormal{GP}-\omega^2_2)^2 + \omega^2_2\omega_\textnormal{GP}^2/Q^2_2},
\end{align}\\
\textnormal{for which we applied the reparametrization using the hyperparameters}\\
\begin{align}
    Q_1      & = 0.5 + Q_0 + \delta Q\\
    \omega_1 & = \frac{4\pi Q_1}{P_\textnormal{rot}\sqrt{4Q_1^2 -1}}\\
    S_1     & = \frac{\sigma_\textnormal{GP}^2}{(1+f)\omega_1 Q_1}\\
    Q_2     & = 0.5 + Q_0\\
    \omega_2 & = 2\omega_1  = \frac{8\pi Q_1}{P_\textnormal{rot}\sqrt{4Q_1^2 -1}}\\
    S_2 & = \frac{f \sigma_\textnormal{GP}^2}{(1+f)\omega_2 Q_2},
\end{align}
\end{subequations}

\noindent and where $\sigma_\textnormal{GP}$ is the amplitude of the Gaussian process (GP) kernel, $P_\textnormal{rot}$, is the primary period of the variability, $Q_0$ is the quality factor for the secondary oscillation, $\delta Q$ is the difference between the quality factors of the first and second oscillations, and $f$ represents the fractional amplitude of the secondary oscillation with respect to the primary one. These stated hyperparameters were used in the analysis and are listed in Table~\ref{tab:priors_rvonly}. Such a kernel choice is well suited to represent stellar signals modulated by the rotation period of the star because of its flexible nature and smooth variations \citep[e.g.,][]{Medina2020}. We adopted the double SHO (dSHO) because the SHO alone lacks the ability to model the presence of more than one stellar spot \citep{JeffersKeller2009}. 

\begin{figure}
    \centering
    \includegraphics[width=\linewidth]{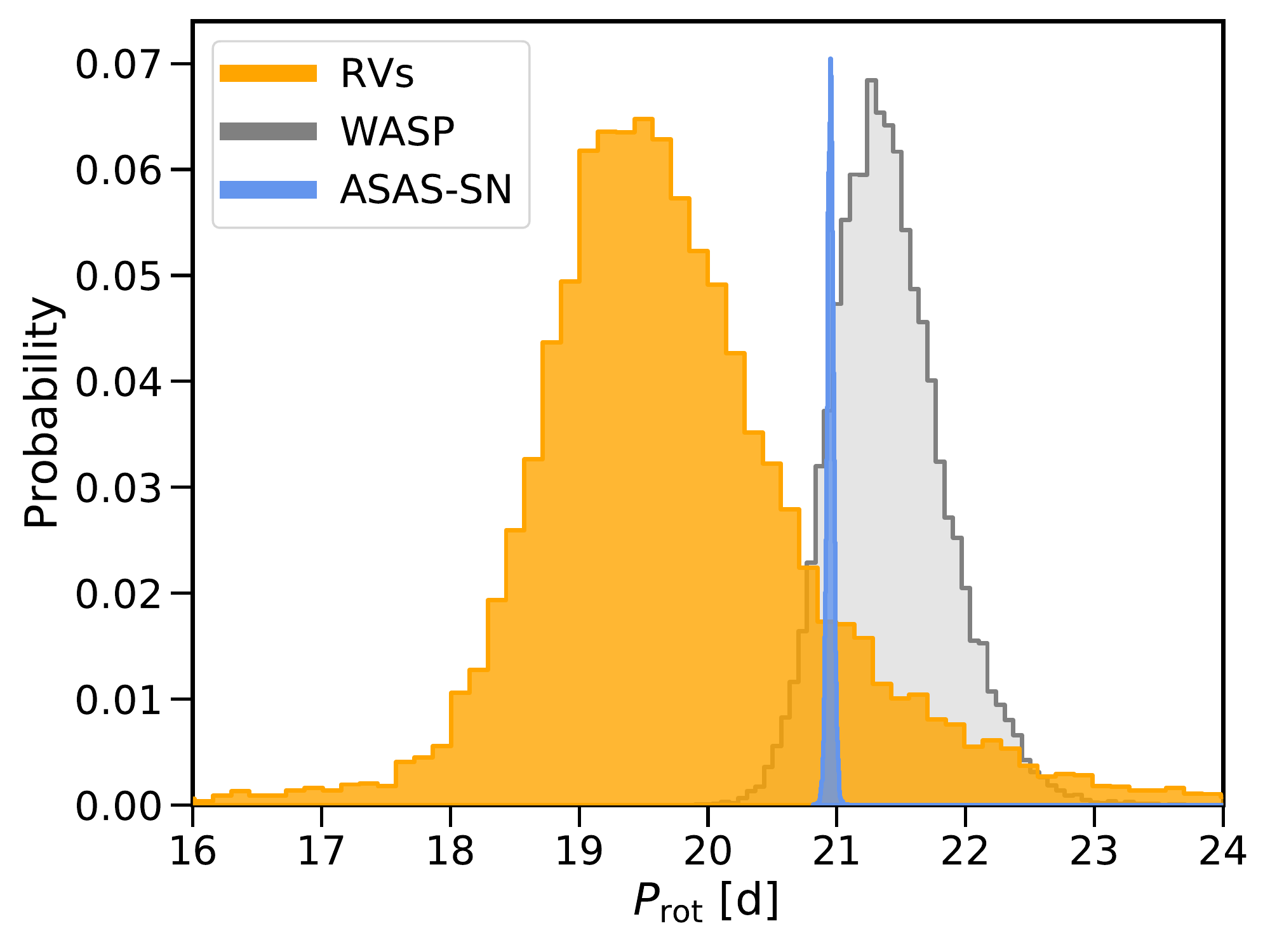}
    \caption[Posterior probability density for the estimated stellar rotation period for TOI-1201]{Probability density of the posterior samples for the estimated rotational period of the star ($P_\textnormal{rot}$) when using a dSHO-GP to fit the WASP (gray, 2008--2014) and ASAS-SN (blue, 2012--2019) data. The posterior distributions for the RVs (orange, 2019--2020) come from the dSHO-GP component in the final joint fit (Sect.~\ref{sec:jointfit}).%Note to self: rv prot comes from final joint fit transit + RV 2pl+dshogp-19d, asas from dshogp, wasp from dshogp
    }
    \label{fig:protposteriors}
\end{figure}

The WASP data set consists of 11 seasons of data, where each was considered to be an independent instrument with its own instrumental offset and jitter term. The data of each instrument were nightly binned in order to cut down on computational time, which does not create any problems because we are searching for signals on the order of a few days. 
From the posterior results (Fig.~\ref{fig:protposteriors}), we obtained a photometric stellar rotation period of $P_\textnormal{rot} = 21.37\pm0.46$\,d. However, as already mentioned in Sect.~\ref{sec:longtermphot}, both stars fall on the same pixel making it impossible to distinguish whether the signal belongs to TOI-1201 or its companion, or even a combination of both. However, the posterior distributions from the RVs suggest it to be an individual signal originating from TOI-1201. 

Moving on, we followed the same analysis as on the WASP data set for the ASAS-SN data. There were four different cameras (ba, bb, be, bf), which we treated each as individual instruments with their own offsets and jitter terms, and the GP hyperparameters were shared. The fit produces a GP rotational period of $P_\textnormal{rot} = 20.951\pm 0.025$\,d. The quoted uncertainty is likely much less than the true error given that star-spot patterns evolve over time and, therefore, the rotational modulation is not coherent.
Like before, the data support a signal at 21\,d but that cannot be attributed to one of the two objects. In addition, the periodogram shows a sharp peak at $\sim$400\,d that might be attributed to a magnetic cycle of the star, though this was not further investigated. 

We also considered the \tess\ light curve before being corrected for systematics (\texttt{SAP} flux) and simply divided the \tess\ sector 4 light curve into three chunks, which we shifted using a sinusoidal signal with a period of 20.5$\pm$0.5\,d and amplitude variations of $\sim$6.7\,ppt (Fig.~\ref{fig:longtermtess}). These data also support the 21\,d signal, though yet again, the \tess\ pixel (21\,arcsec) contains both objects. 

\paragraph{Spectroscopy.} 
We additionally inspected the various activity indicators from the spectroscopic observations to further affirm the stellar activity presence. 
TOI-1201 is a weakly H$\alpha$ active star (see Table~\ref{tab:stellarparams}), but only shows a very moderate level of activity that appears to be relatively stable.

\begin{figure*}
    \centering
    \includegraphics[width=1\linewidth]{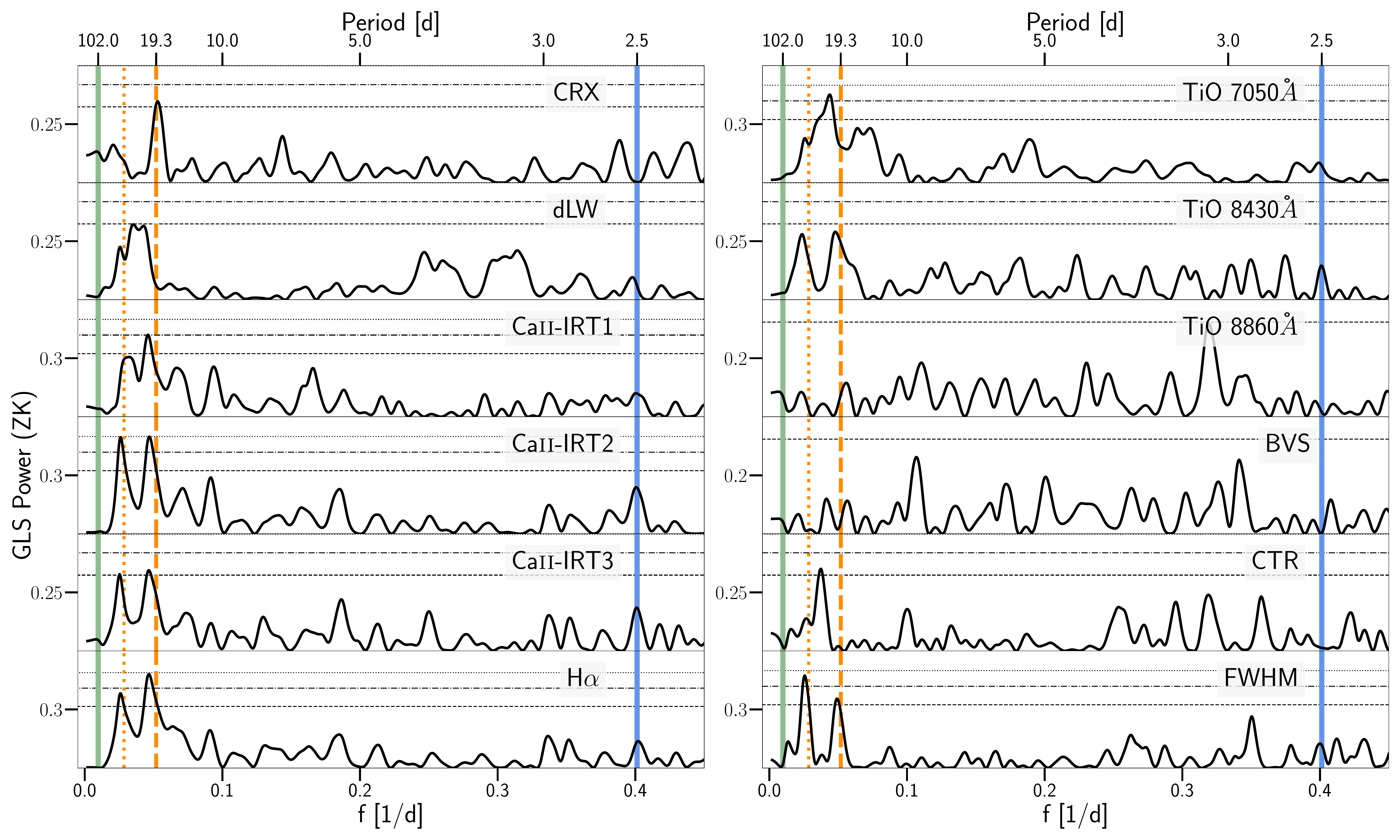}
    \caption[GLS periodograms of the various stellar activity indicators from the CARMENES spectroscopic data for TOI-1201.]{
    GLS periodograms of the various stellar activity indicators from the CARMENES spectroscopic data for TOI-1201. The vertical solid blue line corresponds to the transiting planet (2.49\,d) and the green one to the long-term signal ($\sim$102\,d) found in the RVs (Sect.~\ref{sec:rvonlymodeling}). The vertical dashed and dotted orange lines correspond to the rotational period present in the RVs ($\sim$19\,d) and its alias ($\sim$35\,d), respectively. The horizontal dotted, dot-dashed, and dashed lines represent the 10\,\%, 1\,\%, and 0.1\,\% FAP levels. The window function for the data set is displayed in Fig.~\ref{fig:rvgls}.
    %note: this is not using bootstrap but pyperiod powerlevel function aka theoretical levels as in Zechmeister2009.
    }
    \label{fig:glsperiodogramtoi1201activity}
\end{figure*}

Correlations between the RVs and all activity indicators were computed using Pearson's p-probability and no strong or moderate correlations were found.
Notably, the chromospheric indicators (H$\alpha$ index, Ca~\textsc{ii}~IRT have the tendency to show similar periods.
The periodicities in the generalized Lomb-Scargle  (GLS) periodograms show power ranging from 19\,d to 23\,d (Fig.~\ref{fig:glsperiodogramtoi1201activity}), hinting toward the 21\,d signal found in the photometry. 
The CRX, dLW, H$\alpha$ index, Ca~\textsc{ii}~IRT, FWHM, TiO~$\lambda\ 7050\,\AA$, and TiO~$\lambda\ 8430\,\AA$ all indicate some power in the range 19--23\,d, with power of sometimes $>$1\,\% false alarm probability (FAP), or broadly around $\sim$35--40\,d, corresponding to an alias of the $\sim$21\,d signal due to the sampling frequency of 41\,d found in the window function. The FAP levels were computed following the theoretical levels \citep[Eq.~24 in][]{Zechmeister2009_GLS}.
The remaining activity indicators do not show any peaks of interest.
This is consistent with the results of \cite{Lafarga2021}, who found that periodicities from activity indices depend on the mass and activity level of the star. They reported that, for less active stars, chromospheric lines are more likely to indicate the true rotation period.

Additionally, the RVs themselves exhibit a peak at around 19\,d (Fig.~\ref{fig:rvgls}), which is coincident with the CRX periodicity. When constructing our final model described in Sect.~\ref{sec:rvonlymodeling}, we took all the mentioned evidence above into consideration to assume this signal to have quasi-periodic behavior. This produced a periodicity of $P_\textnormal{rot,\ RV} = 19.62^{+1.10}_{-0.81}$\,d and is plotted in comparison to the long-term photometric results in Fig.~\ref{fig:protposteriors}.

To summarize, we see a $\sim$21\,d signal in the photometry, a corresponding 19\,d signal in the RVs, and a number of activity indicators peaking between 19--23\,d.
We, therefore, establish the rotational period of TOI-1201 to be 19--23\,d.

\section{Determination of planetary parameters for TOI-1201~b} \label{sec:results}

\subsection{Transit-only modeling} \label{sec:transitonlymodeling}

To first obtain refined values for the orbital period and transit time of the transiting planet, we performed fits on just the given transit photometry mentioned earlier in Sects.~\ref{sec:tessphot} and~\ref{sec:followuptransit}.  

All modeling fits for this paper were done using \juliet\footnote{\url{https://juliet.readthedocs.io/en/latest/}} \citep{juliet}, a python fitting package for joint-modeling (transit and RV) that uses nested samplers to explore the prior volume in order to efficiently compute the Bayesian model log evidence, $\ln \mathcal{Z}$. 
Though there are a variety of nested samplers, we employed the dynamic nested sampling algorithm provided by \texttt{dynesty}\footnote{\url{https://github.com/joshspeagle/dynesty}} \citep{dynesty,dynesty2020} due to the large parameter space of our models. The \juliet\ tool makes use of the established python packages \texttt{radvel}\footnote{\url{https://radvel.readthedocs.io/en/latest/}} \citep{radvel} and \texttt{batman}\footnote{\url{https://www.cfa.harvard.edu/~lkreidberg/batman/}} \citep{batman} to model the RVs and transits, respectively. The GP kernels are implemented via \texttt{george}\footnote{\url{https://george.readthedocs.io/en/latest/}} \citep{george} and \texttt{celerite}\footnote{\url{https://celerite.readthedocs.io/en/stable/}} \citep{celerite}. 
For model comparison, we followed the general rule that if $\Delta \ln \mathcal{Z}$ $\lesssim$ 2.5, then the two models are indistinguishable and neither is preferred so the simpler model would then be chosen \citep[e.g.,][]{Trotta2008,Feroz2011}. For any $\Delta \ln \mathcal{Z}$ that is greater than 2.5, the one with the larger Bayesian log evidence is moderately favored. Values greater than 5 indicate an even stronger evidence advocating for the winning model. 

For the transit model, we applied the following parameter transformations. 
Instead of fitting directly for the planet-to-star radius ratio, $p \equiv R_p/R_\star$, and the impact parameter of the orbit, $b$, we fit for the parameters $r_1$ and $r_2$ \citep[introduced by][]{Espinoza2018_pb} to ensure uniform sampling of the $b$-$p$ plane. 
Additionally, we took advantage of our derived stellar parameters coming from newer and more precise data (i.e., from \gaia\ EDR3) to better constrain $a/R_\star$ from the transiting light curve. Therefore, the scaled semimajor axis, $a/R_\star$, was replaced and re-transformed by the stellar density, $\rho_\star$, as given in Table~\ref{tab:stellarparams}. 
The \tess\ data were modeled jointly with a quadratic limb-darkening law (i.e., the $q_1$ and $q_2$ parameters were shared among both \tess\ sectors), while the ground-based instruments were assigned linear limb-darkening laws, both parametrized with the uniform sampling scheme of \citet{Kipping2013_ldc}.
The decision to use a two-parameter law for \tess\ and a linear one for the ground-based instruments was based on the work of \cite{EspinozaJordan2016}.
While other studies recommended the use of alternative limb-darkening  laws, especially for M dwarfs \citep{Morello2017,Maxted2018}, we  performed some numerical tests showing that the choice of law has a negligible impact relative to the parameters precision ($<$0.2 $\sigma$).
Instrumental offsets were considered, as well as instrumental jitter terms, which were added in quadrature to the given instrumental measurement uncertainty. The dilution factor, or the amount that a light curve is diluted due to neighboring light pollution, was fixed to one (i.e., no significant flux contamination) for each instrument. The \tess\ light curve provided by the SPOC pipeline (Sect.~\ref{sec:tessphot}) already takes this factor into account and the apertures of the ground-based instruments were not affected substantially by the neighboring bright companion. We performed fits with a free dilution factor to verify this assumption ($\Delta\ln{\mathcal{Z}} = \ln{\mathcal{Z}}_{D=1.0} - \ln{\mathcal{Z}}_{D \neq 1.0}>6$). 

To detrend time-correlated noise in the \tess\ sector 31 data, we adopted the squared-exponential GP kernel
\begin{equation*}\label{eqn:squaredexponentialkernel}
k_{i,j}(\tau)=\sigma^2_\textnormal{GP}\exp\left(-\frac{\tau}{T_\textnormal{GP}}\right) 
\end{equation*}
\noindent where $\tau = |t_{i} - t_{j}|$ is the temporal distance between two points, $\sigma_\textnormal{GP}$ is the amplitude of the GP modulation, and $T_\textnormal{GP}$ is the characteristic timescale. This kernel is characterized as being smooth since it is infinitely differentiable, which is sufficient for the \tess\ sector 31 data but not for the data from \tess\ sector 4. 
To account for the roughness in the \tess\ sector 4 data, we applied the (approximated) M\'atern-3/2 kernel provided by \texttt{celerite}
\begin{equation*}\label{eqn:matern32kernel}
k_{i,j}(\tau)=\sigma^2_\textnormal{GP}\left(1 + \frac{\sqrt{3}\tau}{\rho_\textnormal{GP}}\right)\exp\left(-\frac{\sqrt{3}\tau}{\rho_\textnormal{GP}}\right),
\end{equation*}
where $\tau = |t_{i} - t_{j}|$ is again the temporal distance between two points and $\sigma_\textnormal{GP}$ is the amplitude of the GP, but now $\rho_\textnormal{GP}$ is the length scale of the GP modulations to vary the smoothness of the return functions. Both \tess\ sectors had their own respective GP amplitude ($\sigma_\textnormal{GP}$). As for the ground-based transit photometry, we detrended the LCO-SAAO and LCO-SSO light curves with airmass using a linear term. Based on the peaks using the transit-least-squares\footnote{\url{https://github.com/hippke/tls}} method \citep{Hippke2019}, we set up the prior on the period to be uniform, $\mathcal{U}(2.45,2.55)$, and the ephemeris to be focused on the last transit in the data, $\mathcal{U}(2459169.2,2459169.3)$. 
When combining all the transit observations, we determined $P_b = 2.49198582\pm0.0000029$\,d and $t_{0,b}$ (BJD-TDB) $= 2459169.23219^{+0.00049}_{- 0.00047}$ (barycentric Julian date-temps dynamique barycentrique). These posterior values then served as a guide for the priors for the final joint fit in Sect.~\ref{sec:jointfit}.

\subsection{RV-only modeling} \label{sec:rvonlymodeling}

\begin{figure}%[hbt!]
    \centering
    \includegraphics[width=1\linewidth]{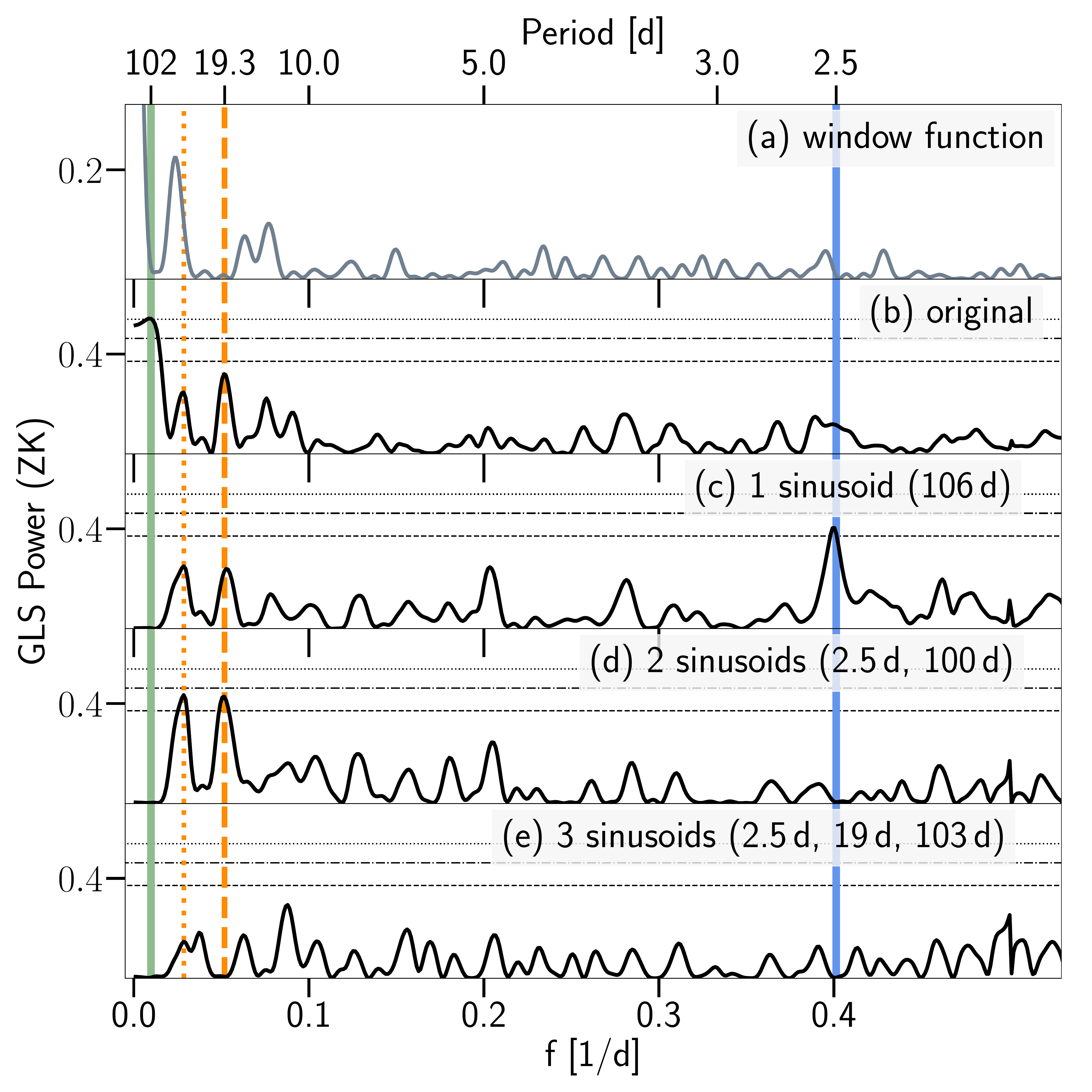}
    \caption[GLS periodograms of the RV residuals after subtracting different models for TOI-1201.]{
    GLS periodograms of the RV residuals after subtracting different models for TOI-1201. The vertical solid blue line corresponds to the transiting planet (2.5\,d) and the green one to the long-term signal ($\sim$102\,d). The vertical dashed and dotted orange lines correspond to the rotational period picked up in the RVs ($\sim$19\,d) and its alias ($\sim$35\,d), respectively. Panel (a): Window function of the data set. Panel (b): No signal fitted, just the original RVs. Panel (c): Residuals after subtracting the long-term signal at $\sim$102\,d. Panel (d): Residuals after subtracting a simultaneous model fit of two signals at 102\,d and 2.5\,d. Panel (e): Residuals after subtracting a simultaneous model fit of three signals at 102\,d, 19\,d, and 2.5\,d. The horizontal dashed gray lines represent the FAP levels of 10\,\%, 1\,\%, and 0.1\,\% (from bottom to top).}
    \label{fig:rvgls}
\end{figure}

To search for signals within the RV data, we first calculated the GLS periodograms \citep{Zechmeister2009_GLS}, approaching the data as if we had no prior information on the transiting planet. 
We initially employed the \exostriker\footnote{\url{https://github.com/3fon3fonov/exostriker}} \citep{exostriker} to identify potential combinations of the signals present in the GLS that describe the data appropriately. 
We used this information to build the priors for our RV-only \juliet\ runs.
Fig.~\ref{fig:rvgls} shows a sequence of these GLS periodograms after subtracting an increasing number of sinusoidal signals. 

The 2.5\,d signal only significantly appears after subtracting out the dominating long-term signal at around $\sim$102\,d (Fig.~\ref{fig:rvgls}~b). The residuals from simultaneously fitting two sinusoids for the 2.5\,d and $\sim$102\,d signals then show two prominent, although not significant, peaks at 19\,d and 35\,d, which are aliases of one another (induced by a peak at 41\,d found in the window function). We attempted to determine which one of the two is the true signal using \texttt{AliasFinder}\footnote{\url{https://github.com/JonasKemmer/AliasFinder}} \citep{Stock2020_aliasfinder,Stock2020_yzceti}. The algorithm follows the alias-testing method described by \cite{DawsonFabrycky2010}, in which the periodograms of simulated time series, injecting either of the two aliasing signals, are compared to the periodogram of the original data set. The injected signal that can best reproduce the data is deemed as the true period. Still, it was not conclusive which one was the correct signal. % Both m=1 and m=2 were tested out 
Additionally, we performed some \juliet\ runs using both 19\,d or 35\,d as the true period, but these were also inconclusive, as expected. 
Nonetheless, as was already presented in Sect.~\ref{sec:rotperiod}, the 19\,d signal is consistent with the rotational period of the star based on ground-based photometry, as well as other stellar activity indicators. After removing this 19\,d signal with a sinusoid fit, the resulting residuals show no further peaks in the GLS periodogram that would indicate the presence of additional signals (Fig.~\ref{fig:rvgls}). 

To sum up, the CARMENES RV data show three significant signals: the transiting planet (2.5\,d), the stellar rotation period ($\sim$19\,d), and a long-term signal ($\sim$102\,d). The next step was to compare various models to test whether circular or eccentric Keplerian orbits were preferred, if the stellar activity indicators (e.g., CRX, dLW) should be included as linear terms or not, and to check what kind of impact a GP kernel may have.
The runs are listed in Table~\ref{tab:modelcomparison} and the priors in Table~\ref{tab:priors_rvonly}.

\begin{figure*}
\centering
\begin{minipage}{0.92\textwidth}
  \centering
  \includegraphics[width=1\linewidth]{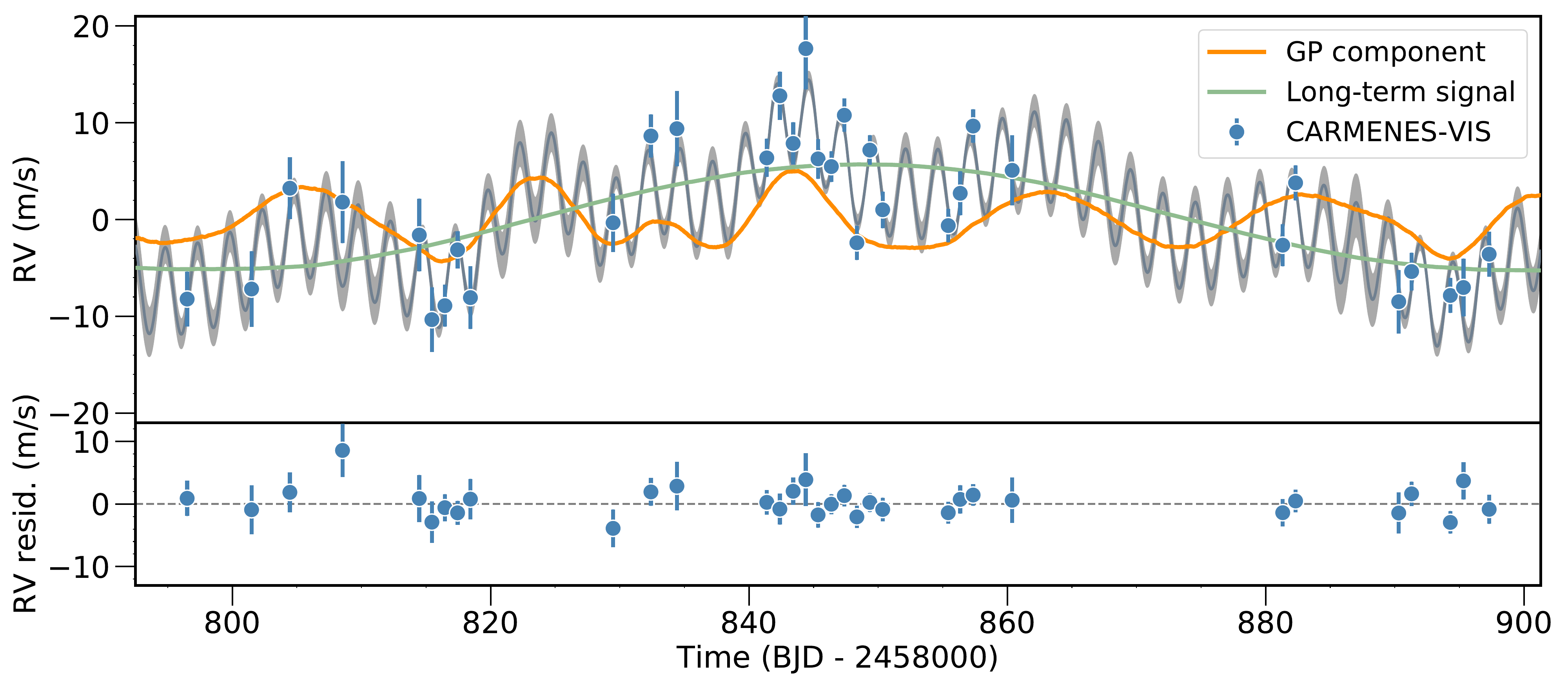}
\end{minipage}
\begin{minipage}{.45\textwidth}
  \centering
  \includegraphics[width=1\linewidth]{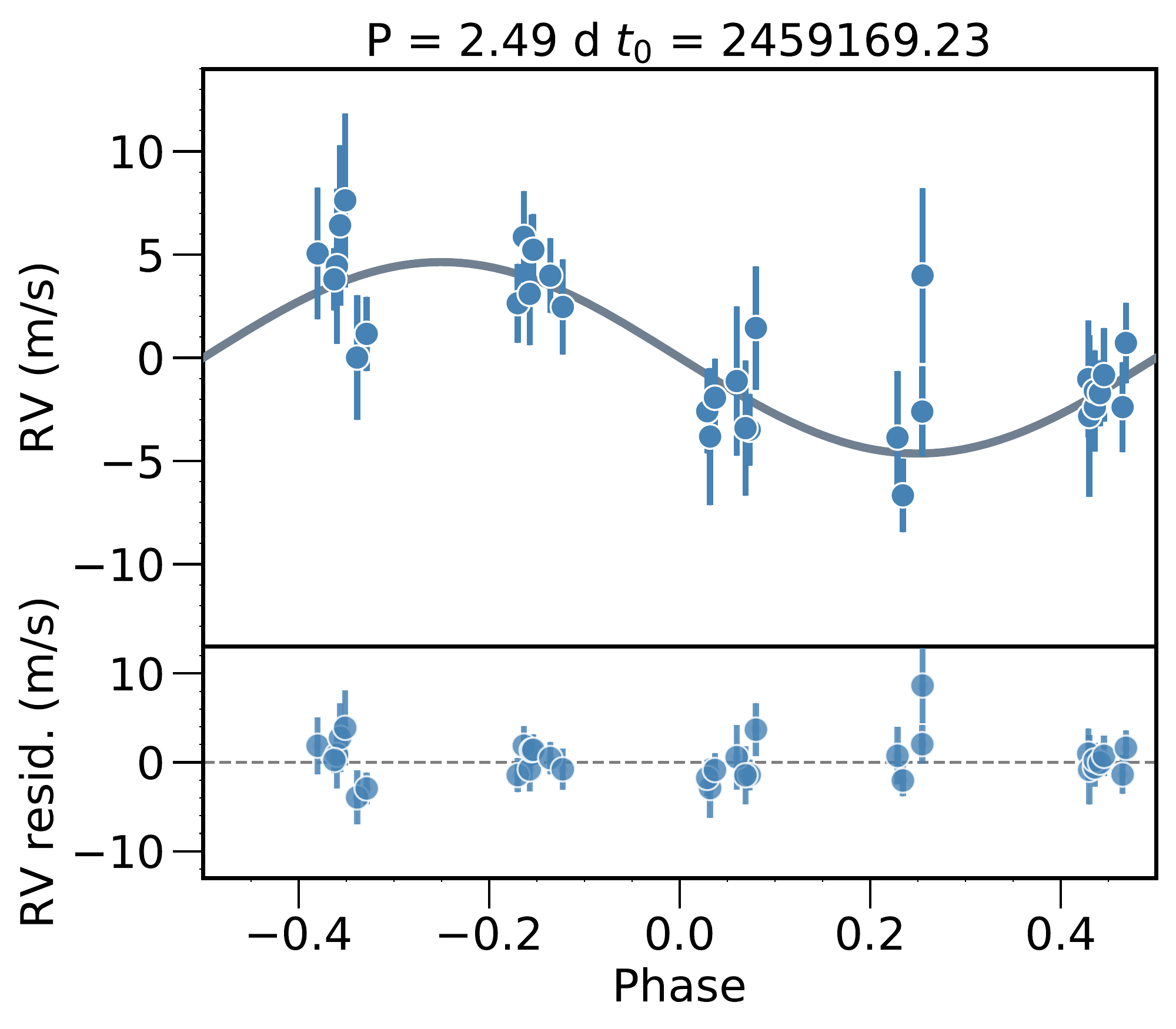}
\end{minipage}
\begin{minipage}{.45\textwidth}
  \centering
  \includegraphics[width=1\linewidth]{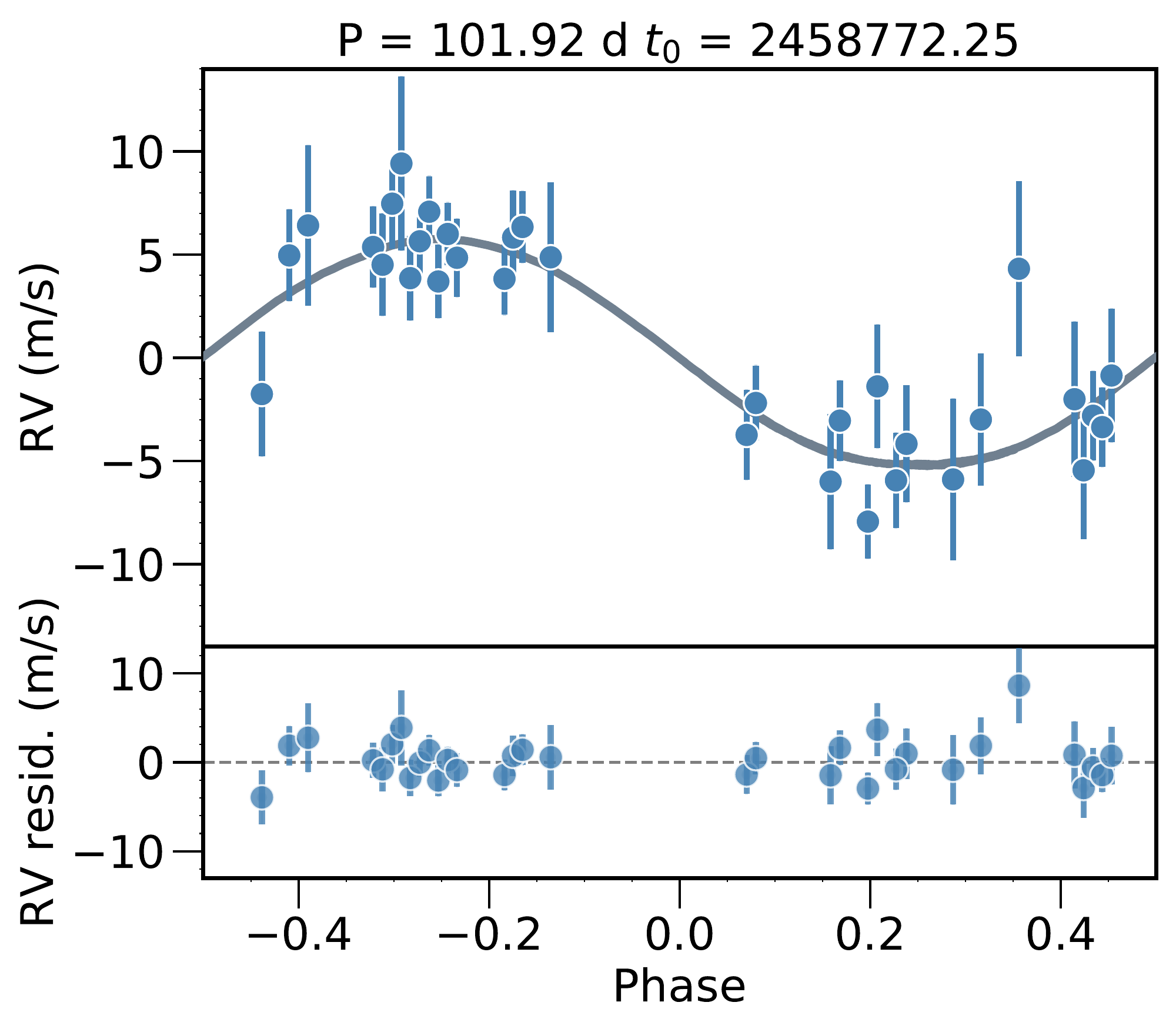}
\end{minipage}
\caption[RV time series and phase-folded RVs for TOI-1201~b]{ CARMENES RV data for TOI-1201 with the best-fit model from the joint fit overplotted (dark gray), the dSHO-GP component (orange), and the long-term signal (green). \textit{Top panel:} RV time series. The light gray band represents the 68\,\% credibility interval. \textit{Bottom panels:} RVs phase-folded to the periods of the transiting planet (\textit{left}) and long-term signal (\textit{right}). The bottom panel of each plot shows the residuals after the model is subtracted.}
\label{fig:rvs}
\end{figure*}

\paragraph{Transiting mini-Neptune.}
For the transiting planet, TOI-1201~b, we fixed the priors for the period and the transit time to the median values based on the posteriors from the transit-only fits (Sect.~\ref{sec:transitonlymodeling}) since the transit data provide us with very well-determined values, such that the precision of the RV data would not be able to. An eccentric orbit was indistinguishable from a circular one, where most of the posterior samples were consistent with a zero eccentricity. Therefore, we continued to model the transiting planet with a circular Keplerian. Nevertheless, the current phase coverage (see Fig.~\ref{fig:rvs}) might not be able to adequately recognize an eccentricity, which could affect certain parameters (e.g., the semi-amplitude $K_b$, and subsequently, the planetary mass). Filling in the phase coverage gaps with future, carefully planned RV measurements will help constrain the eccentricity, though a circular orbit is currently a reasonable assumption, given the short orbital period.

\paragraph{Stellar rotation period.}
We tested fitting the stellar rotation signal at 19\,d with a sinusoid and a dSHO-GP kernel. 
The dSHO-GP kernel used was the same one as when determining $P_\textnormal{rot}$ (Sect.~\ref{sec:rotperiod}) and introduced in Eq.~\ref{eq:dsho}. 
We implemented a narrow prior for the period, $\mathcal{U}(15\,d,25\,d)$, which we named GP$_\textnormal{19d}$. 
The motivation for the narrow prior centered around 19\,d was to avoid picking up periods (i.e., the alias at 35\,d) not associated with the stellar rotation. 

The models between using a sinusoid and a GP were indistinguishable ($\Delta\ln{\mathcal{Z}} \sim 2$).
Therefore, we decided to use the GP to describe the nature of the 19\,d signal. Given our prior knowledge that this signal is physically produced by the stellar rotation period, the quasi-periodic behavior of the GP better explains the underlying incoherent behavior of a stellar activity signal that may not be exhibited with the current data given the relatively short time span. Even if we choose to model the stellar rotation period with a sinusoid, the minimum mass of the transiting planet is not drastically affected, though the uncertainties are slightly enlarged (Fig.~\ref{fig:boxplot_rvmodels}, specifically focusing on 1 Kep + 1 Sin + GP$_\textnormal{19d}$ versus 1 Kep + 2 Sin). 

In addition to the dSHO-GP, we also experimented with the quasi-periodic GP kernel (QP-GP), as introduced in \texttt{george} and presented in Appendix~\ref{appendix:gpwide}. Both GP kernels were indistinguishable in their log evidence, so we chose the dSHO-GP, as it serves our purpose to fit the quasi-periodic stellar activity signal very well and, at the same time, is computationally much faster compared to the QP-GP.

\paragraph{Long-term signal.}
To account for the $\sim$102\,d signal, we experimented with a quadratic trend, a sinusoidal signal, and, for completeness, also a Keplerian model.
We used a wide, uniform prior for the period, $\mathcal{U}(60\,\textnormal{d},150\,\textnormal{d})$, in all our models. For the sinusoidal and the Keplerian model, the time of transit center was additionally set uniform, $\mathcal{U}(2458730,2458840)$.
Though the prior was kept consistent for all attempted models, the median value of the posterior distribution for the period varied to some extent due to different models, finding slightly distinct optimized configurations depending on how the other parameters were adjusted (Table~\ref{tab:modelcomparison}).
Compared to the quadratic model, the sinusoidal signal is preferred ($\Delta\ln{\mathcal{Z}} = \ln{\mathcal{Z}}_\textnormal{sinusoid} - \ln{\mathcal{Z}}_\textnormal{quadratic} \sim 19$). 
The Keplerian model, with eccentricity varying freely, has a comparable evidence to the purely sinusoidal signal. However, the posteriors favor in this case a periodicity at $\sim$80\,d with a relatively high eccentricity of $\sim$0.5--0.6. 
The suggested high eccentricity is likely a result of only one cycle or less being observed and the phase not being completely sampled (Fig.~\ref{fig:rvs}). With the current baseline, it is no longer possible to distinguish between 80\,d and 102\,d considering that 400\,d of observations would be necessary (i.e., $\Delta t_\textnormal{baseline} = $ (1/80\,d -- 1/102\,d)$^{-1}$).
Therefore, we continued to model the $\sim$102\,d signal with the simplest, and better constrained, model of a sinusoidal variation.

\paragraph{Final model choice.}
We had the prior physical knowledge that there is a transiting planet at 2.5\,d and a stellar rotation signal at $\sim$19\,d. For this reason, we expected these two signals to appear in the RVs for which we modeled them with a Keplerian and a GP, respectively. The most prominent signal in the data was, however, the $\sim$102\,d signal, which we could not ignore. To acknowledge the signal, we modeled it with a sinusoid. Moreover, we did experiment what happens if we were to apply a QP-GP with a wider prior for $P_\textnormal{rot}$ (detailed in Appendix.~\ref{appendix:gpwide}). 

Nonetheless, to ensure that the parameters of the transiting planet were not drastically affected by our model choice, we examined the minimum mass derived from the RV-only fits (see Fig.~\ref{fig:boxplot_rvmodels}). All considered models agree within their uncertainties. Furthermore, no stellar activity indicator is significant in power at the frequencies of the transiting planet ($\sim$2.5\,d) or of the $\sim$102\,d signal (Fig.~\ref{fig:glsperiodogramtoi1201activity}).
Based on these results, we continued the analysis with the 1 circular Kep (2.5\,d) + 1 Sin ($\sim$102\,d) + dSHO-GP$_\textnormal{19d}$ as the favored model. 
The final model choice consists of 13 free parameters to the 33 RV data points.
The transiting planet follows a circular Keplerian model, the stellar rotation period is represented with a dSHO-GP centered around the period of interest, and the most significant signal at $\sim$102\,d is modeled as a sinusoidal variation.
We emphasize that we do not claim the 102\,d signal to be a planet candidate or a photometric variability cycle.
Considering that we only observed about one period or less, we cannot elaborate on the nature of this signal and further monitoring is needed in order to determine its origin.

\begin{table}
\caption{RV-only model comparison using the Bayesian log evidence for TOI-1201. }
\label{tab:modelcomparison}
\centering
\footnotesize
\begin{tabular}{l l S[table-format=1] S}
\hline \hline
\noalign{\smallskip}
Model & $P$ (d) & \multicolumn{1}{c}{$\ln{\mathcal{Z}}$} & \multicolumn{1}{c}{$\Delta\ln{\mathcal{Z}}$} \\
\noalign{\smallskip}
\hline 
\noalign{\smallskip}
Flat & \dots & -116.70 & -18.31 \\
\textbf{1 Kep + 1 Sin + dSHO-GP$_\textnormal{19d}$} & $\mathbf{2.5, 102}$ & $\mathbf{-98.39}$ & $\mathbf{0.0}$ \\
1 Kep + 1 Sin + dSHO-GP$_{\rm 19d}$ & 2.5\tablefootmark{*}, 102 & -99.38 & -0.99 \\
2 Kep + dSHO-GP$_{\rm 19d}$ & 2.5, 88\tablefootmark{*} & -98.55 & -0.16 \\
1 Kep + Quad + dSHO-GP$_{\rm 19d}$ & 2.5 & -116.99 & -18.61 \\
1 Kep + 2 Sin & 2.5, 19, 104 & -100.24 & -1.85 \\
\noalign{\smallskip}
\hline
\end{tabular}
\tablefoot{A larger, positive $\Delta\ln{\mathcal{Z}}$ indicates a better model. In the model names, ``Kep'' refers to a Keplerian orbit and ``Sin'' to a sinusoidal signal. The model that we used for the final joint fit was the 1 Kep (2.5\,d) + 1 Sin (102\,d) + dSHO-GP$_\textnormal{19d}$, indicated by the bold-faced row. 
The model names correspond to those in Fig.~\ref{fig:boxplot_rvmodels}. \tablefoottext{*}{An eccentric orbit was used.}}
\end{table}

\subsection{Joint modeling} \label{sec:jointfit}

We finally combined all the data from the \tess\ observations, the ground-based transit follow-ups, and the CARMENES RV data to produce the most precise planetary parameters. Our final model comprises the 2.5\,d transiting planet, a stellar rotational period component at 19\,d, and a long-term 102\,d signal, respectively modeled as a circular Keplerian, with a dSHO-GP centered on 19\,d, and a sinusoid (Sect.~\ref{sec:rvonlymodeling}).
We used the results from the posteriors of the transit-only and RV-only fits to set up the priors for the final joint fit, as discussed in detail and justified by \cite{Kemmer2020_toi488}. 
These priors can be found in Table~\ref{tab:priors_joint}.

Our findings from the joint fit, including the derived planetary parameters, are presented in Table~\ref{tab:posteriors_planet}. The best model fits for the transit photometry and the RVs are shown in Figs.~\ref{fig:lc} and~\ref{fig:rvs}, respectively. Table~\ref{tab:posteriors_full} is a comprehensive posterior summary including all of the model parameters, where the posterior probability densities are presented in Figs.~\ref{fig:cornerplot_p1}, \ref{fig:cornerplot_gp}, and~\ref{fig:cornerplot_p2}. 

\begin{table}
\centering
\caption{Derived planetary parameters for TOI-1201~b.}
\label{tab:posteriors_planet}
\begin{tabular}{lcl}
\hline
\hline
\noalign{\smallskip}
Parameter & \multicolumn{1}{c}{Posterior} & Unit \\
 & TOI-1201~b & \\
\noalign{\smallskip}
\hline
\noalign{\smallskip}
$P_p$ & $2.4919863^{+0.0000030}_{-0.0000031}$ & d\\[0.1 cm]
$t_{0,p}$ (BJD) & $2459169.23222^{+0.00052}_{-0.00054}$ & d\\[0.1 cm]
$r_{1,p}$ & $0.603^{+0.048}_{-0.055}$ & $\dots$\\[0.1 cm]
$r_{2,p}$ & $0.04383^{+0.00096}_{-0.00110}$ & $\dots$\\[0.1 cm]
$K_p$ & $4.65^{+0.60}_{-0.64}$ & \ms\\[0.1 cm]
$e_p$ & $0.0$ (fixed) & $\dots$\\[0.1 cm]
$\omega_p$ & $90.0$ (fixed) & $\dots$\\[0.1 cm]
$p = R_{p}/R_\star$ & $0.04383^{+0.00096}_{-0.00110}$ & $\dots$\\[0.1 cm]
$b = (a_{p}/R_\star) \cos (i_{p})$ & $0.404^{+0.071}_{-0.082}$ & $\dots$\\[0.1 cm]
$a_{p}/R_\star$ & $12.23^{+0.36}_{-0.36}$ & $\dots$\\[0.1 cm]
\noalign{\smallskip}\noalign{\smallskip}
\hline
\noalign{\smallskip}\noalign{\smallskip}
$i_p$ & $88.11^{+0.42}_{-0.40}$ & deg\\[0.1 cm]
$t_T$ & $1.747^{+0.096}_{-0.091}$ & h\\[0.1 cm]
$M_p$ & $6.28^{+0.84}_{-0.88}$ & $M_\oplus$\\[0.1 cm]
$R_p$ & $2.415^{+0.091}_{-0.090}$ & $R_\oplus$\\[0.1 cm]
$\rho_p$ & $2.45^{+0.48}_{-0.42}$ & g cm$^{-3}$\\[0.1 cm]
$g_p$ & $10.5^{+1.8}_{-1.6}$ & m s$^{-2}$\\[0.1 cm]
$a_p$ & $0.0287^{+0.0012}_{-0.0012}$ & \au\\[0.1 cm]
$T_\textnormal{eq}$ & $703^{+15}_{-14}$ & K\\[0.1 cm]
$S_p$ & $40.6^{+3.6}_{-3.2}$ & $S_\oplus$\\[0.1 cm]
\noalign{\smallskip}
\hline
\end{tabular}
\end{table}
\section{Radial velocities of PM~J02489--1432E} \label{sec:companionrvs}

Most observations for the 8\,arcsecond-wide companion of TOI-1201 were acquired on the same nights as for the host star. As a result, with the high-resolution spectra from CARMENES, we were able to compute very precise stellar parameters for both stars in the binary system (Sect.~\ref{sec:stellarprops}).
We then performed a complementary analysis on the RVs and various stellar activity indicators of the companion to search for potentially interesting signals. The GLS periodogram of the RVs (second panel of Fig.~\ref{fig:compgls}) shows a significant peak at 27\,d (above 10\,\% FAP level). 
Doing \juliet\ runs, we compared the Bayesian log evidence of a flat model, linear model, and a sinusoidal (i.e., circular Keplerian) model. We found the best one to be the sinusoidal model at 28.5\,d with a semi-amplitude of $K$ = 3.42$\pm$0.89\,\ms\ (see Table~\ref{tab:modelcomparison_comp} for the results and Fig.~\ref{fig:rvs_comp} for the RVs themselves). To investigate the nature of the signal, we considered the same stellar activity indicators as mentioned in Sect.~\ref{sec:rotperiod}. While no signals are significant, some power excess at $\sim$16\,d and $\sim$30\,d, which are the 41\,d aliases of one another, hint that the 28.5\,d signal in the RVs might be somehow related to stellar activity (Fig.~\ref{fig:compgls}). 
Given the similarity between the two stars, one would expect similar $P_\textnormal{rot}$, too.
However, given the small number of data points, its origin remains inconclusive. Thus given the current data at hand, additional RV data would be necessary for the companion of TOI-1201 to determine the origin of this signal.
Additionally, to test whether we could detect a similar planet to TOI-1201~b orbiting PM~J02489--1432E, we first subtracted the 28\,d periodicity and then followed the procedure as outlined by \citet{Bonfils2013}. We estimate that we can exclude planets above $\sim$6\,\mearth\ orbiting the companion with a period shorter than 10\,d (Fig.~\ref{fig:detectionlimits}). 

\begin{table}
\caption{RV-only model comparison using the Bayesian log evidence for PM~J02489--1432E (TOI-393).}
\label{tab:modelcomparison_comp}
\centering
\begin{tabular}{l S[table-format=1] S[table-format=1]}
\hline \hline
\noalign{\smallskip}
Model &\multicolumn{1}{c}{$\ln{\mathcal{Z}}$} & \multicolumn{1}{c}{$\Delta\ln{\mathcal{Z}}$} \\
\noalign{\smallskip}
\hline 
\noalign{\smallskip}
Flat & -63.40 & -4.13 \\
Linear & -71.38 & -12.11 \\
\textbf{1 Sin (28.5\,d)} & $\mathbf{-59.27}$ & $\mathbf{0.0}$ \\
\noalign{\smallskip}
\hline
\end{tabular}
\end{table}

% RV timeseries for TOI-1201

\section{Discussion and future prospects} \label{sec:discussion}

We validate and characterize the exoplanet TOI-1201~b. The relatively hot mini-Neptune orbits its host star every 2.49\,d, has a radius of $2.415^{+0.091}_{-0.090}\,R_\oplus$, a mass of $6.28^{+0.84}_{-0.88}\,M_{\oplus}$, and a bulk density of $2.45^{+0.48}_{-0.42}$\,g\,cm$^{-3}$. A summary of all derived planetary parameters and their uncertainties can be found in Table~\ref{tab:posteriors_planet}.
An additional signal at 102\,d is revealed in the RV data, whose origin is still unknown. If we were to assume that it has a planetary origin, we would then obtain a period of $P=102^{+21}_{-15}$\,d and a semi-amplitude of $K=5.84^{+0.91}_{-0.87}$\,\ms\ such that the minimum mass would be $27.0^{+5.6}_{-4.5}\,M_\oplus$. 
TOI-1201 could thus be a multi-planetary system (see Appendix~\ref{appendix:twoplanetmodel}), making further monitoring quite appealing.
Additionally, the RV data show a 19\,d signal corresponding to the $\sim$21\,d rotational period of the star as inferred by photometry.
Below we discuss the plentiful potential that TOI-1201~b has for future follow-up observations and characterizations of the system.

\paragraph{Spin-orbit alignment.} 
Following \cite{Boue2013}, the expected Rossiter-McLaughlin \citep[RM;][]{Rossiter1924,McLaughlin1924} amplitude is in the range $K_{RM} \sim$ 1.5--3.0\,\ms, since the $v\sin i$ of the star (<\,2\,k\ms) and the spin-orbit angle are not exactly defined. Collecting roughly ten data points within the transit (duration of $1.8\pm0.1$\,h) to achieve a precision better than the RM amplitude is well within reach for the current state-of-the-art spectrographs with sub-\ms\ precision such as ESPRESSO \citep{espresso}, MAROON-X \citep{maroonx}, or EXPRES \citep{expres}. 

For this reason, the mini-Neptune TOI-1201~b presents itself as a promising candidate for measuring the spin-orbit angle between the stellar spin axis and orbit of the transiting planet, which can be determined by the RM effect. Studying this effect can shed light on orbital architectures of planetary systems specifically around low-mass stars. It is expected that close-orbiting giant planets around such stars are aligned \citep{GaudiWinn2007_rm,Winn2010_rm,MunozPerets2018,Louden2021} because of strong tidal interactions with the stellar convective envelope. 
To date, there are only six other planets around M dwarfs with measured obliquities, namely \object{GJ~436}~b \citep{Bourrier2018_RM_GJ436b}, \object{TRAPPIST-1}~b,~e,~f \citep{Hirano2020_RM_TRAPPIST1}, \object{AU~Mic}~b \citep{Addison2020_RM_AUMic, Hirano2020_RM_AUMic, Palle2020_RM_AUMic}, and \object{K2-25}~b \citep{Stefansson2020_RM_K2-25b}. 
Determining the spin-orbit angle of TOI-1201~b is especially interesting in the context of the companion (320\,\au\ away) because it could hint possible interaction.

\begin{figure*}
% https://lweb.cfa.harvard.edu/~lzeng/planetmodels.html#mrrelation
    \centering
    \includegraphics[width=1\linewidth]{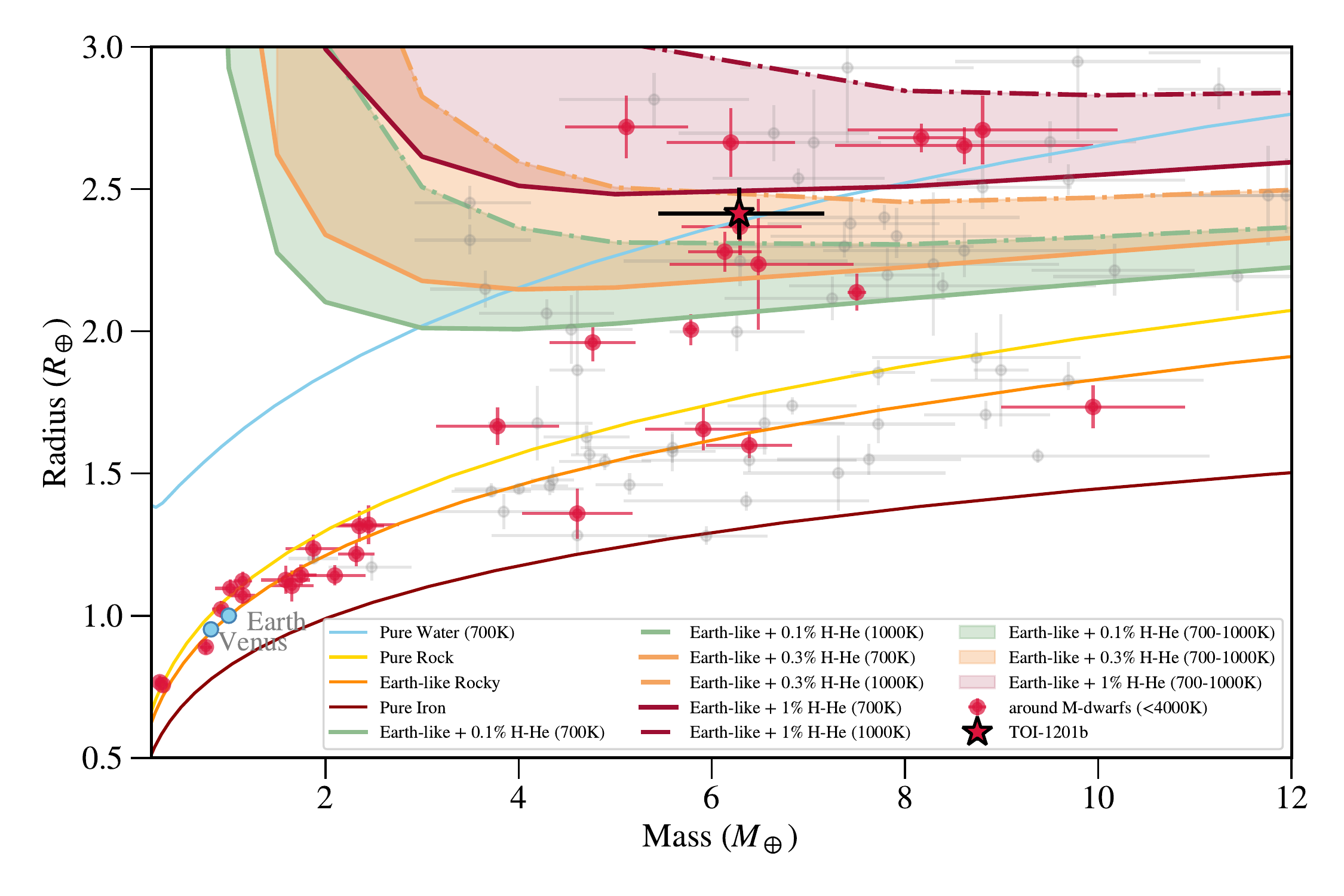}
    \caption[Mass-radius diagram for currently known transiting exoplanets around M dwarfs with a measured mass.]{Mass-radius diagram based on the TEPcat catalog \citep[][visited on 10 August 2021]{Southworth2011}. TOI-1201~b is indicated as a red star outlined in black. The gray points represent the other planets ($R < 3\,R_\oplus$ and $M < 12\,M_\oplus$) with masses and radii measured to precision better than 20\,\%. The red points are those planets around M dwarfs ($T_\textnormal{eff} < 4000$\,K). All theoretical composition models are taken from \cite{Zeng2019}. The lines represent models for cores composed of pure iron (100\,\% Fe), Earth-like composition (32.5\,\% Fe + 67.5\,\% MgSiO$_3$), pure rock (100\,\% MgSiO$_3$), and pure water (100\,\% H$_2$O at 700\,K). The area between the solid and dash-dotted lines of the same color represent the theoretical composition models for Earth-like rocky cores with varying sizes of H-He envelopes (0.1\,\%, 0.3\,\%, 1\,\%) for equilibrium temperatures of 700\,K and 1000\,K, respectively. The plot is inspired by \cite{VanEylen2021}, though presented here in linear scale to emphasize the mini-Neptune population.}
    \label{fig:mrd}
\end{figure*}

\paragraph{Transmission spectroscopy.}
% table 1 1.5<re<2.75re - scalefactor=1.26 and cutoff for followup efforts is 92
% TSM for TOI-1201~b:  128.4140230833101
% ESM for TOI-1201~b:  11.556980523569585
All the currently known transiting exoplanets with a measured mass (with precision better than 20\,\%) are displayed in Fig.~\ref{fig:mrd}.
TOI-1201~b falls in the realm of Neptune-sized planets above the M-dwarf radius valley \citep[Fig.~7 in][$R$ = 1.4--1.7\,\rearth]{VanEylen2021}.
In this regard, the density of $\rho_b = 2.45^{+0.48}_{-0.42}$\,g\,cm$^{-3}$ suggests that its composition is in line with an Earth-like rocky core with an H-He envelope of 0.3\,\% by mass, following the theoretical composition models from \cite{Zeng2019}\footnote{\url{https://lweb.cfa.harvard.edu/~lzeng/planetmodels.html}}. Likewise, it is consistent with being a water world, signifying that its internal composition is rather dominated with H$_2$O in the form of ices and fluids, which is also a very plausible scenario in the realm of planet formation \citep[e.g.,][]{Bitsch2019,Venturini2020}.
Atmosphere characterization via transmission spectroscopy paired with precise mass and radius determinations and a good set of interior models could aid in breaking this ambiguity of its interior architecture.

\begin{figure}
    \centering
    \includegraphics[width=1\linewidth]{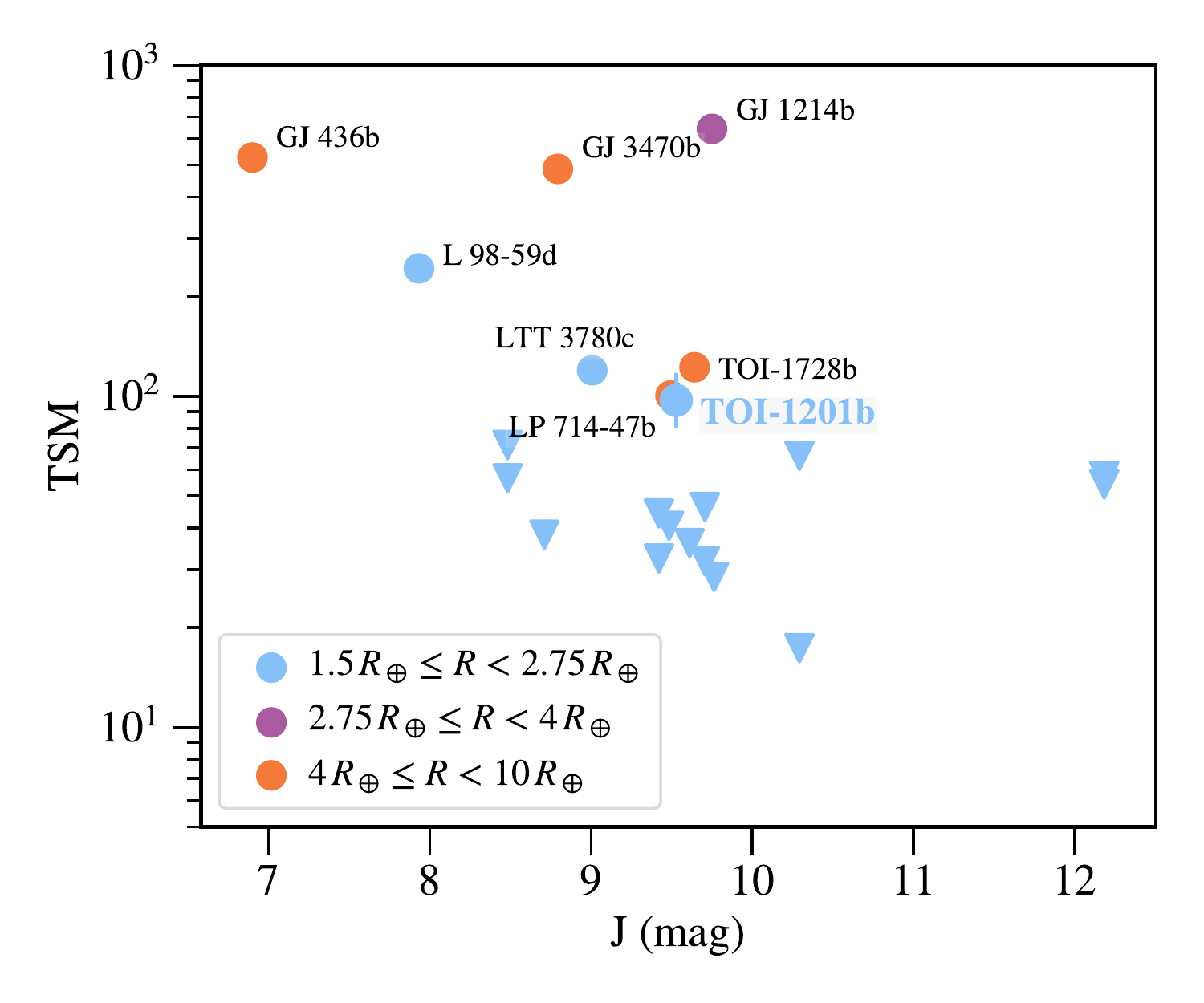}
    \caption[Transmission spectroscopy metric (TSM) as a function of $J$ magnitude for currently known transiting Neptune-sized exoplanets around M dwarfs with a measured mass.]{Transmission spectroscopy metric (TSM) as a function of $J$ magnitude for currently known transiting Neptune-sized planets around M dwarfs with a measured mass. 
    The color of each point corresponds to its respective radius bin as defined by \citet{Kempton2018}. 
    Furthermore, promising targets that are above their respective radius-bin cutoff are labeled and marked with a circle, and those below the cutoff are represented with an upsidedown triangle. The uncertainties are only computed for TOI-1201~b.
} 
    \label{fig:tmp_jmagvstsm}
\end{figure}

Following \cite{Kempton2018}, the transmission spectroscopy metric (TSM) of TOI-1201~b turns out to be $97^{+21}_{-16}$, which is slightly above the cutoff of 92 with the \jwst/NIRISS bandpass (for planet radii of 1.5\,$R_\oplus$\,<\,$R_p$\,<\,2.75\,$R_\oplus$).  
Among other transiting planets with masses measured via RVs or transit time variations around M dwarfs \citep{Trifonov2021}\footnote{Using the continuously updated table: \url{https://carmenes.caha.es/ext/tmp/}. Last updated on 09 April 2021.}, TOI-1201~b is ranked high and one of just a few suitable targets known thus far (Fig.~\ref{fig:tmp_jmagvstsm}). It is besides an intriguing candidate for comparison with GJ~1214~b, which has similar mass and radius but a slightly lower equilibrium temperature \citep[$8.17\pm0.43$\mearth, $2.742\pm0.053$\rearth, $596\pm19$\,K;][]{Cloutier2021}. GJ~1214~b is known to have high-altitude clouds or haze \citep{Kreidberg2014}, but the higher temperatures of TOI-1201~b may lead to less cloudy or hazy atmospheres \citep{CrossfieldKreidberg2017}. 
In preparation for the presumed launch of \jwst\ in late 2021, our transit ephemeris from the joint fit (see Table~\ref{tab:posteriors_planet}) results in an uncertainty of $\sim$60\,s toward the beginning of 2022. The uncertainty then increases to $\sim$100\,s and $\sim$140\,s for the beginning of 2023 and 2024, respectively. Further ground-based observations could help refine the ephemeris and further improve the planet radius. 
Moreover, TOI-1201 is a viable target for space-based follow-up observations with \cheops\ \citep{cheops}. The visibility is good given its ecliptic latitude, allowing it to be observed for an accumulated time of 60 days in one year. Blending is expected because the size of the PSF of \cheops\ is approximately 16\,arcsec. Using the typical aperture radius of 20--25\,arcsec for a star with the brightness of TOI-1201, however, the final extracted flux would simply be the added flux of both stars and can be easily mediated with a dilution factor taken into account when using \juliet\ \citep{juliet}.

Likewise, TOI-1201~b is ideal for low-to-mid-resolution ground-based transit spectroscopy covering the optical regime, serving as a great complement to the expected near- and mid-IR wavelength range measurements from \jwst. These ground-based observations are typically challenging due to strong atmospheric and instrumental variations \citep[e.g.,][]{Nikolov2016,Huitson2017,Diamond-Lowe2018,Espinoza2019,Wilson2021,Chen2021_transmission}. Companion stars that are close in brightness and space to the target star help mitigate these effects since both stars experience a similar path through the Earth's atmosphere and the instrument, allowing the effects from the target star to be monitored and removed. In this regard, the binary companion \object{PM~J02489--1432E}, which has very similar stellar properties (see Table~\ref{tab:stellarparams}), is a perfect star to detrend atmospheric and instrumental effects from TOI-1201's light curve.

Furthermore, the relatively young age of the system (600--800\,Myr) makes TOI-1201~b an intriguing object for studying the future preservation or evaporation of the atmosphere. Planets found above the radius gap are in the process of losing their atmospheres through two conceivable mechanisms with corresponding timescales of $\sim$1\,Gyr due to core-powered mass loss \citep{GuptaSchlichting2020} or of $\sim$100\,Myr due to photoevaporation \citep{OwenWu2017}. 
Other currently known transiting mini-Neptunes planets with mass determinations in young systems orbit stellar hosts are too faint for atmospheric characterization ($J >$ 11\,mag), for example, K2-25 \citep{Mann2016_K2-25, Stefansson2020_RM_K2-25b,Thao2020_TS_K2-25}, \object{EPIC-247589423} \citep{Mann2018_epic}, K2-264 \citep{Rizzuto2018_K2-264,Livingston2019_K2-264}, and K2-101, K2-101, K2-103, K2-104, EPIC-211901114, and K2-95 \citep{Mann2017}.
The only suitable candidate orbiting a sufficiently bright host star for this selective target sample is \object{K2-100}~b \citep{Barragan2019_K2-100,Gaidos2020}. 
Thus, the relative brightness ($J \approx$ 9.5\,mag) and intermediate age of the TOI-1201 system can help constrain the empirical timescale for evaporation and provide a test case for the two possible theories.

TOI-1201~b is, therefore, a very compelling candidate for atmospheric characterization using transmission spectroscopy.
Determining the atmospheric compositions will allow for a better understanding of the formation histories of mini-Neptunes and aid in characterizing them.
 
% to find the uncertainty for \jwst\
% import numpy as np
% import juliet
% period = np.random.normal(2.4919863,0.0000031,10000)
% t0 = np.random.normal(2459169.23222,0.00054,10000)
% t0new = t0+period*165 # for dec 31 2021
% val,valup,valdown = juliet.utils.get_quantiles(t0new)
% print(val,(valup-val)*24*3600)
% # 2459580.409946563 63.545018434524536
% t0new = t0+period*310 # for dec 27 2022
% val,valup,valdown = juliet.utils.get_quantiles(t0new)
% print(val,(valup-val)*24*3600)
% # 2459941.7479580473 99.8326912522316
% t0new = t0+period*458 # for dec 31 2023
% val,valup,valdown = juliet.utils.get_quantiles(t0new)
% print(val,(valup-val)*24*3600)
% # 2460310.561919404 136.57978624105453

\paragraph{Planet occurrence rates around M-dwarf binary systems.}
Most surveys are strongly biased against stars in known binary or multiple systems, introducing a selection bias against them. As already mentioned in Sect.~\ref{sec:intro}, many planets discovered around M-dwarf binary systems are in systems with a wide separation. 
The same is true for TOI-1201 ($s\sim$\,320\,\au). Therefore, the TOI-1201 system most likely does not face significant gravitational interactions with its companion to hinder planet formation. In this respect, it is essentially identical to a system around a single star \citep{DesideraBarbieri2007,Roell2012}.
Interestingly enough, in all of these systems, the stars are quite different in spectral type, with the exception of GJ~338~ABb \citep[M0\,V+M0\,V;][]{Gonzalez2020_gj338B}, TOI-1201 (M2.0\,V+M2.5\,V; this paper), and LTT~1445~AbBC \citep[M-dwarf trio of similar masses;][]{Winters2019_ltt1445}. 

Unlike most other systems, where spectroscopic measurements and precise stellar parameters are available only for the primary, PM~J02489–-1432E, the stellar companion of TOI-1201 has also been well characterized in this paper. No signal indicative of a planet was found for this star (Sect.~\ref{sec:companionrvs}). For comparison, this is also the case for the stellar companion of GJ~338~B.
Even though the two stars in each system are very similar in mass (and age), they have different planetary architecture. 
As for the \object{LTT~1445} system, RV measurements were carried out only for the host star providing an upper mass limit for the planet, though further observations are underway in determining its mass.

\section{Conclusions}\label{sec:conclusions}

In this paper we presented \tess\ and ground-based photometric observations, together with CARMENES spectroscopic measurements, of the early M dwarf TOI-1201. We confirmed the transiting planet with a period of about 2.5\,d and provided a mass determination using RV follow-up to obtain the following derived planetary parameters: $2.415^{+0.091}_{-0.090}\, R_{\oplus}$ for the radius and $6.28^{+0.84}_{-0.88}\,M_{\oplus}$ for the mass. This transiting planet then carries a density of $2.45^{+0.48}_{-0.42}$\,g\,cm$^{-3}$, classifying it as a mini-Neptune. The planet is a favorable target for further studies, namely: (i) spin-orbit alignment using the RM effect, which is achievable with the current spectroscopic instruments, and (ii) atmospheric studies using transmission spectroscopy with ground-based facilities (which is possible thanks to its nearly identical, close-by companion) and with upcoming space-based missions, such as \jwst. The relatively young age of the system ($\sim$600--800\,Myr) gives it the additional advantage of potentially constraining the timescales of currently accepted atmospheric mass-loss mechanisms.

The RVs also exhibited a long-term signal with a high semi-amplitude at $\sim$102\,d. 
If the signal is due to an additional planet in the system, then its minimum mass would be $27.0^{+5.6}_{-4.5}\,M_{\oplus}$. Such system architectures are commonly reproduced in core accretion formation models \citep[see Appendix~\ref{appendix:twoplanetmodel} and][]{Schlecker2020}.
Further RV measurements are, however, necessary to falsify or to prove a planetary origin of the signal, which could provide more insight into the architecture of multi-planetary systems.

We were able to narrow down the stellar rotation period for TOI-1201 to 19--23\,d using archival long-term photometry and stellar activity indicators provided by the spectral information. The signature of the stellar rotation period also presented itself in the RVs, which we take into account in our final model. 
The stellar host is the primary of a wide ($s\sim$ 320\,\au) binary system of two nearly identical M2.0--2.5 dwarfs. We obtained CARMENES RV measurements for the secondary as well, providing precise stellar parameters.
One significant signal at around 27\,d, most likely related to the stellar rotation period, was detected around the companion. 

All things considered, the TOI-1201 system contains two low-mass stars at relatively short separation, and around one of them there is at least one transiting mini-Neptune with a precise mass determination and with favorable opportunities for atmospheric characterization.
However, despite the high quality data on hand and the comprehensive analyses performed, there remain unanswered questions, the answers to which require further monitoring. 

\begin{acknowledgements}
Part of this work was supported by the German {Deut\-sche For\-schungs\-ge\-mein\-schaft (DFG)} project number Ts~17/2--1.

CARMENES is an instrument at the Centro Astron\'omico Hispano-Alem\'an (CAHA) at Calar Alto (Almer\'{\i}a, Spain), operated jointly by the Junta de Andaluc\'ia and the Instituto de Astrof\'isica de Andaluc\'ia (CSIC).
  
CARMENES was funded by the Max-Planck-Gesellschaft (MPG), the Consejo Superior de Investigaciones Cient\'{\i}ficas (CSIC),
  the Ministerio de Econom\'ia y Competitividad (MINECO) and the European Regional Development Fund (ERDF) through projects FICTS-2011-02, ICTS-2017-07-CAHA-4, and CAHA16-CE-3978, 
  and the members of the CARMENES Consortium 
  (Max-Planck-Institut f\"ur Astronomie,
  Instituto de Astrof\'{\i}sica de Andaluc\'{\i}a,
  Landessternwarte K\"onigstuhl,
  Institut de Ci\`encies de l'Espai,
  Institut f\"ur Astrophysik G\"ottingen,
  Universidad Complutense de Madrid,
  Th\"uringer Landessternwarte Tautenburg,
  Instituto de Astrof\'{\i}sica de Canarias,
  Hamburger Sternwarte,
  Centro de Astrobiolog\'{\i}a and
  Centro Astron\'omico Hispano-Alem\'an), 
  with additional contributions by the MINECO, 
  the Deutsche Forschungsgemeinschaft through the Major Research Instrumentation Programme and Research Unit FOR2544 ``Blue Planets around Red Stars'', 
  the Klaus Tschira Stiftung, 
  the states of Baden-W\"urttemberg and Niedersachsen, 
  and by the Junta de Andaluc\'{\i}a.
  
  This work was based on data from the CARMENES data archive at CAB (CSIC-INTA).
  
 We acknowledge financial support from the Agencia Estatal de Investigaci\'on of the Ministerio de Ciencia, Innovaci\'on y Universidades and the ERDF through projects 
  PID2019-109522GB-C5[1:4]/AEI/10.13039/501100011033    % CAB+IAA+IAC+UCM
  %PGC2018-098153-B-C33                                 % ICE
  %AYA2018-84089,                                       % CAB SVO
  %ESP2017-87676-C5-1-R,                                % CAB PLATO  
and the Centre of Excellence ``Severo Ochoa'' and ``Mar\'ia de Maeztu'' awards to the Instituto de Astrof\'isica de Canarias (SEV-2015-0548), Instituto de Astrof\'isica de Andaluc\'ia (SEV-2017-0709), and Centro de Astrobiolog\'ia (MDM-2017-0737),
the European Research Council under the Horizon 2020 Framework Program (ERC Advanced Grant Origins 83 24 28), 
the Generalitat de Catalunya/CERCA programme,
the DFG priority program SPP 1992 ``Exploring the Diversity of Extrasolar Planets (JE 701/5-1)'',
the European Research Council under the Horizon 2020 Framework Program via ERC Advanced Grant Origins 832428 and under Marie Sk\l{}odowska-Curie grant 895525.

Funding for the \tess\ mission is provided by NASA's Science Mission Directorate. This paper includes data collected by the \tess\ mission that are publicly available from the Mikulski Archive for Space Telescopes (MAST). We acknowledge the use of public \tess\ data from pipelines at the \tess\ Science Office and at the TESS Science Processing Operations Center. Resources supporting this work were provided by the NASA High-End Computing (HEC) Program through the NASA Advanced Supercomputing (NAS) Division at Ames Research Center for the production of the SPOC data products. 
This work makes use of observations from the LCOGT network. Part of the LCOGT telescope time was granted by NOIRLab through the Mid-Scale Innovations Program (MSIP). MSIP is funded by NSF.

We thank Brian Mason for providing us WDS data.

\end{acknowledgements}

\bibliographystyle{aa} % style aa.bst
\bibliography{biblio} % your references biblio.bib

\begin{thebibliography}{192}
\expandafter\ifx\csname natexlab\endcsname\relax\def\natexlab#1{#1}\fi

\bibitem[{{Addison} {et~al.}(2020){Addison}, {Horner}, {Wittenmyer},
  {Plavchan}, {Wright}, {Nicholson}, {Marshall}, {Clark}, {Kane}, {Hirano},
  {Kielkopf}, {Shporer}, {Tinney}, {Zhang}, {Ballard}, {Bedding}, {Bowler},
  {Mengel}, {Okumura}, \& {Gaidos}}]{Addison2020_RM_AUMic}
{Addison}, B.~C., {Horner}, J., {Wittenmyer}, R.~A., {et~al.} 2020, arXiv
  e-prints, arXiv:2006.13675

\bibitem[{{Alam} {et~al.}(2015){Alam}, {Albareti}, {Allende Prieto}, {Anders},
  {Anderson}, {Anderton}, {Andrews}, {Armengaud}, {Aubourg}, {Bailey}, {Basu},
  {Bautista}, {Beaton}, {Beers}, {Bender}, {Berlind}, {Beutler}, {Bhardwaj},
  {Bird}, {Bizyaev}, {Blake}, {Blanton}, {Blomqvist}, {Bochanski}, {Bolton},
  {Bovy}, {Shelden Bradley}, {Brandt}, {Brauer}, {Brinkmann}, {Brown},
  {Brownstein}, {Burden}, {Burtin}, {Busca}, {Cai}, {Capozzi}, {Carnero
  Rosell}, {Carr}, {Carrera}, {Chambers}, {Chaplin}, {Chen}, {Chiappini},
  {Chojnowski}, {Chuang}, {Clerc}, {Comparat}, {Covey}, {Croft}, {Cuesta},
  {Cunha}, {da Costa}, {Da Rio}, {Davenport}, {Dawson}, {De Lee}, {Delubac},
  {Deshpande}, {Dhital}, {Dutra-Ferreira}, {Dwelly}, {Ealet}, {Ebelke},
  {Edmondson}, {Eisenstein}, {Ellsworth}, {Elsworth}, {Epstein}, {Eracleous},
  {Escoffier}, {Esposito}, {Evans}, {Fan}, {Fern{\'a}ndez-Alvar}, {Feuillet},
  {Filiz Ak}, {Finley}, {Finoguenov}, {Flaherty}, {Fleming}, {Font-Ribera},
  {Foster}, {Frinchaboy}, {Galbraith-Frew}, {Garc{\'\i}a},
  {Garc{\'\i}a-Hern{\'a}ndez}, {Garc{\'\i}a P{\'e}rez}, {Gaulme}, {Ge},
  {G{\'e}nova-Santos}, {Georgakakis}, {Ghezzi}, {Gillespie}, {Girardi},
  {Goddard}, {Gontcho}, {Gonz{\'a}lez Hern{\'a}ndez}, {Grebel}, {Green},
  {Grieb}, {Grieves}, {Gunn}, {Guo}, {Harding}, {Hasselquist}, {Hawley},
  {Hayden}, {Hearty}, {Hekker}, {Ho}, {Hogg}, {Holley-Bockelmann}, {Holtzman},
  {Honscheid}, {Huber}, {Huehnerhoff}, {Ivans}, {Jiang}, {Johnson},
  {Kinemuchi}, {Kirkby}, {Kitaura}, {Klaene}, {Knapp}, {Kneib}, {Koenig},
  {Lam}, {Lan}, {Lang}, {Laurent}, {Le Goff}, {Leauthaud}, {Lee}, {Lee},
  {Licquia}, {Liu}, {Long}, {L{\'o}pez-Corredoira}, {Lorenzo-Oliveira},
  {Lucatello}, {Lundgren}, {Lupton}, {Mack}, {Mahadevan}, {Maia}, {Majewski},
  {Malanushenko}, {Malanushenko}, {Manchado}, {Manera}, {Mao}, {Maraston},
  {Marchwinski}, {Margala}, {Martell}, {Martig}, {Masters}, {Mathur},
  {McBride}, {McGehee}, {McGreer}, {McMahon}, {M{\'e}nard}, {Menzel},
  {Merloni}, {M{\'e}sz{\'a}ros}, {Miller}, {Miralda-Escud{\'e}}, {Miyatake},
  {Montero-Dorta}, {More}, {Morganson}, {Morice-Atkinson}, {Morrison},
  {Mosser}, {Muna}, {Myers}, {Nandra}, {Newman}, {Neyrinck}, {Nguyen},
  {Nichol}, {Nidever}, {Noterdaeme}, {Nuza}, {O'Connell}, {O'Connell},
  {O'Connell}, {Ogando}, {Olmstead}, {Oravetz}, {Oravetz}, {Osumi}, {Owen},
  {Padgett}, {Padmanabhan}, {Paegert}, {Palanque-Delabrouille}, {Pan},
  {Parejko}, {P{\^a}ris}, {Park}, {Pattarakijwanich}, {Pellejero-Ibanez},
  {Pepper}, {Percival}, {P{\'e}rez-Fournon}, {P{\textasciiacute}rez-Ra`fols},
  {Petitjean}, {Pieri}, {Pinsonneault}, {Porto de Mello}, {Prada}, {Prakash},
  {Price-Whelan}, {Protopapas}, {Raddick}, {Rahman}, {Reid}, {Rich}, {Rix},
  {Robin}, {Rockosi}, {Rodrigues}, {Rodr{\'\i}guez-Torres}, {Roe}, {Ross},
  {Ross}, {Rossi}, {Ruan}, {Rubi{\~n}o-Mart{\'\i}n}, {Rykoff},
  {Salazar-Albornoz}, {Salvato}, {Samushia}, {S{\'a}nchez}, {Santiago},
  {Sayres}, {Schiavon}, {Schlegel}, {Schmidt}, {Schneider}, {Schultheis},
  {Schwope}, {Sc{\'o}ccola}, {Scott}, {Sellgren}, {Seo}, {Serenelli}, {Shane},
  {Shen}, {Shetrone}, {Shu}, {Silva Aguirre}, {Sivarani}, {Skrutskie},
  {Slosar}, {Smith}, {Sobreira}, {Souto}, {Stassun}, {Steinmetz}, {Stello},
  {Strauss}, {Streblyanska}, {Suzuki}, {Swanson}, {Tan}, {Tayar}, {Terrien},
  {Thakar}, {Thomas}, {Thomas}, {Thompson}, {Tinker}, {Tojeiro}, {Troup},
  {Vargas-Maga{\~n}a}, {Vazquez}, {Verde}, {Viel}, {Vogt}, {Wake}, {Wang},
  {Weaver}, {Weinberg}, {Weiner}, {White}, {Wilson}, {Wisniewski},
  {Wood-Vasey}, {Ye`che}, {York}, {Zakamska}, {Zamora}, {Zasowski}, {Zehavi},
  {Zhao}, {Zheng}, {Zhou}, {Zhou}, {Zou}, \& {Zhu}}]{Alam2015}
{Alam}, S., {Albareti}, F.~D., {Allende Prieto}, C., {et~al.} 2015, \apjs, 219,
  12

\bibitem[{{Aller} {et~al.}(2020){Aller}, {Lillo-Box}, {Jones}, {Miranda}, \&
  {Barcel{\'o} Forteza}}]{Aller2020_tpf}
{Aller}, A., {Lillo-Box}, J., {Jones}, D., {Miranda}, L.~F., \& {Barcel{\'o}
  Forteza}, S. 2020, \aap, 635, A128

\bibitem[{{Ambikasaran} {et~al.}(2015){Ambikasaran}, {Foreman-Mackey},
  {Greengard}, {Hogg}, \& {O'Neil}}]{george}
{Ambikasaran}, S., {Foreman-Mackey}, D., {Greengard}, L., {Hogg}, D.~W., \&
  {O'Neil}, M. 2015, IEEE Transactions on Pattern Analysis and Machine
  Intelligence, 38, 252

\bibitem[{{Anderson} {et~al.}(1990){Anderson}, {Duvall}, \&
  {Jefferies}}]{Anderson1990}
{Anderson}, E.~R., {Duvall}, Thomas~L., J., \& {Jefferies}, S.~M. 1990, \apj,
  364, 699

\bibitem[{{Barnes} {et~al.}(2009){Barnes}, {Jackson}, {Raymond}, {West}, \&
  {Greenberg}}]{Barnes2009}
{Barnes}, R., {Jackson}, B., {Raymond}, S.~N., {West}, A.~A., \& {Greenberg},
  R. 2009, \apj, 695, 1006

\bibitem[{{Barrag{\'a}n} {et~al.}(2019){Barrag{\'a}n}, {Aigrain}, {Kubyshkina},
  {Gandolfi}, {Livingston}, {Fridlund}, {Fossati}, {Korth}, {Parviainen},
  {Malavolta}, {Palle}, {Deeg}, {Nowak}, {Rajpaul}, {Zicher}, {Antoniciello},
  {Narita}, {Albrecht}, {Bedin}, {Cabrera}, {Cochran}, {de Leon},
  {Eigm{\"u}ller}, {Fukui}, {Granata}, {Grziwa}, {Guenther}, {Hatzes},
  {Kusakabe}, {Latham}, {Libralato}, {Luque},
  {Monta{\~n}{\'e}s-Rodr{\'\i}guez}, {Murgas}, {Nardiello}, {Pagano}, {Piotto},
  {Persson}, {Redfield}, \& {Tamura}}]{Barragan2019_K2-100}
{Barrag{\'a}n}, O., {Aigrain}, S., {Kubyshkina}, D., {et~al.} 2019, \mnras,
  490, 698

\bibitem[{{Bauer} {et~al.}(2020){Bauer}, {Zechmeister}, {Kaminski},
  {Rodr{\'\i}guez L{\'o}pez}, {Caballero}, {Azzaro}, {Stahl}, {Kossakowski},
  {Quirrenbach}, {Becerril Jarque}, {Rodr{\'\i}guez}, {Amado}, {Seifert},
  {Reiners}, {Sch{\"a}fer}, {Ribas}, {B{\'e}jar}, {Cort{\'e}s-Contreras},
  {Dreizler}, {Hatzes}, {Henning}, {Jeffers}, {K{\"u}rster}, {Lafarga},
  {Montes}, {Morales}, {Schmitt}, {Schweitzer}, \& {Solano}}]{Bauer2020}
{Bauer}, F.~F., {Zechmeister}, M., {Kaminski}, A., {et~al.} 2020, \aap, 640,
  A50

\bibitem[{{Bean} {et~al.}(2021){Bean}, {Raymond}, \& {Owen}}]{Bean2021}
{Bean}, J.~L., {Raymond}, S.~N., \& {Owen}, J.~E. 2021, JGR (Planets), 126,
  e06639

\bibitem[{{Benneke} \& {Seager}(2013)}]{BennekeSeager2013}
{Benneke}, B. \& {Seager}, S. 2013, \apj, 778, 153

\bibitem[{{Benz} {et~al.}(2021){Benz}, {Broeg}, {Fortier}, {Rando}, {Beck},
  {Beck}, {Queloz}, {Ehrenreich}, {Maxted}, {Isaak}, {Billot}, {Alibert},
  {Alonso}, {Ant{\'o}nio}, {Asquier}, {Bandy}, {B{\'a}rczy}, {Barrado},
  {Barros}, {Baumjohann}, {Bekkelien}, {Bergomi}, {Biondi}, {Bonfils},
  {Borsato}, {Brandeker}, {Busch}, {Cabrera}, {Cessa}, {Charnoz}, {Chazelas},
  {Collier Cameron}, {Corral Van Damme}, {Cortes}, {Davies}, {Deleuil},
  {Deline}, {Delrez}, {Demangeon}, {Demory}, {Erikson}, {Farinato}, {Fossati},
  {Fridlund}, {Futyan}, {Gandolfi}, {Garcia Munoz}, {Gillon}, {Guterman},
  {Gutierrez}, {Hasiba}, {Heng}, {Hernandez}, {Hoyer}, {Kiss}, {Kovacs},
  {Kuntzer}, {Laskar}, {Lecavelier des Etangs}, {Lendl}, {L{\'o}pez}, {Lora},
  {Lovis}, {L{\"u}ftinger}, {Magrin}, {Malvasio}, {Marafatto}, {Michaelis}, {de
  Miguel}, {Modrego}, {Munari}, {Nascimbeni}, {Olofsson}, {Ottacher},
  {Ottensamer}, {Pagano}, {Palacios}, {Pall{\'e}}, {Peter}, {Piazza}, {Piotto},
  {Pizarro}, {Pollaco}, {Ragazzoni}, {Ratti}, {Rauer}, {Ribas}, {Rieder},
  {Rohlfs}, {Safa}, {Salatti}, {Santos}, {Scandariato}, {S{\'e}gransan},
  {Simon}, {Smith}, {Sordet}, {Sousa}, {Steller}, {Szab{\'o}}, {Szoke},
  {Thomas}, {Tschentscher}, {Udry}, {Van Grootel}, {Viotto}, {Walter},
  {Walton}, {Wildi}, \& {Wolter}}]{cheops}
{Benz}, W., {Broeg}, C., {Fortier}, A., {et~al.} 2021, Experimental Astronomy,
  51, 109

\bibitem[{{Bianchi} {et~al.}(2011){Bianchi}, {Herald}, {Efremova}, {Girardi},
  {Zabot}, {Marigo}, {Conti}, \& {Shiao}}]{Bianchi2011}
{Bianchi}, L., {Herald}, J., {Efremova}, B., {et~al.} 2011, \apss, 335, 161

\bibitem[{{Bitsch} {et~al.}(2019){Bitsch}, {Raymond}, \&
  {Izidoro}}]{Bitsch2019}
{Bitsch}, B., {Raymond}, S.~N., \& {Izidoro}, A. 2019, \aap, 624, A109

\bibitem[{{Bonfils} {et~al.}(2013){Bonfils}, {Delfosse}, {Udry}, {Forveille},
  {Mayor}, {Perrier}, {Bouchy}, {Gillon}, {Lovis}, {Pepe}, {Queloz}, {Santos},
  {S{\'e}gransan}, \& {Bertaux}}]{Bonfils2013}
{Bonfils}, X., {Delfosse}, X., {Udry}, S., {et~al.} 2013, \aap, 549, A109

\bibitem[{{Borucki} {et~al.}(2010){Borucki}, {Koch}, {Basri}, {Batalha},
  {Brown}, {Caldwell}, {Caldwell}, {Christensen-Dalsgaard}, {Cochran},
  {DeVore}, {Dunham}, {Dupree}, {Gautier}, {Geary}, {Gilliland}, {Gould},
  {Howell}, {Jenkins}, {Kondo}, {Latham}, {Marcy}, {Meibom}, {Kjeldsen},
  {Lissauer}, {Monet}, {Morrison}, {Sasselov}, {Tarter}, {Boss}, {Brownlee},
  {Owen}, {Buzasi}, {Charbonneau}, {Doyle}, {Fortney}, {Ford}, {Holman},
  {Seager}, {Steffen}, {Welsh}, {Rowe}, {Anderson}, {Buchhave}, {Ciardi},
  {Walkowicz}, {Sherry}, {Horch}, {Isaacson}, {Everett}, {Fischer}, {Torres},
  {Johnson}, {Endl}, {MacQueen}, {Bryson}, {Dotson}, {Haas}, {Kolodziejczak},
  {Van Cleve}, {Chandrasekaran}, {Twicken}, {Quintana}, {Clarke}, {Allen},
  {Li}, {Wu}, {Tenenbaum}, {Verner}, {Bruhweiler}, {Barnes}, \&
  {Prsa}}]{Borucki2010_Kepler}
{Borucki}, W.~J., {Koch}, D., {Basri}, G., {et~al.} 2010, Science, 327, 977

\bibitem[{{Bou{\'e}} {et~al.}(2013){Bou{\'e}}, {Montalto}, {Boisse}, {Oshagh},
  \& {Santos}}]{Boue2013}
{Bou{\'e}}, G., {Montalto}, M., {Boisse}, I., {Oshagh}, M., \& {Santos}, N.~C.
  2013, \aap, 550, A53

\bibitem[{{Bourrier} {et~al.}(2018){Bourrier}, {Lovis}, {Beust}, {Ehrenreich},
  {Henry}, {Astudillo-Defru}, {Allart}, {Bonfils}, {S{\'e}gransan}, {Delfosse},
  {Cegla}, {Wyttenbach}, {Heng}, {Lavie}, \& {Pepe}}]{Bourrier2018_RM_GJ436b}
{Bourrier}, V., {Lovis}, C., {Beust}, H., {et~al.} 2018, \nat, 553, 477

\bibitem[{Brown {et~al.}(2013)Brown, Baliber, Bianco, Bowman, Burleson, Conway,
  Crellin, Depagne, Vera, Dilday, Dragomir, Dubberley, Eastman, Elphick,
  Falarski, Foale, Ford, Fulton, Garza, Gomez, Graham, Greene, Haldeman,
  Hawkins, Haworth, Haynes, Hidas, Hjelstrom, Howell, Hygelund, Lister,
  Lobdill, Martinez, Mullins, Norbury, Parrent, Paulson, Petry, Pickles,
  Posner, Rosing, Ross, Sand, Saunders, Shobbrook, Shporer, Street, Thomas,
  Tsapras, Tufts, Valenti, Horst, Walker, White, \& Willis}]{Brown2013_LCO}
Brown, T.~M., Baliber, N., Bianco, F.~B., {et~al.} 2013, PASP, 125, 1031

\bibitem[{{Burn} {et~al.}(2021){Burn}, {Schlecker}, {Mordasini}, {Emsenhuber},
  {Alibert}, {Henning}, {Klahr}, \& {Benz}}]{Burn2021}
{Burn}, R., {Schlecker}, M., {Mordasini}, C., {et~al.} 2021, arXiv e-prints,
  arXiv:2105.04596

\bibitem[{{Caballero}(2010)}]{Caballero2010}
{Caballero}, J.~A. 2010, \aap, 514, A18

\bibitem[{{Caballero} {et~al.}(2016{\natexlab{a}}){Caballero},
  {Cort{\'e}s-Contreras}, {Alonso-Floriano}, {Montes}, {Quirrenbach}, {Amado},
  {Ribas}, {Reiners}, {Abellan}, {B{\'e}jar}, {Brinkm{\"o}ller}, {Czesla},
  {Dorda}, {Gallardo}, {Gonz{\'a}lez-{\'A}lvarez}, {Hidalgo}, {Holgado},
  {Jeffers}, {Kim}, {Klutsch}, {Lamert}, {Llamas}, {L{\'o}pez-Santiago},
  {Mart{\'\i}nez-Rodr{\'\i}guez}, {Morales}, {Mundt}, {Passegger},
  {Sch{\"o}fer}, {Seifert}, \& {Zechmeister}}]{Caballero2016}
{Caballero}, J.~A., {Cort{\'e}s-Contreras}, M., {Alonso-Floriano}, F.~J.,
  {et~al.} 2016{\natexlab{a}}, in 19th Cambridge Workshop on Cool Stars,
  Stellar Systems, and the Sun (CS19), 148

\bibitem[{{Caballero} {et~al.}(2016{\natexlab{b}}){Caballero}, {Gu{\`a}rdia},
  {L{\'o}pez del Fresno}, {Zechmeister}, {de Juan}, {Alonso-Floriano}, {Amado},
  {Colom{\'e}}, {Cort{\'e}s-Contreras}, {Garc{\'\i}a-Piquer}, {Gesa}, {de
  Guindos}, {Hagen}, {Helmling}, {Hern{\'a}ndez Casta{\~n}o}, {K{\"u}rster},
  {L{\'o}pez-Santiago}, {Montes}, {Morales Mu{\~n}oz}, {Pavlov}, {Quirrenbach},
  {Reiners}, {Ribas}, {Seifert}, \& {Solano}}]{Caballero2016_SPIE}
{Caballero}, J.~A., {Gu{\`a}rdia}, J., {L{\'o}pez del Fresno}, M., {et~al.}
  2016{\natexlab{b}}, in Society of Photo-Optical Instrumentation Engineers
  (SPIE) Conference Series, Vol. 9910, Observatory Operations: Strategies,
  Processes, and Systems VI, ed. A.~B. {Peck}, R.~L. {Seaman}, \& C.~R. {Benn},
  99100E

\bibitem[{{Chen} {et~al.}(2021){Chen}, {Pall{\'e}}, {Parviainen}, {Wang}, {van
  Boekel}, {Murgas}, {Yan}, {B{\'e}jar}, {Casasayas-Barris}, {Crouzet},
  {Esparza-Borges}, {Fukui}, {Garai}, {Kawauchi}, {Kurita}, {Kusakabe}, {de
  Leon}, {Livingston}, {Luque}, {Madrigal-Aguado}, {Mori}, {Narita},
  {Nishiumi}, {Oshagh}, {S{\'a}nchez-Benavente}, {Tamura}, {Terada}, \&
  {Watanabe}}]{Chen2021_transmission}
{Chen}, G., {Pall{\'e}}, E., {Parviainen}, H., {et~al.} 2021, \mnras, 500, 5420

\bibitem[{{Cifuentes} {et~al.}(2020){Cifuentes}, {Caballero},
  {Cort{\'e}s-Contreras}, {Montes}, {Abell{\'a}n}, {Dorda}, {Holgado},
  {Zapatero Osorio}, {Morales}, {Amado}, {Passegger}, {Quirrenbach}, {Reiners},
  {Ribas}, {Sanz-Forcada}, {Schweitzer}, {Seifert}, \&
  {Solano}}]{Cifuentes2020}
{Cifuentes}, C., {Caballero}, J.~A., {Cort{\'e}s-Contreras}, M., {et~al.} 2020,
  \aap, 642, A115

\bibitem[{{Cloutier} {et~al.}(2021){Cloutier}, {Charbonneau}, {Deming},
  {Bonfils}, \& {Astudillo-Defru}}]{Cloutier2021}
{Cloutier}, R., {Charbonneau}, D., {Deming}, D., {Bonfils}, X., \&
  {Astudillo-Defru}, N. 2021, arXiv e-prints, arXiv:2107.14732

\bibitem[{{Cloutier} {et~al.}(2020){Cloutier}, {Eastman}, {Rodriguez},
  {Astudillo-Defru}, {Bonfils}, {Mortier}, {Watson}, {Stalport}, {Pinamonti},
  {Lienhard}, {Harutyunyan}, {Damasso}, {Latham}, {Collins}, {Massey}, {Irwin},
  {Winters}, {Charbonneau}, {Ziegler}, {Matthews}, {Crossfield}, {Kreidberg},
  {Quinn}, {Ricker}, {Vanderspek}, {Seager}, {Winn}, {Jenkins}, {Vezie},
  {Udry}, {Twicken}, {Tenenbaum}, {Sozzetti}, {S{\'e}gransan}, {Schlieder},
  {Sasselov}, {Santos}, {Rice}, {Rackham}, {Poretti}, {Piotto}, {Phillips},
  {Pepe}, {Molinari}, {Mignon}, {Micela}, {Melo}, {de Medeiros}, {Mayor},
  {Matson}, {Martinez Fiorenzano}, {Mann}, {Magazz{\'u}}, {Lovis},
  {L{\'o}pez-Morales}, {Lopez}, {Lissauer}, {L{\'e}pine}, {Law}, {Kielkopf},
  {Johnson}, {Jensen}, {Howell}, {Gonzales}, {Ghedina}, {Forveille},
  {Figueira}, {Dumusque}, {Dressing}, {Doyon}, {D{\'\i}az}, {Fabrizio},
  {Delfosse}, {Cosentino}, {Conti}, {Collins}, {Cameron}, {Ciardi}, {Caldwell},
  {Burke}, {Buchhave}, {Brice{\~n}o}, {Boyd}, {Bouchy}, {Beichman}, {Artigau},
  \& {Almenara}}]{Cloutier2020_toi732}
{Cloutier}, R., {Eastman}, J.~D., {Rodriguez}, J.~E., {et~al.} 2020, \aj, 160,
  3

\bibitem[{{Cloutier} \& {Menou}(2020)}]{CloutierMenou2020}
{Cloutier}, R. \& {Menou}, K. 2020, \aj, 159, 211

\bibitem[{{Collins} {et~al.}(2017){Collins}, {Kielkopf}, {Stassun}, \&
  {Hessman}}]{Collins:2017}
{Collins}, K.~A., {Kielkopf}, J.~F., {Stassun}, K.~G., \& {Hessman}, F.~V.
  2017, \aj, 153, 77

\bibitem[{{Cort{\'e}s-Contreras} {et~al.}(2017){Cort{\'e}s-Contreras},
  {B{\'e}jar}, {Caballero}, {Gauza}, {Montes}, {Alonso-Floriano}, {Jeffers},
  {Morales}, {Reiners}, {Ribas}, {Sch{\"o}fer}, {Quirrenbach}, {Amado},
  {Mundt}, \& {Seifert}}]{Cortes-Contreras2017}
{Cort{\'e}s-Contreras}, M., {B{\'e}jar}, V.~J.~S., {Caballero}, J.~A., {et~al.}
  2017, \aap, 597, A47

\bibitem[{{Crossfield} \& {Kreidberg}(2017)}]{CrossfieldKreidberg2017}
{Crossfield}, I. J.~M. \& {Kreidberg}, L. 2017, \aj, 154, 261

\bibitem[{{Curtis} {et~al.}(2019){Curtis}, {Ag{\"u}eros}, {Mamajek}, {Wright},
  \& {Cummings}}]{Curtis2019}
{Curtis}, J.~L., {Ag{\"u}eros}, M.~A., {Mamajek}, E.~E., {Wright}, J.~T., \&
  {Cummings}, J.~D. 2019, \aj, 158, 77

\bibitem[{{Cutri} \& {et al.}(2012)}]{Cutri2012}
{Cutri}, R.~M. \& {et al.} 2012, VizieR Online Data Catalog, 2311

\bibitem[{{Cutri} \& {et al.}(2014)}]{Cutri2014}
{Cutri}, R.~M. \& {et al.} 2014, VizieR Online Data Catalog, 2328

\bibitem[{{Dawson} \& {Fabrycky}(2010)}]{DawsonFabrycky2010}
{Dawson}, R.~I. \& {Fabrycky}, D.~C. 2010, \apj, 722, 937

\bibitem[{{Deacon} {et~al.}(2016){Deacon}, {Kraus}, {Mann}, {Magnier},
  {Chambers}, {Wainscoat}, {Tonry}, {Kaiser}, {Waters}, {Flewelling}, {Hodapp},
  \& {Burgett}}]{Deacon2016}
{Deacon}, N.~R., {Kraus}, A.~L., {Mann}, A.~W., {et~al.} 2016, \mnras, 455,
  4212

\bibitem[{{Desidera} \& {Barbieri}(2007)}]{DesideraBarbieri2007}
{Desidera}, S. \& {Barbieri}, M. 2007, \aap, 462, 345

\bibitem[{Diamond-Lowe {et~al.}(2018)Diamond-Lowe, Berta-Thompson, Charbonneau,
  \& Kempton}]{Diamond-Lowe2018}
Diamond-Lowe, H., Berta-Thompson, Z., Charbonneau, D., \& Kempton, E. M.-R.
  2018, The Astronomical Journal, 156, 42

\bibitem[{{Douglas} {et~al.}(2019){Douglas}, {Curtis}, {Ag{\"u}eros},
  {Cargile}, {Brewer}, {Meibom}, \& {Jansen}}]{Douglas2019}
{Douglas}, S.~T., {Curtis}, J.~L., {Ag{\"u}eros}, M.~A., {et~al.} 2019, \apj,
  879, 100

\bibitem[{{Dressing} \& {Charbonneau}(2013)}]{DressingCharbonneau2013}
{Dressing}, C.~D. \& {Charbonneau}, D. 2013, \apj, 767, 95

\bibitem[{{Dressing} \& {Charbonneau}(2015)}]{DressingCharbonneau2015}
{Dressing}, C.~D. \& {Charbonneau}, D. 2015, \apj, 807, 45

\bibitem[{{Duquennoy} \& {Mayor}(1991)}]{DuquennoyMayor1991}
{Duquennoy}, A. \& {Mayor}, M. 1991, \aap, 500, 337

\bibitem[{{Eggen}(1958)}]{Eggen1958}
{Eggen}, O.~J. 1958, \mnras, 118, 65

\bibitem[{{Eggen}(1984)}]{Eggen1984}
{Eggen}, O.~J. 1984, \aj, 89, 1358

\bibitem[{{El-Badry} \& {Rix}(2018)}]{El-BradyRix2018}
{El-Badry}, K. \& {Rix}, H.-W. 2018, \mnras, 480, 4884

\bibitem[{Emsenhuber {et~al.}(2020)Emsenhuber, Mordasini, Burn, Alibert, Benz,
  \& Asphaug}]{Emsenhuber2020}
Emsenhuber, A., Mordasini, C., Burn, R., {et~al.} 2020
  [\eprint[arXiv]{2007.05561}]

\bibitem[{{Epchtein} {et~al.}(1997){Epchtein}, {de Batz}, {Capoani},
  {Chevallier}, {Copet}, {Fouqu{\'e}}, {Lacombe}, {Le Bertre}, {Pau}, {Rouan},
  {Ruphy}, {Simon}, {Tiph{\`e}ne}, {Burton}, {Bertin}, {Deul}, {Habing},
  {Borsenberger}, {Dennefeld}, {Guglielmo}, {Loup}, {Mamon}, {Ng}, {Omont},
  {Provost}, {Renault}, {Tanguy}, {Kimeswenger}, {Kienel}, {Garzon}, {Persi},
  {Ferrari-Toniolo}, {Robin}, {Paturel}, {Vauglin}, {Forveille}, {Delfosse},
  {Hron}, {Schultheis}, {Appenzeller}, {Wagner}, {Balazs}, {Holl},
  {L{\'e}pine}, {Boscolo}, {Picazzio}, {Duc}, \& {Mennessier}}]{denis}
{Epchtein}, N., {de Batz}, B., {Capoani}, L., {et~al.} 1997, The Messenger, 87,
  27

\bibitem[{Espinoza(2018)}]{Espinoza2018_pb}
Espinoza, N. 2018, Research Notes of the AAS, 2, 209

\bibitem[{{Espinoza} \& {Jord{\'a}n}(2016)}]{EspinozaJordan2016}
{Espinoza}, N. \& {Jord{\'a}n}, A. 2016, \mnras, 457, 3573

\bibitem[{{Espinoza} {et~al.}(2019{\natexlab{a}}){Espinoza}, {Kossakowski}, \&
  {Brahm}}]{juliet}
{Espinoza}, N., {Kossakowski}, D., \& {Brahm}, R. 2019{\natexlab{a}}, \mnras,
  490, 2262

\bibitem[{{Espinoza} {et~al.}(2019{\natexlab{b}}){Espinoza}, {Rackham},
  {Jord{\'a}n}, {Apai}, {L{\'o}pez-Morales}, {Osip}, {Grimm}, {Hoeijmakers},
  {Wilson}, {Bixel}, {McGruder}, {Rodler}, {Weaver}, {Lewis}, {Fortney}, \&
  {Fraine}}]{Espinoza2019}
{Espinoza}, N., {Rackham}, B.~V., {Jord{\'a}n}, A., {et~al.}
  2019{\natexlab{b}}, \mnras, 482, 2065

\bibitem[{{Fang} {et~al.}(2018){Fang}, {Zhao}, {Zhao}, \& {Bharat
  Kumar}}]{Fang2018}
{Fang}, X.-S., {Zhao}, G., {Zhao}, J.-K., \& {Bharat Kumar}, Y. 2018, \mnras,
  476, 908

\bibitem[{{Feroz} {et~al.}(2011){Feroz}, {Balan}, \& {Hobson}}]{Feroz2011}
{Feroz}, F., {Balan}, S.~T., \& {Hobson}, M.~P. 2011, \mnras, 415, 3462

\bibitem[{{Foreman-Mackey}(2018)}]{celerite2}
{Foreman-Mackey}, D. 2018, RNAAS, 2, 31

\bibitem[{{Foreman-Mackey} {et~al.}(2017){Foreman-Mackey}, {Agol},
  {Ambikasaran}, \& {Angus}}]{celerite}
{Foreman-Mackey}, D., {Agol}, E., {Ambikasaran}, S., \& {Angus}, R. 2017, \aj,
  154, 220

\bibitem[{{Frith} {et~al.}(2013){Frith}, {Pinfield}, {Jones}, {Barnes},
  {Pavlenko}, {Martin}, {Brown}, {Kuznetsov}, {Marocco}, {Tata}, \&
  {Cappetta}}]{Frith2013}
{Frith}, J., {Pinfield}, D.~J., {Jones}, H.~R.~A., {et~al.} 2013, \mnras, 435,
  2161

\bibitem[{{Fulton} \& {Petigura}(2018)}]{FultonPetigura2018}
{Fulton}, B.~J. \& {Petigura}, E.~A. 2018, \aj, 156, 264

\bibitem[{{Fulton} {et~al.}(2018){Fulton}, {Petigura}, {Blunt}, \&
  {Sinukoff}}]{radvel}
{Fulton}, B.~J., {Petigura}, E.~A., {Blunt}, S., \& {Sinukoff}, E. 2018, \pasp,
  130, 044504

\bibitem[{{Fulton} {et~al.}(2017){Fulton}, {Petigura}, {Howard}, {Isaacson},
  {Marcy}, {Cargile}, {Hebb}, {Weiss}, {Johnson}, {Morton}, {Sinukoff},
  {Crossfield}, \& {Hirsch}}]{Fulton2017}
{Fulton}, B.~J., {Petigura}, E.~A., {Howard}, A.~W., {et~al.} 2017, \aj, 154,
  109

\bibitem[{{Gaia Collaboration} {et~al.}(2018){Gaia Collaboration}, {Brown},
  {Vallenari}, {Prusti}, {de Bruijne}, {Babusiaux}, {Bailer-Jones}, {Biermann},
  {Evans}, {Eyer}, {Jansen}, {Jordi}, {Klioner}, {Lammers}, {Lindegren},
  {Luri}, {Mignard}, {Panem}, {Pourbaix}, {Randich}, {Sartoretti}, {Siddiqui},
  {Soubiran}, {van Leeuwen}, {Walton}, {Arenou}, {Bastian}, {Cropper},
  {Drimmel}, {Katz}, {Lattanzi}, {Bakker}, {Cacciari}, {Casta{\~n}eda},
  {Chaoul}, {Cheek}, {De Angeli}, {Fabricius}, {Guerra}, {Holl}, {Masana},
  {Messineo}, {Mowlavi}, {Nienartowicz}, {Panuzzo}, {Portell}, {Riello},
  {Seabroke}, {Tanga}, {Th{\'e}venin}, {Gracia-Abril}, {Comoretto},
  {Garcia-Reinaldos}, {Teyssier}, {Altmann}, {Andrae}, {Audard},
  {Bellas-Velidis}, {Benson}, {Berthier}, {Blomme}, {Burgess}, {Busso},
  {Carry}, {Cellino}, {Clementini}, {Clotet}, {Creevey}, {Davidson}, {De
  Ridder}, {Delchambre}, {Dell'Oro}, {Ducourant},
  {Fern{\'a}ndez-Hern{\'a}ndez}, {Fouesneau}, {Fr{\'e}mat}, {Galluccio},
  {Garc{\'\i}a-Torres}, {Gonz{\'a}lez-N{\'u}{\~n}ez}, {Gonz{\'a}lez-Vidal},
  {Gosset}, {Guy}, {Halbwachs}, {Hambly}, {Harrison}, {Hern{\'a}ndez},
  {Hestroffer}, {Hodgkin}, {Hutton}, {Jasniewicz}, {Jean-Antoine-Piccolo},
  {Jordan}, {Korn}, {Krone-Martins}, {Lanzafame}, {Lebzelter}, {L{\"o}ffler},
  {Manteiga}, {Marrese}, {Mart{\'\i}n-Fleitas}, {Moitinho}, {Mora}, {Muinonen},
  {Osinde}, {Pancino}, {Pauwels}, {Petit}, {Recio-Blanco}, {Richards},
  {Rimoldini}, {Robin}, {Sarro}, {Siopis}, {Smith}, {Sozzetti}, {S{\"u}veges},
  {Torra}, {van Reeven}, {Abbas}, {Abreu Aramburu}, {Accart}, {Aerts},
  {Altavilla}, {{\'A}lvarez}, {Alvarez}, {Alves}, {Anderson}, {Andrei},
  {Anglada Varela}, {Antiche}, {Antoja}, {Arcay}, {Astraatmadja}, {Bach},
  {Baker}, {Balaguer-N{\'u}{\~n}ez}, {Balm}, {Barache}, {Barata}, {Barbato},
  {Barblan}, {Barklem}, {Barrado}, {Barros}, {Barstow}, {Bartholom{\'e}
  Mu{\~n}oz}, {Bassilana}, {Becciani}, {Bellazzini}, {Berihuete}, {Bertone},
  {Bianchi}, {Bienaym{\'e}}, {Blanco-Cuaresma}, {Boch}, {Boeche}, {Bombrun},
  {Borrachero}, {Bossini}, {Bouquillon}, {Bourda}, {Bragaglia}, {Bramante},
  {Breddels}, {Bressan}, {Brouillet}, {Br{\"u}semeister}, {Brugaletta},
  {Bucciarelli}, {Burlacu}, {Busonero}, {Butkevich}, {Buzzi}, {Caffau},
  {Cancelliere}, {Cannizzaro}, {Cantat-Gaudin}, {Carballo}, {Carlucci},
  {Carrasco}, {Casamiquela}, {Castellani}, {Castro-Ginard}, {Charlot},
  {Chemin}, {Chiavassa}, {Cocozza}, {Costigan}, {Cowell}, {Crifo}, {Crosta},
  {Crowley}, {Cuypers}, {Dafonte}, {Damerdji}, {Dapergolas}, {David}, {David},
  {de Laverny}, {De Luise}, {De March}, {de Martino}, {de Souza}, {de Torres},
  {Debosscher}, {del Pozo}, {Delbo}, {Delgado}, {Delgado}, {Di Matteo},
  {Diakite}, {Diener}, {Distefano}, {Dolding}, {Drazinos}, {Dur{\'a}n},
  {Edvardsson}, {Enke}, {Eriksson}, {Esquej}, {Eynard Bontemps}, {Fabre},
  {Fabrizio}, {Faigler}, {Falc{\~a}o}, {Farr{\`a}s Casas}, {Federici},
  {Fedorets}, {Fernique}, {Figueras}, {Filippi}, {Findeisen}, {Fonti},
  {Fraile}, {Fraser}, {Fr{\'e}zouls}, {Gai}, {Galleti}, {Garabato},
  {Garc{\'\i}a-Sedano}, {Garofalo}, {Garralda}, {Gavel}, {Gavras}, {Gerssen},
  {Geyer}, {Giacobbe}, {Gilmore}, {Girona}, {Giuffrida}, {Glass}, {Gomes},
  {Granvik}, {Gueguen}, {Guerrier}, {Guiraud}, {Guti{\'e}rrez-S{\'a}nchez},
  {Haigron}, {Hatzidimitriou}, {Hauser}, {Haywood}, {Heiter}, {Helmi}, {Heu},
  {Hilger}, {Hobbs}, {Hofmann}, {Holland}, {Huckle}, {Hypki}, {Icardi},
  {Jan{\ss}en}, {Jevardat de Fombelle}, {Jonker}, {Juh{\'a}sz}, {Julbe},
  {Karampelas}, {Kewley}, {Klar}, {Kochoska}, {Kohley}, {Kolenberg},
  {Kontizas}, {Kontizas}, {Koposov}, {Kordopatis}, {Kostrzewa-Rutkowska},
  {Koubsky}, {Lambert}, {Lanza}, {Lasne}, {Lavigne}, {Le Fustec}, {Le
  Poncin-Lafitte}, {Lebreton}, {Leccia}, {Leclerc}, {Lecoeur-Taibi},
  {Lenhardt}, {Leroux}, {Liao}, {Licata}, {Lindstr{\o}m}, {Lister}, {Livanou},
  {Lobel}, {L{\'o}pez}, {Managau}, {Mann}, {Mantelet}, {Marchal}, {Marchant},
  {Marconi}, {Marinoni}, {Marschalk{\'o}}, {Marshall}, {Martino}, {Marton},
  {Mary}, {Massari}, {Matijevi{\v{c}}}, {Mazeh}, {McMillan}, {Messina},
  {Michalik}, {Millar}, {Molina}, {Molinaro}, {Moln{\'a}r}, {Montegriffo},
  {Mor}, {Morbidelli}, {Morel}, {Morris}, {Mulone}, {Muraveva}, {Musella},
  {Nelemans}, {Nicastro}, {Noval}, {O'Mullane}, {Ord{\'e}novic},
  {Ord{\'o}{\~n}ez-Blanco}, {Osborne}, {Pagani}, {Pagano}, {Pailler},
  {Palacin}, {Palaversa}, {Panahi}, {Pawlak}, {Piersimoni}, {Pineau}, {Plachy},
  {Plum}, {Poggio}, {Poujoulet}, {Pr{\v{s}}a}, {Pulone}, {Racero}, {Ragaini},
  {Rambaux}, {Ramos-Lerate}, {Regibo}, {Reyl{\'e}}, {Riclet}, {Ripepi}, {Riva},
  {Rivard}, {Rixon}, {Roegiers}, {Roelens}, {Romero-G{\'o}mez}, {Rowell},
  {Royer}, {Ruiz-Dern}, {Sadowski}, {Sagrist{\`a} Sell{\'e}s}, {Sahlmann},
  {Salgado}, {Salguero}, {Sanna}, {Santana-Ros}, {Sarasso}, {Savietto},
  {Schultheis}, {Sciacca}, {Segol}, {Segovia}, {S{\'e}gransan}, {Shih},
  {Siltala}, {Silva}, {Smart}, {Smith}, {Solano}, {Solitro}, {Sordo}, {Soria
  Nieto}, {Souchay}, {Spagna}, {Spoto}, {Stampa}, {Steele},
  {Steidelm{\"u}ller}, {Stephenson}, {Stoev}, {Suess}, {Surdej}, {Szabados},
  {Szegedi-Elek}, {Tapiador}, {Taris}, {Tauran}, {Taylor}, {Teixeira},
  {Terrett}, {Teyssandier}, {Thuillot}, {Titarenko}, {Torra Clotet}, {Turon},
  {Ulla}, {Utrilla}, {Uzzi}, {Vaillant}, {Valentini}, {Valette}, {van Elteren},
  {Van Hemelryck}, {van Leeuwen}, {Vaschetto}, {Vecchiato}, {Veljanoski},
  {Viala}, {Vicente}, {Vogt}, {von Essen}, {Voss}, {Votruba}, {Voutsinas},
  {Walmsley}, {Weiler}, {Wertz}, {Wevers}, {Wyrzykowski}, {Yoldas},
  {{\v{Z}}erjal}, {Ziaeepour}, {Zorec}, {Zschocke}, {Zucker}, {Zurbach}, \&
  {Zwitter}}]{gaiadr2summary}
{Gaia Collaboration}, {Brown}, A.~G.~A., {Vallenari}, A., {et~al.} 2018, \aap,
  616, A1

\bibitem[{{Gaia Collaboration} {et~al.}(2016){Gaia Collaboration}, {Prusti},
  {de Bruijne}, {Brown}, {Vallenari}, {Babusiaux}, {Bailer-Jones}, {Bastian},
  {Biermann}, {Evans}, {Eyer}, {Jansen}, {Jordi}, {Klioner}, {Lammers},
  {Lindegren}, {Luri}, {Mignard}, {Milligan}, {Panem}, {Poinsignon},
  {Pourbaix}, {Randich}, {Sarri}, {Sartoretti}, {Siddiqui}, {Soubiran},
  {Valette}, {van Leeuwen}, {Walton}, {Aerts}, {Arenou}, {Cropper}, {Drimmel},
  {H{\o}g}, {Katz}, {Lattanzi}, {O'Mullane}, {Grebel}, {Holland}, {Huc},
  {Passot}, {Bramante}, {Cacciari}, {Casta{\~n}eda}, {Chaoul}, {Cheek}, {De
  Angeli}, {Fabricius}, {Guerra}, {Hern{\'a}ndez}, {Jean-Antoine-Piccolo},
  {Masana}, {Messineo}, {Mowlavi}, {Nienartowicz}, {Ord{\'o}{\~n}ez-Blanco},
  {Panuzzo}, {Portell}, {Richards}, {Riello}, {Seabroke}, {Tanga},
  {Th{\'e}venin}, {Torra}, {Els}, {Gracia-Abril}, {Comoretto},
  {Garcia-Reinaldos}, {Lock}, {Mercier}, {Altmann}, {Andrae}, {Astraatmadja},
  {Bellas-Velidis}, {Benson}, {Berthier}, {Blomme}, {Busso}, {Carry},
  {Cellino}, {Clementini}, {Cowell}, {Creevey}, {Cuypers}, {Davidson}, {De
  Ridder}, {de Torres}, {Delchambre}, {Dell'Oro}, {Ducourant}, {Fr{\'e}mat},
  {Garc{\'\i}a-Torres}, {Gosset}, {Halbwachs}, {Hambly}, {Harrison}, {Hauser},
  {Hestroffer}, {Hodgkin}, {Huckle}, {Hutton}, {Jasniewicz}, {Jordan},
  {Kontizas}, {Korn}, {Lanzafame}, {Manteiga}, {Moitinho}, {Muinonen},
  {Osinde}, {Pancino}, {Pauwels}, {Petit}, {Recio-Blanco}, {Robin}, {Sarro},
  {Siopis}, {Smith}, {Smith}, {Sozzetti}, {Thuillot}, {van Reeven}, {Viala},
  {Abbas}, {Abreu Aramburu}, {Accart}, {Aguado}, {Allan}, {Allasia},
  {Altavilla}, {{\'A}lvarez}, {Alves}, {Anderson}, {Andrei}, {Anglada Varela},
  {Antiche}, {Antoja}, {Ant{\'o}n}, {Arcay}, {Atzei}, {Ayache}, {Bach},
  {Baker}, {Balaguer-N{\'u}{\~n}ez}, {Barache}, {Barata}, {Barbier}, {Barblan},
  {Baroni}, {Barrado y Navascu{\'e}s}, {Barros}, {Barstow}, {Becciani},
  {Bellazzini}, {Bellei}, {Bello Garc{\'\i}a}, {Belokurov}, {Bendjoya},
  {Berihuete}, {Bianchi}, {Bienaym{\'e}}, {Billebaud}, {Blagorodnova},
  {Blanco-Cuaresma}, {Boch}, {Bombrun}, {Borrachero}, {Bouquillon}, {Bourda},
  {Bouy}, {Bragaglia}, {Breddels}, {Brouillet}, {Br{\"u}semeister},
  {Bucciarelli}, {Budnik}, {Burgess}, {Burgon}, {Burlacu}, {Busonero}, {Buzzi},
  {Caffau}, {Cambras}, {Campbell}, {Cancelliere}, {Cantat-Gaudin}, {Carlucci},
  {Carrasco}, {Castellani}, {Charlot}, {Charnas}, {Charvet}, {Chassat},
  {Chiavassa}, {Clotet}, {Cocozza}, {Collins}, {Collins}, {Costigan}, {Crifo},
  {Cross}, {Crosta}, {Crowley}, {Dafonte}, {Damerdji}, {Dapergolas}, {David},
  {David}, {De Cat}, {de Felice}, {de Laverny}, {De Luise}, {De March}, {de
  Martino}, {de Souza}, {Debosscher}, {del Pozo}, {Delbo}, {Delgado},
  {Delgado}, {di Marco}, {Di Matteo}, {Diakite}, {Distefano}, {Dolding}, {Dos
  Anjos}, {Drazinos}, {Dur{\'a}n}, {Dzigan}, {Ecale}, {Edvardsson}, {Enke},
  {Erdmann}, {Escolar}, {Espina}, {Evans}, {Eynard Bontemps}, {Fabre},
  {Fabrizio}, {Faigler}, {Falc{\~a}o}, {Farr{\`a}s Casas}, {Faye}, {Federici},
  {Fedorets}, {Fern{\'a}ndez-Hern{\'a}ndez}, {Fernique}, {Fienga}, {Figueras},
  {Filippi}, {Findeisen}, {Fonti}, {Fouesneau}, {Fraile}, {Fraser}, {Fuchs},
  {Furnell}, {Gai}, {Galleti}, {Galluccio}, {Garabato}, {Garc{\'\i}a-Sedano},
  {Gar{\'e}}, {Garofalo}, {Garralda}, {Gavras}, {Gerssen}, {Geyer}, {Gilmore},
  {Girona}, {Giuffrida}, {Gomes}, {Gonz{\'a}lez-Marcos},
  {Gonz{\'a}lez-N{\'u}{\~n}ez}, {Gonz{\'a}lez-Vidal}, {Granvik}, {Guerrier},
  {Guillout}, {Guiraud}, {G{\'u}rpide}, {Guti{\'e}rrez-S{\'a}nchez}, {Guy},
  {Haigron}, {Hatzidimitriou}, {Haywood}, {Heiter}, {Helmi}, {Hobbs},
  {Hofmann}, {Holl}, {Holland}, {Hunt}, {Hypki}, {Icardi}, {Irwin}, {Jevardat
  de Fombelle}, {Jofr{\'e}}, {Jonker}, {Jorissen}, {Julbe}, {Karampelas},
  {Kochoska}, {Kohley}, {Kolenberg}, {Kontizas}, {Koposov}, {Kordopatis},
  {Koubsky}, {Kowalczyk}, {Krone-Martins}, {Kudryashova}, {Kull}, {Bachchan},
  {Lacoste-Seris}, {Lanza}, {Lavigne}, {Le Poncin-Lafitte}, {Lebreton},
  {Lebzelter}, {Leccia}, {Leclerc}, {Lecoeur-Taibi}, {Lemaitre}, {Lenhardt},
  {Leroux}, {Liao}, {Licata}, {Lindstr{\o}m}, {Lister}, {Livanou}, {Lobel},
  {L{\"o}ffler}, {L{\'o}pez}, {Lopez-Lozano}, {Lorenz}, {Loureiro},
  {MacDonald}, {Magalh{\~a}es Fernandes}, {Managau}, {Mann}, {Mantelet},
  {Marchal}, {Marchant}, {Marconi}, {Marie}, {Marinoni}, {Marrese},
  {Marschalk{\'o}}, {Marshall}, {Mart{\'\i}n-Fleitas}, {Martino}, {Mary},
  {Matijevi{\v{c}}}, {Mazeh}, {McMillan}, {Messina}, {Mestre}, {Michalik},
  {Millar}, {Miranda}, {Molina}, {Molinaro}, {Molinaro}, {Moln{\'a}r},
  {Moniez}, {Montegriffo}, {Monteiro}, {Mor}, {Mora}, {Morbidelli}, {Morel},
  {Morgenthaler}, {Morley}, {Morris}, {Mulone}, {Muraveva}, {Musella},
  {Narbonne}, {Nelemans}, {Nicastro}, {Noval}, {Ord{\'e}novic},
  {Ordieres-Mer{\'e}}, {Osborne}, {Pagani}, {Pagano}, {Pailler}, {Palacin},
  {Palaversa}, {Parsons}, {Paulsen}, {Pecoraro}, {Pedrosa}, {Pentik{\"a}inen},
  {Pereira}, {Pichon}, {Piersimoni}, {Pineau}, {Plachy}, {Plum}, {Poujoulet},
  {Pr{\v{s}}a}, {Pulone}, {Ragaini}, {Rago}, {Rambaux}, {Ramos-Lerate},
  {Ranalli}, {Rauw}, {Read}, {Regibo}, {Renk}, {Reyl{\'e}}, {Ribeiro},
  {Rimoldini}, {Ripepi}, {Riva}, {Rixon}, {Roelens}, {Romero-G{\'o}mez},
  {Rowell}, {Royer}, {Rudolph}, {Ruiz-Dern}, {Sadowski}, {Sagrist{\`a}
  Sell{\'e}s}, {Sahlmann}, {Salgado}, {Salguero}, {Sarasso}, {Savietto},
  {Schnorhk}, {Schultheis}, {Sciacca}, {Segol}, {Segovia}, {Segransan},
  {Serpell}, {Shih}, {Smareglia}, {Smart}, {Smith}, {Solano}, {Solitro},
  {Sordo}, {Soria Nieto}, {Souchay}, {Spagna}, {Spoto}, {Stampa}, {Steele},
  {Steidelm{\"u}ller}, {Stephenson}, {Stoev}, {Suess}, {S{\"u}veges}, {Surdej},
  {Szabados}, {Szegedi-Elek}, {Tapiador}, {Taris}, {Tauran}, {Taylor},
  {Teixeira}, {Terrett}, {Tingley}, {Trager}, {Turon}, {Ulla}, {Utrilla},
  {Valentini}, {van Elteren}, {Van Hemelryck}, {van Leeuwen}, {Varadi},
  {Vecchiato}, {Veljanoski}, {Via}, {Vicente}, {Vogt}, {Voss}, {Votruba},
  {Voutsinas}, {Walmsley}, {Weiler}, {Weingrill}, {Werner}, {Wevers},
  {Whitehead}, {Wyrzykowski}, {Yoldas}, {{\v{Z}}erjal}, {Zucker}, {Zurbach},
  {Zwitter}, {Alecu}, {Allen}, {Allende Prieto}, {Amorim},
  {Anglada-Escud{\'e}}, {Arsenijevic}, {Azaz}, {Balm}, {Beck}, {Bernstein},
  {Bigot}, {Bijaoui}, {Blasco}, {Bonfigli}, {Bono}, {Boudreault}, {Bressan},
  {Brown}, {Brunet}, {Bunclark}, {Buonanno}, {Butkevich}, {Carret}, {Carrion},
  {Chemin}, {Ch{\'e}reau}, {Corcione}, {Darmigny}, {de Boer}, {de Teodoro}, {de
  Zeeuw}, {Delle Luche}, {Domingues}, {Dubath}, {Fodor}, {Fr{\'e}zouls},
  {Fries}, {Fustes}, {Fyfe}, {Gallardo}, {Gallegos}, {Gardiol}, {Gebran},
  {Gomboc}, {G{\'o}mez}, {Grux}, {Gueguen}, {Heyrovsky}, {Hoar}, {Iannicola},
  {Isasi Parache}, {Janotto}, {Joliet}, {Jonckheere}, {Keil}, {Kim},
  {Klagyivik}, {Klar}, {Knude}, {Kochukhov}, {Kolka}, {Kos}, {Kutka}, {Lainey},
  {LeBouquin}, {Liu}, {Loreggia}, {Makarov}, {Marseille}, {Martayan},
  {Martinez-Rubi}, {Massart}, {Meynadier}, {Mignot}, {Munari}, {Nguyen},
  {Nordlander}, {Ocvirk}, {O'Flaherty}, {Olias Sanz}, {Ortiz}, {Osorio},
  {Oszkiewicz}, {Ouzounis}, {Palmer}, {Park}, {Pasquato}, {Peltzer}, {Peralta},
  {P{\'e}turaud}, {Pieniluoma}, {Pigozzi}, {Poels}, {Prat}, {Prod'homme},
  {Raison}, {Rebordao}, {Risquez}, {Rocca-Volmerange}, {Rosen}, {Ruiz-Fuertes},
  {Russo}, {Sembay}, {Serraller Vizcaino}, {Short}, {Siebert}, {Silva},
  {Sinachopoulos}, {Slezak}, {Soffel}, {Sosnowska}, {Strai{\v{z}}ys}, {ter
  Linden}, {Terrell}, {Theil}, {Tiede}, {Troisi}, {Tsalmantza}, {Tur},
  {Vaccari}, {Vachier}, {Valles}, {Van Hamme}, {Veltz}, {Virtanen}, {Wallut},
  {Wichmann}, {Wilkinson}, {Ziaeepour}, \& {Zschocke}}]{gaiadr1}
{Gaia Collaboration}, {Prusti}, T., {de Bruijne}, J.~H.~J., {et~al.} 2016,
  \aap, 595, A1

\bibitem[{{Gaia Collaboration} {et~al.}(2021){Gaia Collaboration}, {Smart},
  {Sarro}, {Rybizki}, {Reyl{\'e}}, {Robin}, {Hambly}, {Abbas}, {Barstow}, {de
  Bruijne}, \& et~al.}]{GaiaEDR3}
{Gaia Collaboration}, {Smart}, R.~L., {Sarro}, L.~M., {et~al.} 2021, \aap, 649,
  A6

\bibitem[{{Gaidos} {et~al.}(2020){Gaidos}, {Hirano}, {Mann}, {Owens}, {Berger},
  {France}, {Vanderburg}, {Harakawa}, {Hodapp}, {Ishizuka}, {Jacobson},
  {Konishi}, {Kotani}, {Kudo}, {Kurokawa}, {Kuzuhara}, {Nishikawa}, {Omiya},
  {Serizawa}, {Tamura}, \& {Ueda}}]{Gaidos2020}
{Gaidos}, E., {Hirano}, T., {Mann}, A.~W., {et~al.} 2020, \mnras, 495, 650

\bibitem[{{Gardner} {et~al.}(2009){Gardner}, {Mather}, {Clampin}, {Doyon},
  {Flanagan}, {Franx}, {Greenhouse}, {Hammel}, {Hutchings}, {Jakobsen},
  {Lilly}, {Lunine}, {McCaughrean}, {Mountain}, {Rieke}, {Rieke}, {Sonneborn},
  {Stiavelli}, {Windhorst}, \& {Wright}}]{jwst}
{Gardner}, J.~P., {Mather}, J.~C., {Clampin}, M., {et~al.} 2009, Astrophysics
  and Space Science Proceedings, 10, 1

\bibitem[{{Gaudi} \& {Winn}(2007)}]{GaudiWinn2007_rm}
{Gaudi}, B.~S. \& {Winn}, J.~N. 2007, \apj, 655, 550

\bibitem[{{Ginzburg} {et~al.}(2016){Ginzburg}, {Schlichting}, \&
  {Sari}}]{Ginzburg2016}
{Ginzburg}, S., {Schlichting}, H.~E., \& {Sari}, R. 2016, \apj, 825, 29

\bibitem[{{Ginzburg} {et~al.}(2018){Ginzburg}, {Schlichting}, \&
  {Sari}}]{Ginzburg2018}
{Ginzburg}, S., {Schlichting}, H.~E., \& {Sari}, R. 2018, \mnras, 476, 759

\bibitem[{{Gonz{\'a}lez-{\'A}lvarez} {et~al.}(2020){Gonz{\'a}lez-{\'A}lvarez},
  {Zapatero Osorio}, {Caballero}, {Sanz-Forcada}, {B{\'e}jar},
  {Gonz{\'a}lez-Cuesta}, {Dreizler}, {Bauer}, {Rodr{\'\i}guez}, {Tal-Or},
  {Zechmeister}, {Montes}, {L{\'o}pez-Gonz{\'a}lez}, {Ribas}, {Reiners},
  {Quirrenbach}, {Amado}, {Anglada-Escud{\'e}}, {Azzaro},
  {Cort{\'e}s-Contreras}, {Hatzes}, {Henning}, {Jeffers}, {Kaminski},
  {K{\"u}rster}, {Lafarga}, {Morales}, {Pall{\'e}}, {Perger}, \&
  {Schmitt}}]{Gonzalez2020_gj338B}
{Gonz{\'a}lez-{\'A}lvarez}, E., {Zapatero Osorio}, M.~R., {Caballero}, J.~A.,
  {et~al.} 2020, \aap, 637, A93

\bibitem[{{Goodwin} {et~al.}(2007){Goodwin}, {Kroupa}, {Goodman}, \&
  {Burkert}}]{Goodwin2007}
{Goodwin}, S.~P., {Kroupa}, P., {Goodman}, A., \& {Burkert}, A. 2007, in
  Protostars and Planets V, ed. B.~{Reipurth}, D.~{Jewitt}, \& K.~{Keil}, 133

\bibitem[{{Goodwin} {et~al.}(2008){Goodwin}, {Nutter}, {Kroupa},
  {Ward-Thompson}, \& {Whitworth}}]{Goodwin2008}
{Goodwin}, S.~P., {Nutter}, D., {Kroupa}, P., {Ward-Thompson}, D., \&
  {Whitworth}, A.~P. 2008, \aap, 477, 823

\bibitem[{{Gupta} \& {Schlichting}(2019)}]{Gupta2019}
{Gupta}, A. \& {Schlichting}, H.~E. 2019, \mnras, 487, 24

\bibitem[{{Gupta} \& {Schlichting}(2020)}]{GuptaSchlichting2020}
{Gupta}, A. \& {Schlichting}, H.~E. 2020, \mnras, 493, 792

\bibitem[{{Gupta} \& {Schlichting}(2021)}]{Gupta2021}
{Gupta}, A. \& {Schlichting}, H.~E. 2021, \mnras, 504, 4634

\bibitem[{{Hambly} {et~al.}(2001){Hambly}, {MacGillivray}, {Read}, {Tritton},
  {Thomson}, {Kelly}, {Morgan}, {Smith}, {Driver}, {Williamson}, {Parker},
  {Hawkins}, {Williams}, \& {Lawrence}}]{Hambly2001}
{Hambly}, N.~C., {MacGillivray}, H.~T., {Read}, M.~A., {et~al.} 2001, \mnras,
  326, 1279

\bibitem[{{Henry} {et~al.}(2006){Henry}, {Jao}, {Subasavage}, {Beaulieu},
  {Ianna}, {Costa}, \& {M{\'e}ndez}}]{Henry2006}
{Henry}, T.~J., {Jao}, W.-C., {Subasavage}, J.~P., {et~al.} 2006, \aj, 132,
  2360

\bibitem[{Hippke \& Heller(2019)}]{Hippke2019}
Hippke, M. \& Heller, R. 2019, \aap, 623, A39

\bibitem[{{Hirano} {et~al.}(2020{\natexlab{a}}){Hirano}, {Gaidos}, {Winn},
  {Dai}, {Fukui}, {Kuzuhara}, {Kotani}, {Tamura}, {Hjorth}, {Albrecht},
  {Huber}, {Bolmont}, {Harakawa}, {Hodapp}, {Ishizuka}, {Jacobson}, {Konishi},
  {Kudo}, {Kurokawa}, {Nishikawa}, {Omiya}, {Serizawa}, {Ueda}, \&
  {Weiss}}]{Hirano2020_RM_TRAPPIST1}
{Hirano}, T., {Gaidos}, E., {Winn}, J.~N., {et~al.} 2020{\natexlab{a}}, \apjl,
  890, L27

\bibitem[{{Hirano} {et~al.}(2020{\natexlab{b}}){Hirano}, {Krishnamurthy},
  {Gaidos}, {Flewelling}, {Mann}, {Narita}, {Plavchan}, {Kotani}, {Tamura},
  {Harakawa}, {Hodapp}, {Ishizuka}, {Jacobson}, {Konishi}, {Kudo}, {Kurokawa},
  {Kuzuhara}, {Nishikawa}, {Omiya}, {Serizawa}, {Ueda}, \&
  {Vievard}}]{Hirano2020_RM_AUMic}
{Hirano}, T., {Krishnamurthy}, V., {Gaidos}, E., {et~al.} 2020{\natexlab{b}},
  \apjl, 899, L13

\bibitem[{{Hobson} {et~al.}(2018){Hobson}, {Jofr{\'e}}, {Garc{\'\i}a},
  {Petrucci}, \& {G{\'o}mez}}]{Hobson2018}
{Hobson}, M.~J., {Jofr{\'e}}, E., {Garc{\'\i}a}, L., {Petrucci}, R., \&
  {G{\'o}mez}, M. 2018, RMxAA, 54, 65

\bibitem[{{Howard} {et~al.}(2014){Howard}, {Marcy}, {Fischer}, {Isaacson},
  {Muirhead}, {Henry}, {Boyajian}, {von Braun}, {Becker}, {Wright}, \&
  {Johnson}}]{Howard2014_GJ15A}
{Howard}, A.~W., {Marcy}, G.~W., {Fischer}, D.~A., {et~al.} 2014, \apj, 794, 51

\bibitem[{{Howell} {et~al.}(2014){Howell}, {Sobeck}, {Haas}, {Still},
  {Barclay}, {Mullally}, {Troeltzsch}, {Aigrain}, {Bryson}, {Caldwell},
  {Chaplin}, {Cochran}, {Huber}, {Marcy}, {Miglio}, {Najita}, {Smith},
  {Twicken}, \& {Fortney}}]{Howell2014_K2}
{Howell}, S.~B., {Sobeck}, C., {Haas}, M., {et~al.} 2014, \pasp, 126, 398

\bibitem[{{Huitson} {et~al.}(2017){Huitson}, {D{\'e}sert}, {Bean}, {Fortney},
  {Stevenson}, \& {Bergmann}}]{Huitson2017}
{Huitson}, C.~M., {D{\'e}sert}, J.~M., {Bean}, J.~L., {et~al.} 2017, \aj, 154,
  95

\bibitem[{{Janson} {et~al.}(2014){Janson}, {Bergfors}, {Brandner},
  {Kudryavtseva}, {Hormuth}, {Hippler}, \& {Henning}}]{Janson2014}
{Janson}, M., {Bergfors}, C., {Brandner}, W., {et~al.} 2014, \apj, 789, 102

\bibitem[{{Janson} {et~al.}(2012){Janson}, {Hormuth}, {Bergfors}, {Brandner},
  {Hippler}, {Daemgen}, {Kudryavtseva}, {Schmalzl}, {Schnupp}, \&
  {Henning}}]{Janson2012}
{Janson}, M., {Hormuth}, F., {Bergfors}, C., {et~al.} 2012, \apj, 754, 44

\bibitem[{{Jeffers} \& {Keller}(2009)}]{JeffersKeller2009}
{Jeffers}, S.~V. \& {Keller}, C.~U. 2009, in American Institute of Physics
  Conference Series, Vol. 1094, 15th Cambridge Workshop on Cool Stars, Stellar
  Systems, and the Sun, ed. E.~{Stempels}, 664--667

\bibitem[{{Jeffers} {et~al.}(2018){Jeffers}, {Sch{\"o}fer}, {Lamert},
  {Reiners}, {Montes}, {Caballero}, {Cort{\'e}s-Contreras}, {Marvin},
  {Passegger}, {Zechmeister}, {Quirrenbach}, {Alonso-Floriano}, {Amado},
  {Bauer}, {Casal}, {Diez Alonso}, {Herrero}, {Morales}, {Mundt}, {Ribas}, \&
  {Sarmiento}}]{Jeffers2018}
{Jeffers}, S.~V., {Sch{\"o}fer}, P., {Lamert}, A., {et~al.} 2018, \aap, 614,
  A76

\bibitem[{{Jenkins} {et~al.}(2016){Jenkins}, {Twicken}, {McCauliff},
  {Campbell}, {Sanderfer}, {Lung}, {Mansouri-Samani}, {Girouard}, {Tenenbaum},
  {Klaus}, {Smith}, {Caldwell}, {Chacon}, {Henze}, {Heiges}, {Latham},
  {Morgan}, {Swade}, {Rinehart}, \& {Vanderspek}}]{Jenkins2016}
{Jenkins}, J.~M., {Twicken}, J.~D., {McCauliff}, S., {et~al.} 2016, in Society
  of Photo-Optical Instrumentation Engineers (SPIE) Conference Series, Vol.
  9913, Software and Cyberinfrastructure for Astronomy IV, ed. G.~{Chiozzi} \&
  J.~C. {Guzman}, 99133E

\bibitem[{{Jensen}(2013)}]{Jensen:2013}
{Jensen}, E. 2013, {Tapir: A web interface for transit/eclipse observability},
  Astrophysics Source Code Library

\bibitem[{{Johnson} \& {Soderblom}(1987)}]{JohnsonSoderblom1987}
{Johnson}, D. R.~H. \& {Soderblom}, D.~R. 1987, \aj, 93, 864

\bibitem[{{Johnson} \& {Li}(2012)}]{JohnsonLi2012}
{Johnson}, J.~L. \& {Li}, H. 2012, \apj, 751, 81

\bibitem[{{Jurgenson} {et~al.}(2016){Jurgenson}, {Fischer}, {McCracken},
  {Sawyer}, {Szymkowiak}, {Davis}, {Muller}, \& {Santoro}}]{expres}
{Jurgenson}, C., {Fischer}, D., {McCracken}, T., {et~al.} 2016, in Society of
  Photo-Optical Instrumentation Engineers (SPIE) Conference Series, Vol. 9908,
  Ground-based and Airborne Instrumentation for Astronomy VI, ed. C.~J.
  {Evans}, L.~{Simard}, \& H.~{Takami}, 99086T

\bibitem[{{Kaiser} {et~al.}(2010){Kaiser}, {Burgett}, {Chambers}, {Denneau},
  {Heasley}, {Jedicke}, {Magnier}, {Morgan}, {Onaka}, \& {Tonry}}]{Kaiser2010}
{Kaiser}, N., {Burgett}, W., {Chambers}, K., {et~al.} 2010, in Society of
  Photo-Optical Instrumentation Engineers (SPIE) Conference Series, Vol. 7733,
  Ground-based and Airborne Telescopes III, ed. L.~M. {Stepp}, R.~{Gilmozzi},
  \& H.~J. {Hall}, 77330E

\bibitem[{{Kaminski} {et~al.}(2018){Kaminski}, {Trifonov}, {Caballero},
  {Quirrenbach}, {Ribas}, {Reiners}, {Amado}, {Zechmeister}, {Dreizler},
  {Perger}, {Tal-Or}, {Bonfils}, {Mayor}, {Astudillo-Defru}, {Bauer},
  {B{\'e}jar}, {Cifuentes}, {Colom{\'e}}, {Cort{\'e}s-Contreras}, {Delfosse},
  {D{\'\i}ez-Alonso}, {Forveille}, {Guenther}, {Hatzes}, {Henning}, {Jeffers},
  {K{\"u}rster}, {Lafarga}, {Luque}, {Mandel}, {Montes}, {Morales},
  {Passegger}, {Pedraz}, {Reffert}, {Sadegi}, {Schweitzer}, {Seifert}, {Stahl},
  \& {Udry}}]{Kaminski2018}
{Kaminski}, A., {Trifonov}, T., {Caballero}, J.~A., {et~al.} 2018, \aap, 618,
  A115

\bibitem[{{Kemmer} {et~al.}(2020){Kemmer}, {Stock}, {Kossakowski}, {Kaminski},
  {Molaverdikhani}, {Schlecker}, {Caballero}, {Amado}, {Astudillo-Defru},
  {Bonfils}, {Ciardi}, {Collins}, {Espinoza}, {Fukui}, {Hirano}, {Jenkins},
  {Latham}, {Matthews}, {Narita}, {Pall{\'e}}, {Parviainen}, {Quirrenbach},
  {Reiners}, {Ribas}, {Ricker}, {Schlieder}, {Seager}, {Vanderspek}, {Winn},
  {Almenara}, {B{\'e}jar}, {Bluhm}, {Bouchy}, {Boyd}, {Christiansen},
  {Cifuentes}, {Cloutier}, {Collins}, {Cort{\'e}s-Contreras}, {Crossfield},
  {Crouzet}, {de Leon}, {Della-Rose}, {Delfosse}, {Dreizler}, {Esparza-Borges},
  {Essack}, {Forveille}, {Figueira}, {Galad{\'\i}-Enr{\'\i}quez}, {Gan},
  {Glidden}, {Gonzales}, {Guerra}, {Harakawa}, {Hatzes}, {Henning}, {Herrero},
  {Hodapp}, {Hori}, {Howell}, {Ikoma}, {Isogai}, {Jeffers}, {K{\"u}rster},
  {Kawauchi}, {Kimura}, {Klagyivik}, {Kotani}, {Kurokawa}, {Kusakabe},
  {Kuzuhara}, {Lafarga}, {Livingston}, {Luque}, {Matson}, {Morales}, {Mori},
  {Muirhead}, {Murgas}, {Nishikawa}, {Nishiumi}, {Omiya}, {Reffert},
  {Rodr{\'\i}guez L{\'o}pez}, {Santos}, {Sch{\"o}fer}, {Schwarz}, {Shiao},
  {Tamura}, {Terada}, {Twicken}, {Ueda}, {Vievard}, {Watanabe}, \&
  {Zechmeister}}]{Kemmer2020_toi488}
{Kemmer}, J., {Stock}, S., {Kossakowski}, D., {et~al.} 2020, \aap, 642, A236

\bibitem[{{Kempton} {et~al.}(2018){Kempton}, {Bean}, {Louie}, {Deming}, {Koll},
  {Mansfield}, {Christiansen}, {L{\'o}pez-Morales}, {Swain}, {Zellem},
  {Ballard}, {Barclay}, {Barstow}, {Batalha}, {Beatty}, {Berta-Thompson},
  {Birkby}, {Buchhave}, {Charbonneau}, {Cowan}, {Crossfield}, {de Val-Borro},
  {Doyon}, {Dragomir}, {Gaidos}, {Heng}, {Hu}, {Kane}, {Kreidberg}, {Mallonn},
  {Morley}, {Narita}, {Nascimbeni}, {Pall{\'e}}, {Quintana}, {Rauscher},
  {Seager}, {Shkolnik}, {Sing}, {Sozzetti}, {Stassun}, {Valenti}, \& {von
  Essen}}]{Kempton2018}
{Kempton}, E. M.~R., {Bean}, J.~L., {Louie}, D.~R., {et~al.} 2018, \pasp, 130,
  114401

\bibitem[{{Kipping}(2013)}]{Kipping2013_ldc}
{Kipping}, D.~M. 2013, \mnras, 435, 2152

\bibitem[{{Knapp} \& {Nanson}(2019)}]{KnappNanson2019}
{Knapp}, W. \& {Nanson}, J. 2019, Journal of Double Star Observations, 15, 21

\bibitem[{{Kochanek} {et~al.}(2017){Kochanek}, {Shappee}, {Stanek}, {Holoien},
  {Thompson}, {Prieto}, {Dong}, {Shields}, {Will}, {Britt}, {Perzanowski}, \&
  {Pojma{\'n}ski}}]{Kochanek2017_ASAS}
{Kochanek}, C.~S., {Shappee}, B.~J., {Stanek}, K.~Z., {et~al.} 2017, \pasp,
  129, 104502

\bibitem[{{Kordopatis} {et~al.}(2013){Kordopatis}, {Gilmore}, {Steinmetz},
  {Boeche}, {Seabroke}, {Siebert}, {Zwitter}, {Binney}, {de Laverny},
  {Recio-Blanco}, {Williams}, {Piffl}, {Enke}, {Roeser}, {Bijaoui}, {Wyse},
  {Freeman}, {Munari}, {Carrillo}, {Anguiano}, {Burton}, {Campbell}, {Cass},
  {Fiegert}, {Hartley}, {Parker}, {Reid}, {Ritter}, {Russell}, {Stupar},
  {Watson}, {Bienaym{\'e}}, {Bland-Hawthorn}, {Gerhard}, {Gibson}, {Grebel},
  {Helmi}, {Navarro}, {Conrad}, {Famaey}, {Faure}, {Just}, {Kos},
  {Matijevi{\v{c}}}, {McMillan}, {Minchev}, {Scholz}, {Sharma}, {Siviero}, {de
  Boer}, \& {{\v{Z}}erjal}}]{rave}
{Kordopatis}, G., {Gilmore}, G., {Steinmetz}, M., {et~al.} 2013, \aj, 146, 134

\bibitem[{{Kreidberg}(2015)}]{batman}
{Kreidberg}, L. 2015, \pasp, 127, 1161

\bibitem[{{Kreidberg} {et~al.}(2014){Kreidberg}, {Bean}, {D{\'e}sert},
  {Benneke}, {Deming}, {Stevenson}, {Seager}, {Berta-Thompson}, {Seifahrt}, \&
  {Homeier}}]{Kreidberg2014}
{Kreidberg}, L., {Bean}, J.~L., {D{\'e}sert}, J.-M., {et~al.} 2014, \nat, 505,
  69

\bibitem[{{Lafarga} {et~al.}(2020){Lafarga}, {Ribas}, {Lovis}, {Perger},
  {Zechmeister}, {Bauer}, {K{\"u}rster}, {Cort{\'e}s-Contreras}, {Morales},
  {Herrero}, {Rosich}, {Baroch}, {Reiners}, {Caballero}, {Quirrenbach},
  {Amado}, {Alacid}, {B{\'e}jar}, {Dreizler}, {Hatzes}, {Henning}, {Jeffers},
  {Kaminski}, {Montes}, {Pedraz}, {Rodr{\'\i}guez-L{\'o}pez}, \&
  {Schmitt}}]{Lafarga2020}
{Lafarga}, M., {Ribas}, I., {Lovis}, C., {et~al.} 2020, \aap, 636, A36

\bibitem[{{Lafarga} {et~al.}(2021){Lafarga}, {Ribas}, {Reiners}, {Quirrenbach},
  {Amado}, {Caballero}, {Azzaro}, {B{\'e}jar}, {Cort{\'e}s-Contreras},
  {Dreizler}, {Hatzes}, {Henning}, {Jeffers}, {Kaminski}, {K{\"u}rster},
  {Montes}, {Morales}, {Oshagh}, {Rodr{\'\i}guez-L{\'o}pez}, {Sch{\"o}fer},
  {Schweitzer}, \& {Zechmeister}}]{Lafarga2021}
{Lafarga}, M., {Ribas}, I., {Reiners}, A., {et~al.} 2021, \aap, 652, A28

\bibitem[{{Lawrence} {et~al.}(2007){Lawrence}, {Warren}, {Almaini}, {Edge},
  {Hambly}, {Jameson}, {Lucas}, {Casali}, {Adamson}, {Dye}, {Emerson},
  {Foucaud}, {Hewett}, {Hirst}, {Hodgkin}, {Irwin}, {Lodieu}, {McMahon},
  {Simpson}, {Smail}, {Mortlock}, \& {Folger}}]{Lawrence2007}
{Lawrence}, A., {Warren}, S.~J., {Almaini}, O., {et~al.} 2007, \mnras, 379,
  1599

\bibitem[{{Lehmer} \& {Catling}(2017)}]{LehmerCatling2017}
{Lehmer}, O.~R. \& {Catling}, D.~C. 2017, \apj, 845, 130

\bibitem[{{L{\'e}pine} \& {Gaidos}(2011)}]{LepineGaidos2011}
{L{\'e}pine}, S. \& {Gaidos}, E. 2011, \aj, 142, 138

\bibitem[{{Livingston} {et~al.}(2019){Livingston}, {Dai}, {Hirano}, {Gandolfi},
  {Trani}, {Nowak}, {Cochran}, {Endl}, {Albrecht}, {Barragan}, {Cabrera},
  {Csizmadia}, {de Leon}, {Deeg}, {Eigm{\"u}ller}, {Erikson}, {Fridlund},
  {Fukui}, {Grziwa}, {Guenther}, {Hatzes}, {Korth}, {Kuzuhara}, {Monta{\~n}es},
  {Narita}, {Nespral}, {Palle}, {P{\"a}tzold}, {Persson}, {Prieto-Arranz},
  {Rauer}, {Tamura}, {Van Eylen}, \& {Winn}}]{Livingston2019_K2-264}
{Livingston}, J.~H., {Dai}, F., {Hirano}, T., {et~al.} 2019, \mnras, 484, 8

\bibitem[{{Lodieu} {et~al.}(2019{\natexlab{a}}){Lodieu}, {P{\'e}rez-Garrido},
  {Smart}, \& {Silvotti}}]{Lodieu2019b}
{Lodieu}, N., {P{\'e}rez-Garrido}, A., {Smart}, R.~L., \& {Silvotti}, R.
  2019{\natexlab{a}}, 628, A66

\bibitem[{{Lodieu} {et~al.}(2018){Lodieu}, {Rebolo}, \&
  {P{\'e}rez-Garrido}}]{Lodieu2018b}
{Lodieu}, N., {Rebolo}, R., \& {P{\'e}rez-Garrido}, A. 2018, 615, L12

\bibitem[{{Lodieu} {et~al.}(2019{\natexlab{b}}){Lodieu}, {Smart},
  {P{\'e}rez-Garrido}, \& {Silvotti}}]{Lodieu2019a}
{Lodieu}, N., {Smart}, R.~L., {P{\'e}rez-Garrido}, A., \& {Silvotti}, R.
  2019{\natexlab{b}}, 623, A35

\bibitem[{{L\'opez} \& {Fortney}(2014)}]{LopezFortney2014}
{L\'opez}, E.~D. \& {Fortney}, J.~J. 2014, \apj, 792, 1

\bibitem[{{Louden} {et~al.}(2021){Louden}, {Winn}, {Petigura}, {Isaacson},
  {Howard}, {Masuda}, {Albrecht}, \& {Kosiarek}}]{Louden2021}
{Louden}, E.~M., {Winn}, J.~N., {Petigura}, E.~A., {et~al.} 2021, \aj, 161, 68

\bibitem[{{Makarov} \& {Kaplan}(2005)}]{MakarovKaplan2005}
{Makarov}, V.~V. \& {Kaplan}, G.~H. 2005, \aj, 129, 2420

\bibitem[{{Mann} {et~al.}(2016){Mann}, {Gaidos}, {Mace}, {Johnson}, {Bowler},
  {LaCourse}, {Jacobs}, {Vanderburg}, {Kraus}, {Kaplan}, \&
  {Jaffe}}]{Mann2016_K2-25}
{Mann}, A.~W., {Gaidos}, E., {Mace}, G.~N., {et~al.} 2016, \apj, 818, 46

\bibitem[{{Mann} {et~al.}(2017){Mann}, {Gaidos}, {Vanderburg}, {Rizzuto},
  {Ansdell}, {Medina}, {Mace}, {Kraus}, \& {Sokal}}]{Mann2017}
{Mann}, A.~W., {Gaidos}, E., {Vanderburg}, A., {et~al.} 2017, \aj, 153, 64

\bibitem[{{Mann} {et~al.}(2018){Mann}, {Vanderburg}, {Rizzuto}, {Kraus},
  {Berlind}, {Bieryla}, {Calkins}, {Esquerdo}, {Latham}, {Mace}, {Morris},
  {Quinn}, {Sokal}, \& {Stefanik}}]{Mann2018_epic}
{Mann}, A.~W., {Vanderburg}, A., {Rizzuto}, A.~C., {et~al.} 2018, \aj, 155, 4

\bibitem[{{Marcy} {et~al.}(2014){Marcy}, {Weiss}, {Petigura}, {Isaacson},
  {Howard}, \& {Buchhave}}]{Marcy2014}
{Marcy}, G.~W., {Weiss}, L.~M., {Petigura}, E.~A., {et~al.} 2014, PNAS, 111,
  12655

\bibitem[{{Marocco} {et~al.}(2021){Marocco}, {Eisenhardt}, {Fowler},
  {Kirkpatrick}, {Meisner}, {Schlafly}, {Stanford}, {Garcia}, {Caselden},
  {Cushing}, {Cutri}, {Faherty}, {Gelino}, {Gonzalez}, {Jarrett}, {Koontz},
  {Mainzer}, {Marchese}, {Mobasher}, {Schlegel}, {Stern}, {Teplitz}, \&
  {Wright}}]{Marocco2021}
{Marocco}, F., {Eisenhardt}, P. R.~M., {Fowler}, J.~W., {et~al.} 2021, \apjs,
  253, 8

\bibitem[{{Mart{\'\i}n} {et~al.}(2018){Mart{\'\i}n}, {Lodieu}, {Pavlenko}, \&
  {B{\'e}jar}}]{Martin2018}
{Mart{\'\i}n}, E.~L., {Lodieu}, N., {Pavlenko}, Y., \& {B{\'e}jar}, V. J.~S.
  2018, \apj, 856, 40

\bibitem[{{Mason} {et~al.}(2001){Mason}, {Wycoff}, {Hartkopf}, {Douglass}, \&
  {Worley}}]{Mason2001}
{Mason}, B.~D., {Wycoff}, G.~L., {Hartkopf}, W.~I., {Douglass}, G.~G., \&
  {Worley}, C.~E. 2001, \aj, 122, 3466

\bibitem[{{Mathur} {et~al.}(2017){Mathur}, {Huber}, {Batalha}, {Ciardi},
  {Bastien}, {Bieryla}, {Buchhave}, {Cochran}, {Endl}, {Esquerdo}, {Furlan},
  {Howard}, {Howell}, {Isaacson}, {Latham}, {MacQueen}, \&
  {Silva}}]{Mathur2017}
{Mathur}, S., {Huber}, D., {Batalha}, N.~M., {et~al.} 2017, \apjs, 229, 30

\bibitem[{{Maxted}(2018)}]{Maxted2018}
{Maxted}, P.~F.~L. 2018, \aap, 616, A39

\bibitem[{{McLaughlin}(1924)}]{McLaughlin1924}
{McLaughlin}, D.~B. 1924, \apj, 60, 22

\bibitem[{{Medina} {et~al.}(2020){Medina}, {Winters}, {Irwin}, \&
  {Charbonneau}}]{Medina2020}
{Medina}, A.~A., {Winters}, J.~G., {Irwin}, J.~M., \& {Charbonneau}, D. 2020,
  \apj, 905, 107

\bibitem[{{Miller-Ricci} \& {Fortney}(2010)}]{Miller-RicciFortney2010}
{Miller-Ricci}, E. \& {Fortney}, J.~J. 2010, \apjl, 716, L74

\bibitem[{{Monet} {et~al.}(2003){Monet}, {Levine}, {Canzian}, {Ables}, {Bird},
  {Dahn}, {Guetter}, {Harris}, {Henden}, {Leggett}, {Levison}, {Luginbuhl},
  {Martini}, {Monet}, {Munn}, {Pier}, {Rhodes}, {Riepe}, {Sell}, {Stone},
  {Vrba}, {Walker}, {Westerhout}, {Brucato}, {Reid}, {Schoening}, {Hartley},
  {Read}, \& {Tritton}}]{Monet2003}
{Monet}, D.~G., {Levine}, S.~E., {Canzian}, B., {et~al.} 2003, \aj, 125, 984

\bibitem[{{Monnier} {et~al.}(2018){Monnier}, {Kraus}, {Ireland}, {Baron},
  {Bayo}, {Berger}, {Creech-Eakman}, {Dong}, {Duch{\^e}ne}, {Espaillat},
  {Haniff}, {H{\"o}nig}, {Isella}, {Juhasz}, {Labadie}, {Lacour}, {Leifer},
  {Merand}, {Michael}, {Minardi}, {Mordasini}, {Mozurkewich}, {Olofsson},
  {Paladini}, {Petrov}, {Pott}, {Ridgway}, {Rinehart}, {Stassun}, {Surdej},
  {Brummelaar}, {Turner}, {Tuthill}, {Vahala}, {van Belle}, {Vasisht},
  {Wishnow}, {Young}, \& {Zhu}}]{Monnier2018}
{Monnier}, J.~D., {Kraus}, S., {Ireland}, M.~J., {et~al.} 2018, Experimental
  Astronomy, 46, 517

\bibitem[{{Montes} {et~al.}(2018){Montes}, {Gonz{\'a}lez-Peinado}, {Tabernero},
  {Caballero}, {Marfil}, {Alonso-Floriano}, {Cort{\'e}s-Contreras},
  {Gonz{\'a}lez Hern{\'a}ndez}, {Klutsch}, \& {Moreno-J{\'o}dar}}]{Montes2018}
{Montes}, D., {Gonz{\'a}lez-Peinado}, R., {Tabernero}, H.~M., {et~al.} 2018,
  \mnras, 479, 1332

\bibitem[{{Montes} {et~al.}(2001){Montes}, {L{\'o}pez-Santiago}, {G{\'a}lvez},
  {Fern{\'a}ndez-Figueroa}, {De Castro}, \& {Cornide}}]{Montes2001}
{Montes}, D., {L{\'o}pez-Santiago}, J., {G{\'a}lvez}, M.~C., {et~al.} 2001,
  \mnras, 328, 45

\bibitem[{{Mordasini}(2020)}]{Mordasini2020}
{Mordasini}, C. 2020, \aap, 638, A52

\bibitem[{{Morello} {et~al.}(2017){Morello}, {Tsiaras}, {Howarth}, \&
  {Homeier}}]{Morello2017}
{Morello}, G., {Tsiaras}, A., {Howarth}, I.~D., \& {Homeier}, D. 2017, \aj,
  154, 111

\bibitem[{{Mu{\~n}oz} \& {Perets}(2018)}]{MunozPerets2018}
{Mu{\~n}oz}, D.~J. \& {Perets}, H.~B. 2018, \aj, 156, 253

\bibitem[{{Nikolov} {et~al.}(2016){Nikolov}, {Sing}, {Gibson}, {Fortney},
  {Evans}, {Barstow}, {Kataria}, \& {Wilson}}]{Nikolov2016}
{Nikolov}, N., {Sing}, D.~K., {Gibson}, N.~P., {et~al.} 2016, \apj, 832, 191

\bibitem[{{Nowak} {et~al.}(2020){Nowak}, {Luque}, {Parviainen}, {Pall{\'e}},
  {Molaverdikhani}, {B{\'e}jar}, {Lillo-Box}, {Rodr{\'\i}guez-L{\'o}pez},
  {Caballero}, {Zechmeister}, {Passegger}, {Cifuentes}, {Schweitzer}, {Narita},
  {Cale}, {Espinoza}, {Murgas}, {Hidalgo}, {Zapatero Osorio}, {Pozuelos},
  {Aceituno}, {Amado}, {Barkaoui}, {Barrado}, {Bauer}, {Benkhaldoun},
  {Caldwell}, {Casasayas Barris}, {Chaturvedi}, {Chen}, {Collins}, {Collins},
  {Cort{\'e}s-Contreras}, {Crossfield}, {de Le{\'o}n}, {D{\'\i}ez Alonso},
  {Dreizler}, {El Mufti}, {Esparza-Borges}, {Essack}, {Fukui}, {Gaidos},
  {Gillon}, {Gonzales}, {Guerra}, {Hatzes}, {Henning}, {Herrero}, {Hesse},
  {Hirano}, {Howell}, {Jeffers}, {Jehin}, {Jenkins}, {Kaminski}, {Kemmer},
  {Kielkopf}, {Kossakowski}, {Kotani}, {K{\"u}rster}, {Lafarga}, {Latham},
  {Law}, {Lissauer}, {Lodieu}, {Madrigal-Aguado}, {Mann}, {Massey}, {Matson},
  {Matthews}, {Monta{\~n}{\'e}s-Rodr{\'\i}guez}, {Montes}, {Morales}, {Mori},
  {Nagel}, {Oshagh}, {Pedraz}, {Plavchan}, {Pollacco}, {Quirrenbach},
  {Reffert}, {Reiners}, {Ribas}, {Ricker}, {Rose}, {Schlecker}, {Schlieder},
  {Seager}, {Stangret}, {Stock}, {Tamura}, {Tanner}, {Teske}, {Trifonov},
  {Twicken}, {Vanderspek}, {Watanabe}, {Wittrock}, {Ziegler}, \&
  {Zohrabi}}]{Nowak2020_toi732}
{Nowak}, G., {Luque}, R., {Parviainen}, H., {et~al.} 2020, \aap, 642, A173

\bibitem[{Owen \& Jackson(2012)}]{Owen2012a}
Owen, J.~E. \& Jackson, A.~P. 2012, MNRAS, 425, 2931

\bibitem[{{Owen} \& {Wu}(2013)}]{OwenWu2013}
{Owen}, J.~E. \& {Wu}, Y. 2013, \apj, 775, 105

\bibitem[{{Owen} \& {Wu}(2017)}]{OwenWu2017}
{Owen}, J.~E. \& {Wu}, Y. 2017, \apj, 847, 29

\bibitem[{{Palle} {et~al.}(2020){Palle}, {Oshagh}, {Casasayas-Barris},
  {Hirano}, {Stangret}, {Luque}, {Strachan}, {Gaidos}, {Anglada-Escude},
  {Plavchan}, \& {Addison}}]{Palle2020_RM_AUMic}
{Palle}, E., {Oshagh}, M., {Casasayas-Barris}, N., {et~al.} 2020, \aap, 643,
  A25

\bibitem[{{Passegger} {et~al.}(2020){Passegger}, {Bello-Garc{\'\i}a},
  {Ordieres-Mer{\'e}}, {Caballero}, {Schweitzer}, {Gonz{\'a}lez-Marcos},
  {Ribas}, {Reiners}, {Quirrenbach}, {Amado}, {Azzaro}, {Bauer}, {B{\'e}jar},
  {Cort{\'e}s-Contreras}, {Dreizler}, {Hatzes}, {Henning}, {Jeffers},
  {Kaminski}, {K{\"u}rster}, {Lafarga}, {Marfil}, {Montes}, {Morales}, {Nagel},
  {Sarro}, {Solano}, {Tabernero}, \& {Zechmeister}}]{Passegger2020}
{Passegger}, V.~M., {Bello-Garc{\'\i}a}, A., {Ordieres-Mer{\'e}}, J., {et~al.}
  2020, \aap, 642, A22

\bibitem[{{Passegger} {et~al.}(2019){Passegger}, {Schweitzer}, {Shulyak},
  {Nagel}, {Hauschildt}, {Reiners}, {Amado}, {Caballero},
  {Cort{\'e}s-Contreras}, {Dom{\'\i}nguez-Fern{\'a}ndez}, {Quirrenbach},
  {Ribas}, {Azzaro}, {Anglada-Escud{\'e}}, {Bauer}, {B{\'e}jar}, {Dreizler},
  {Guenther}, {Henning}, {Jeffers}, {Kaminski}, {K{\"u}rster}, {Lafarga},
  {Mart{\'\i}n}, {Montes}, {Morales}, {Schmitt}, \&
  {Zechmeister}}]{Passegger2019}
{Passegger}, V.~M., {Schweitzer}, A., {Shulyak}, D., {et~al.} 2019, \aap, 627,
  A161

\bibitem[{{Pepe} {et~al.}(2021){Pepe}, {Cristiani}, {Rebolo}, {Santos},
  {Dekker}, {Cabral}, {Di Marcantonio}, {Figueira}, {Lo Curto}, {Lovis},
  {Mayor}, {M{\'e}gevand}, {Molaro}, {Riva}, {Zapatero Osorio}, {Amate},
  {Manescau}, {Pasquini}, {Zerbi}, {Adibekyan}, {Abreu}, {Affolter}, {Alibert},
  {Aliverti}, {Allart}, {Allende Prieto}, {{\'A}lvarez}, {Alves}, {Avila},
  {Baldini}, {Bandy}, {Barros}, {Benz}, {Bianco}, {Borsa}, {Bourrier},
  {Bouchy}, {Broeg}, {Calderone}, {Cirami}, {Coelho}, {Conconi}, {Coretti},
  {Cumani}, {Cupani}, {D'Odorico}, {Damasso}, {Deiries}, {Delabre},
  {Demangeon}, {Dumusque}, {Ehrenreich}, {Faria}, {Fragoso}, {Genolet},
  {Genoni}, {G{\'e}nova Santos}, {Gonz{\'a}lez Hern{\'a}ndez}, {Hughes},
  {Iwert}, {Kerber}, {Knudstrup}, {Landoni}, {Lavie}, {Lillo-Box}, {Lizon},
  {Maire}, {Martins}, {Mehner}, {Micela}, {Modigliani}, {Monteiro}, {Monteiro},
  {Moschetti}, {Murphy}, {Nunes}, {Oggioni}, {Oliveira}, {Oshagh}, {Pall{\'e}},
  {Pariani}, {Poretti}, {Rasilla}, {Rebord{\~a}o}, {Redaelli}, {Santana
  Tschudi}, {Santin}, {Santos}, {S{\'e}gransan}, {Schmidt}, {Segovia},
  {Sosnowska}, {Sozzetti}, {Sousa}, {Span{\`o}}, {Su{\'a}rez Mascare{\~n}o},
  {Tabernero}, {Tenegi}, {Udry}, \& {Zanutta}}]{espresso}
{Pepe}, F., {Cristiani}, S., {Rebolo}, R., {et~al.} 2021, \aap, 645, A96

\bibitem[{{Perger} {et~al.}(2019){Perger}, {Scandariato}, {Ribas}, {Morales},
  {Affer}, {Azzaro}, {Amado}, {Anglada-Escud{\'e}}, {Baroch}, {Barrado},
  {Bauer}, {B{\'e}jar}, {Caballero}, {Cort{\'e}s-Contreras}, {Damasso},
  {Dreizler}, {Gonz{\'a}lez-Cuesta}, {Gonz{\'a}lez Hern{\'a}ndez}, {Guenther},
  {Henning}, {Herrero}, {Jeffers}, {Kaminski}, {K{\"u}rster}, {Lafarga},
  {Leto}, {L{\'o}pez-Gonz{\'a}lez}, {Maldonado}, {Micela}, {Montes},
  {Pinamonti}, {Quirrenbach}, {Rebolo}, {Reiners}, {Rodr{\'\i}guez},
  {Rodr{\'\i}guez-L{\'o}pez}, {Schmitt}, {Sozzetti}, {Su{\'a}rez
  Mascare{\~n}o}, {Toledo-Padr{\'o}n}, {Zanmar S{\'a}nchez}, {Zapatero Osorio},
  \& {Zechmeister}}]{Perger2019}
{Perger}, M., {Scandariato}, G., {Ribas}, I., {et~al.} 2019, \aap, 624, A123

\bibitem[{{Pinamonti} {et~al.}(2018){Pinamonti}, {Damasso}, {Marzari},
  {Sozzetti}, {Desidera}, {Maldonado}, {Scandariato}, {Affer}, {Lanza},
  {Bignamini}, {Bonomo}, {Borsa}, {Claudi}, {Cosentino}, {Giacobbe},
  {Gonz{\'a}lez-{\'A}lvarez}, {Gonz{\'a}lez Hern{\'a}ndez}, {Gratton}, {Leto},
  {Malavolta}, {Martinez Fiorenzano}, {Micela}, {Molinari}, {Pagano}, {Pedani},
  {Perger}, {Piotto}, {Rebolo}, {Ribas}, {Su{\'a}rez Mascare{\~n}o}, \&
  {Toledo-Padr{\'o}n}}]{Pinamonti2018_GJ15A}
{Pinamonti}, M., {Damasso}, M., {Marzari}, F., {et~al.} 2018, \aap, 617, A104

\bibitem[{{Pollacco} {et~al.}(2006){Pollacco}, {Skillen}, {Collier Cameron},
  {Christian}, {Hellier}, {Irwin}, {Lister}, {Street}, {West}, {Anderson},
  {Clarkson}, {Deeg}, {Enoch}, {Evans}, {Fitzsimmons}, {Haswell}, {Hodgkin},
  {Horne}, {Kane}, {Keenan}, {Maxted}, {Norton}, {Osborne}, {Parley}, {Ryans},
  {Smalley}, {Wheatley}, \& {Wilson}}]{Pollacco2006_WASP}
{Pollacco}, D.~L., {Skillen}, I., {Collier Cameron}, A., {et~al.} 2006, \pasp,
  118, 1407

\bibitem[{{Quirrenbach} {et~al.}(2014){Quirrenbach}, {Amado}, {Caballero},
  {Mundt}, {Reiners}, {Ribas}, {Seifert}, {Abril}, {Aceituno},
  {Alonso-Floriano}, {Ammler-von Eiff}, {Antona Jim{\'e}nez},
  {Anwand-Heerwart}, {Azzaro}, {Bauer}, {Barrado}, {Becerril}, {B{\'e}jar},
  {Ben{\'{\i}}tez}, {Berdi{\~n}as}, {C{\'a}rdenas}, {Casal}, {Claret},
  {Colom{\'e}}, {Cort{\'e}s-Contreras}, {Czesla}, {Doellinger}, {Dreizler},
  {Feiz}, {Fern{\'a}ndez}, {Galad{\'{\i}}}, {G{\'a}lvez-Ortiz},
  {Garc{\'{\i}}a-Piquer}, {Garc{\'{\i}}a-Vargas}, {Garrido}, {Gesa}, {G{\'o}mez
  Galera}, {Gonz{\'a}lez {\'A}lvarez}, {Gonz{\'a}lez Hern{\'a}ndez},
  {Gr{\"o}zinger}, {Gu{\`a}rdia}, {Guenther}, {de Guindos},
  {Guti{\'e}rrez-Soto}, {Hagen}, {Hatzes}, {Hauschildt}, {Helmling}, {Henning},
  {Hermann}, {Hern{\'a}ndez Casta{\~n}o}, {Herrero}, {Hidalgo}, {Holgado},
  {Huber}, {Huber}, {Jeffers}, {Joergens}, {de Juan}, {Kehr}, {Klein},
  {K{\"u}rster}, {Lamert}, {Lalitha}, {Laun}, {Lemke}, {Lenzen}, {L{\'o}pez del
  Fresno}, {L{\'o}pez Mart{\'{\i}}}, {L{\'o}pez-Santiago}, {Mall}, {Mandel},
  {Mart{\'{\i}}n}, {Mart{\'{\i}}n-Ruiz}, {Mart{\'{\i}}nez-Rodr{\'{\i}}guez},
  {Marvin}, {Mathar}, {Mirabet}, {Montes}, {Morales Mu{\~n}oz}, {Moya},
  {Naranjo}, {Ofir}, {Oreiro}, {Pall{\'e}}, {Panduro}, {Passegger},
  {P{\'e}rez-Calpena}, {P{\'e}rez Medialdea}, {Perger}, {Pluto}, {Ram{\'o}n},
  {Rebolo}, {Redondo}, {Reffert}, {Reinhardt}, {Rhode}, {Rix}, {Rodler},
  {Rodr{\'{\i}}guez}, {Rodr{\'{\i}}guez-L{\'o}pez},
  {Rodr{\'{\i}}guez-P{\'e}rez}, {Rohloff}, {Rosich}, {S{\'a}nchez-Blanco},
  {S{\'a}nchez Carrasco}, {Sanz-Forcada}, {Sarmiento}, {Sch{\"a}fer},
  {Schiller}, {Schmidt}, {Schmitt}, {Solano}, {Stahl}, {Storz}, {St{\"u}rmer},
  {Su{\'a}rez}, {Ulbrich}, {Veredas}, {Wagner}, {Winkler}, {Zapatero Osorio},
  {Zechmeister}, {Abell{\'a}n de Paco}, {Anglada-Escud{\'e}}, {del Burgo},
  {Klutsch}, {Lizon}, {L{\'o}pez-Morales}, {Morales}, {Perryman}, {Tulloch}, \&
  {Xu}}]{CARMENES}
{Quirrenbach}, A., {Amado}, P.~J., {Caballero}, J.~A., {et~al.} 2014, in
  \procspie, Vol. 9147, Ground-based and Airborne Instrumentation for Astronomy
  V, 91471F

\bibitem[{{Quirrenbach} {et~al.}(2018){Quirrenbach}, {Amado}, {Ribas},
  {Reiners}, {Caballero}, {Seifert}, {Aceituno}, {Azzaro}, {Baroch}, {Barrado},
  \& et~al.}]{CARMENES18}
{Quirrenbach}, A., {Amado}, P.~J., {Ribas}, I., {et~al.} 2018, in Society of
  Photo-Optical Instrumentation Engineers (SPIE) Conference Series, Vol. 10702,
  Ground-based and Airborne Instrumentation for Astronomy VII, 107020W

\bibitem[{{Raghavan} {et~al.}(2010){Raghavan}, {McAlister}, {Henry}, {Latham},
  {Marcy}, {Mason}, {Gies}, {White}, \& {ten Brummelaar}}]{Raghavan2010}
{Raghavan}, D., {McAlister}, H.~A., {Henry}, T.~J., {et~al.} 2010, \apjs, 190,
  1

\bibitem[{{Reiners} {et~al.}(2018){Reiners}, {Zechmeister}, {Caballero},
  {Ribas}, {Morales}, {Jeffers}, {Sch{\"o}fer}, {Tal-Or}, {Quirrenbach},
  {Amado}, {Kaminski}, {Seifert}, {Abril}, {Aceituno}, {Alonso-Floriano},
  {Ammler-von Eiff}, {Antona}, {Anglada-Escud{\'e}}, {Anwand-Heerwart},
  {Arroyo-Torres}, {Azzaro}, {Baroch}, {Barrado}, {Bauer}, {Becerril},
  {B{\'e}jar}, {Ben{\'{\i}}tez}, {Berdi{\~n}as}, {Bergond}, {Bl{\"u}mcke},
  {Brinkm{\"o}ller}, {del Burgo}, {Cano}, {C{\'a}rdenas V{\'a}zquez}, {Casal},
  {Cifuentes}, {Claret}, {Colom{\'e}}, {Cort{\'e}s-Contreras}, {Czesla},
  {D{\'{\i}}ez-Alonso}, {Dreizler}, {Feiz}, {Fern{\'a}ndez}, {Ferro},
  {Fuhrmeister}, {Galad{\'{\i}}-Enr{\'{\i}}quez}, {Garcia-Piquer},
  {Garc{\'{\i}}a Vargas}, {Gesa}, {G{\'o}mez}, {Galera}, {Gonz{\'a}lez
  Hern{\'a}ndez}, {Gonz{\'a}lez-Peinado}, {Gr{\"o}zinger}, {Grohnert},
  {Gu{\`a}rdia}, {Guenther}, {Guijarro}, {de Guindos}, {Guti{\'e}rrez-Soto},
  {Hagen}, {Hatzes}, {Hauschildt}, {Hedrosa}, {Helmling}, {Henning}, {Hermelo},
  {Hern{\'a}ndez Arab{\'{\i}}}, {Hern{\'a}ndez Casta{\~n}o}, {Hern{\'a}ndez
  Hernando}, {Herrero}, {Huber}, {Huke}, {Johnson}, {de Juan}, {Kim}, {Klein},
  {Kl{\"u}ter}, {Klutsch}, {K{\"u}rster}, {Lafarga}, {Lamert}, {Lamp{\'o}n},
  {Lara}, {Laun}, {Lemke}, {Lenzen}, {Launhardt}, {L{\'o}pez del Fresno},
  {L{\'o}pez-Gonz{\'a}lez}, {L{\'o}pez-Puertas}, {L{\'o}pez Salas},
  {L{\'o}pez-Santiago}, {Luque}, {Mag{\'a}n Madinabeitia}, {Mall}, {Mancini},
  {Mandel}, {Marfil}, {Mar{\'{\i}}n Molina}, {Maroto}, {Fern{\'a}ndez},
  {Mart{\'{\i}}n}, {Mart{\'{\i}}n-Ruiz}, {Marvin}, {Mathar}, {Mirabet},
  {Montes}, {Moreno-Raya}, {Moya}, {Mundt}, {Nagel}, {Naranjo}, {Nortmann},
  {Nowak}, {Ofir}, {Oreiro}, {Pall{\'e}}, {Panduro}, {Pascual}, {Passegger},
  {Pavlov}, {Pedraz}, {P{\'e}rez-Calpena}, {P{\'e}rez Medialdea}, {Perger},
  {Perryman}, {Pluto}, {Rabaza}, {Ram{\'o}n}, {Rebolo}, {Redondo}, {Reffert},
  {Reinhart}, {Rhode}, {Rix}, {Rodler}, {Rodr{\'{\i}}guez},
  {Rodr{\'{\i}}guez-L{\'o}pez}, {Rodr{\'{\i}}guez Trinidad}, {Rohloff},
  {Rosich}, {Sadegi}, {S{\'a}nchez-Blanco}, {S{\'a}nchez Carrasco},
  {S{\'a}nchez-L{\'o}pez}, {Sanz-Forcada}, {Sarkis}, {Sarmiento},
  {Sch{\"a}fer}, {Schmitt}, {Schiller}, {Schweitzer}, {Solano}, {Stahl},
  {Strachan}, {St{\"u}rmer}, {Su{\'a}rez}, {Tabernero}, {Tala}, {Trifonov},
  {Tulloch}, {Ulbrich}, {Veredas}, {Vico Linares}, {Vilardell}, {Wagner},
  {Winkler}, {Wolthoff}, {Xu}, {Yan}, \& {Zapatero Osorio}}]{Reiners2018b}
{Reiners}, A., {Zechmeister}, M., {Caballero}, J.~A., {et~al.} 2018, \aap, 612,
  A49

\bibitem[{{Reyl{\'e}} {et~al.}(2021){Reyl{\'e}}, {Jardine}, {Fouqu{\'e}},
  {Caballero}, {Smart}, \& {Sozzetti}}]{Reyle2021}
{Reyl{\'e}}, C., {Jardine}, K., {Fouqu{\'e}}, P., {et~al.} 2021, \aap, 650,
  A201

\bibitem[{{Ricker} {et~al.}(2015){Ricker}, {Winn}, {Vanderspek}, {Latham},
  {Bakos}, {Bean}, {Berta-Thompson}, {Brown}, {Buchhave}, {Butler}, {Butler},
  {Chaplin}, {Charbonneau}, {Christensen-Dalsgaard}, {Clampin}, {Deming},
  {Doty}, {De Lee}, {Dressing}, {Dunham}, {Endl}, {Fressin}, {Ge}, {Henning},
  {Holman}, {Howard}, {Ida}, {Jenkins}, {Jernigan}, {Johnson}, {Kaltenegger},
  {Kawai}, {Kjeldsen}, {Laughlin}, {Levine}, {Lin}, {Lissauer}, {MacQueen},
  {Marcy}, {McCullough}, {Morton}, {Narita}, {Paegert}, {Palle}, {Pepe},
  {Pepper}, {Quirrenbach}, {Rinehart}, {Sasselov}, {Sato}, {Seager},
  {Sozzetti}, {Stassun}, {Sullivan}, {Szentgyorgyi}, {Torres}, {Udry}, \&
  {Villasenor}}]{Ricker2015}
{Ricker}, G.~R., {Winn}, J.~N., {Vanderspek}, R., {et~al.} 2015, JATIS, 1,
  014003

\bibitem[{{Rizzuto} {et~al.}(2018){Rizzuto}, {Vanderburg}, {Mann}, {Kraus},
  {Dressing}, {Ag{\"u}eros}, {Douglas}, \& {Krolikowski}}]{Rizzuto2018_K2-264}
{Rizzuto}, A.~C., {Vanderburg}, A., {Mann}, A.~W., {et~al.} 2018, \aj, 156, 195

\bibitem[{{Roell} {et~al.}(2012){Roell}, {Neuh{\"a}user}, {Seifahrt}, \&
  {Mugrauer}}]{Roell2012}
{Roell}, T., {Neuh{\"a}user}, R., {Seifahrt}, A., \& {Mugrauer}, M. 2012, \aap,
  542, A92

\bibitem[{{Rogers}(2015)}]{Rogers2015}
{Rogers}, L.~A. 2015, \apj, 801, 41

\bibitem[{{Rogers} {et~al.}(2011){Rogers}, {Bodenheimer}, {Lissauer}, \&
  {Seager}}]{Rogers2011}
{Rogers}, L.~A., {Bodenheimer}, P., {Lissauer}, J.~J., \& {Seager}, S. 2011,
  \apj, 738, 59

\bibitem[{{Rogers} \& {Seager}(2010)}]{RogersSeager2010_degeneracy}
{Rogers}, L.~A. \& {Seager}, S. 2010, \apj, 712, 974

\bibitem[{{Rossiter}(1924)}]{Rossiter1924}
{Rossiter}, R.~A. 1924, \apj, 60, 15

\bibitem[{{Schlecker} {et~al.}(2020){Schlecker}, {Mordasini}, {Emsenhuber},
  {Klahr}, {Henning}, {Burn}, {Alibert}, \& {Benz}}]{Schlecker2020}
{Schlecker}, M., {Mordasini}, C., {Emsenhuber}, A., {et~al.} 2020, arXiv
  e-prints, arXiv:2007.05563

\bibitem[{{Sch{\"o}fer} {et~al.}(2019){Sch{\"o}fer}, {Jeffers}, {Reiners},
  {Shulyak}, {Fuhrmeister}, {Johnson}, {Zechmeister}, {Ribas}, {Quirrenbach},
  {Amado}, {Caballero}, {Anglada-Escud{\'e}}, {Bauer}, {B{\'e}jar},
  {Cort{\'e}s-Contreras}, {Dreizler}, {Guenther}, {Kaminski}, {K{\"u}rster},
  {Lafarga}, {Montes}, {Morales}, {Pedraz}, \& {Tal-Or}}]{Schoefer2019}
{Sch{\"o}fer}, P., {Jeffers}, S.~V., {Reiners}, A., {et~al.} 2019, \aap, 623,
  A44

\bibitem[{{Schweitzer} {et~al.}(2019){Schweitzer}, {Passegger}, {Cifuentes},
  {B{\'e}jar}, {Cort{\'e}s-Contreras}, {Caballero}, {del Burgo}, {Czesla},
  {K{\"u}rster}, {Montes}, {Zapatero Osorio}, {Ribas}, {Reiners},
  {Quirrenbach}, {Amado}, {Aceituno}, {Anglada-Escud{\'e}}, {Bauer},
  {Dreizler}, {Jeffers}, {Guenther}, {Henning}, {Kaminski}, {Lafarga},
  {Marfil}, {Morales}, {Schmitt}, {Seifert}, {Solano}, {Tabernero}, \&
  {Zechmeister}}]{Schweitzer2019}
{Schweitzer}, A., {Passegger}, V.~M., {Cifuentes}, C., {et~al.} 2019, \aap,
  625, A68

\bibitem[{{Seifahrt} {et~al.}(2018){Seifahrt}, {St{\"u}rmer}, {Bean}, \&
  {Schwab}}]{maroonx}
{Seifahrt}, A., {St{\"u}rmer}, J., {Bean}, J.~L., \& {Schwab}, C. 2018, in
  Society of Photo-Optical Instrumentation Engineers (SPIE) Conference Series,
  Vol. 10702, Ground-based and Airborne Instrumentation for Astronomy VII, ed.
  C.~J. {Evans}, L.~{Simard}, \& H.~{Takami}, 107026D

\bibitem[{{Shappee} {et~al.}(2014){Shappee}, {Prieto}, {Grupe}, {Kochanek},
  {Stanek}, {De Rosa}, {Mathur}, {Zu}, {Peterson}, {Pogge}, {Komossa}, {Im},
  {Jencson}, {Holoien}, {Basu}, {Beacom}, {Szczygie{\l}}, {Brimacombe},
  {Adams}, {Campillay}, {Choi}, {Contreras}, {Dietrich}, {Dubberley},
  {Elphick}, {Foale}, {Giustini}, {Gonzalez}, {Hawkins}, {Howell}, {Hsiao},
  {Koss}, {Leighly}, {Morrell}, {Mudd}, {Mullins}, {Nugent}, {Parrent},
  {Phillips}, {Pojmanski}, {Rosing}, {Ross}, {Sand}, {Terndrup}, {Valenti},
  {Walker}, \& {Yoon}}]{Shappee2014_ASAS}
{Shappee}, B.~J., {Prieto}, J.~L., {Grupe}, D., {et~al.} 2014, \apj, 788, 48

\bibitem[{{Skrutskie} {et~al.}(2006){Skrutskie}, {Cutri}, {Stiening},
  {Weinberg}, {Schneider}, {Carpenter}, {Beichman}, {Capps}, {Chester},
  {Elias}, {Huchra}, {Liebert}, {Lonsdale}, {Monet}, {Price}, {Seitzer},
  {Jarrett}, {Kirkpatrick}, {Gizis}, {Howard}, {Evans}, {Fowler}, {Fullmer},
  {Hurt}, {Light}, {Kopan}, {Marsh}, {McCallon}, {Tam}, {Van Dyk}, \&
  {Wheelock}}]{Skrutskie2006_2MASS}
{Skrutskie}, M.~F., {Cutri}, R.~M., {Stiening}, R., {et~al.} 2006, \aj, 131,
  1163

\bibitem[{{Smith} {et~al.}(2012){Smith}, {Stumpe}, {Van Cleve}, {Jenkins},
  {Barclay}, {Fanelli}, {Girouard}, {Kolodziejczak}, {McCauliff}, {Morris}, \&
  {Twicken}}]{Smith2012}
{Smith}, J.~C., {Stumpe}, M.~C., {Van Cleve}, J.~E., {et~al.} 2012, \pasp, 124,
  1000

\bibitem[{Southworth(2011)}]{Southworth2011}
Southworth, J. 2011, \mnras, 417, 2166

\bibitem[{{Speagle} \& {Barbary}(2018)}]{dynesty}
{Speagle}, J. \& {Barbary}, K. 2018, {dynesty: Dynamic Nested Sampling
  package}, Astrophysics Source Code Library

\bibitem[{{Speagle}(2020)}]{dynesty2020}
{Speagle}, J.~S. 2020, \mnras, 493, 3132

\bibitem[{{Stassun} {et~al.}(2018){Stassun}, {Oelkers}, {Pepper}, {Paegert},
  {De Lee}, {Torres}, {Latham}, {Charpinet}, {Dressing}, {Huber}, {Kane},
  {L{\'e}pine}, {Mann}, {Muirhead}, {Rojas-Ayala}, {Silvotti}, {Fleming},
  {Levine}, \& {Plavchan}}]{Stassun2018}
{Stassun}, K.~G., {Oelkers}, R.~J., {Pepper}, J., {et~al.} 2018, \aj, 156, 102

\bibitem[{{Stefansson} {et~al.}(2020){Stefansson}, {Mahadevan}, {Maney},
  {Ninan}, {Robertson}, {Rajagopal}, {Haase}, {Allen}, {Ford}, {Winn},
  {Wolfgang}, {Dawson}, {Wisniewski}, {Bender}, {Ca{\~n}as}, {Cochran},
  {Diddams}, {Fredrick}, {Halverson}, {Hearty}, {Hebb}, {Kanodia}, {Levi},
  {Metcalf}, {Monson}, {Ramsey}, {Roy}, {Schwab}, {Terrien}, \&
  {Wright}}]{Stefansson2020_RM_K2-25b}
{Stefansson}, G., {Mahadevan}, S., {Maney}, M., {et~al.} 2020, \aj, 160, 192

\bibitem[{{Stock} \& {Kemmer}(2020)}]{Stock2020_aliasfinder}
{Stock}, S. \& {Kemmer}, J. 2020, Journal of Open Source Software, 5(45), 1771

\bibitem[{{Stock} {et~al.}(2020{\natexlab{a}}){Stock}, {Kemmer}, {Reffert},
  {Trifonov}, {Kaminski}, {Dreizler}, {Quirrenbach}, {Caballero}, {Reiners},
  {Jeffers}, {Anglada-Escud{\'e}}, {Ribas}, {Amado}, {Barrado}, {Barnes},
  {Bauer}, {Berdi{\~n}as}, {B{\'e}jar}, {Coleman}, {Cort{\'e}s-Contreras},
  {D{\'\i}ez-Alonso}, {Dom{\'\i}nguez-Fern{\'a}ndez}, {Espinoza}, {Haswell},
  {Hatzes}, {Henning}, {Jenkins}, {Jones}, {Kossakowski}, {K{\"u}rster},
  {Lafarga}, {Lee}, {L{\'o}pez Gonz{\'a}lez}, {Montes}, {Morales}, {Morales},
  {Pall{\'e}}, {Pedraz}, {Rodr{\'\i}guez}, {Rodr{\'\i}guez-L{\'o}pez}, \&
  {Zechmeister}}]{Stock2020_yzceti}
{Stock}, S., {Kemmer}, J., {Reffert}, S., {et~al.} 2020{\natexlab{a}}, \aap,
  636, A119

\bibitem[{{Stock} {et~al.}(2020{\natexlab{b}}){Stock}, {Nagel}, {Kemmer},
  {Passegger}, {Reffert}, {Quirrenbach}, {Caballero}, {Czesla}, {B{\'e}jar},
  {Cardona}, {D{\'\i}ez-Alonso}, {Herrero}, {Lalitha}, {Schlecker}, {Tal-Or},
  {Rodr{\'\i}guez}, {Rodr{\'\i}guez-L{\'o}pez}, {Ribas}, {Reiners}, {Amado},
  {Bauer}, {Bluhm}, {Cort{\'e}s-Contreras}, {Gonz{\'a}lez-Cuesta}, {Dreizler},
  {Hatzes}, {Henning}, {Jeffers}, {Kaminski}, {K{\"u}rster}, {Lafarga},
  {L{\'o}pez-Gonz{\'a}lez}, {Montes}, {Morales}, {Pedraz}, {Sch{\"o}fer},
  {Schweitzer}, {Trifonov}, {Zapatero Osorio}, \&
  {Zechmeister}}]{Stock2020_threesuperearths}
{Stock}, S., {Nagel}, E., {Kemmer}, J., {et~al.} 2020{\natexlab{b}}, \aap, 643,
  A112

\bibitem[{{Stumpe} {et~al.}(2014){Stumpe}, {Smith}, {Catanzarite}, {Van Cleve},
  {Jenkins}, {Twicken}, \& {Girouard}}]{Stumpe2014}
{Stumpe}, M.~C., {Smith}, J.~C., {Catanzarite}, J.~H., {et~al.} 2014, \pasp,
  126, 100

\bibitem[{{Stumpe} {et~al.}(2012){Stumpe}, {Smith}, {Van Cleve}, {Twicken},
  {Barclay}, {Fanelli}, {Girouard}, {Jenkins}, {Kolodziejczak}, {McCauliff}, \&
  {Morris}}]{Stumpe2012}
{Stumpe}, M.~C., {Smith}, J.~C., {Van Cleve}, J.~E., {et~al.} 2012, \pasp, 124,
  985

\bibitem[{{Thao} {et~al.}(2020){Thao}, {Mann}, {Johnson}, {Newton}, {Guo},
  {Kain}, {Rizzuto}, {Charbonneau}, {Dalba}, {Gaidos}, {Irwin}, \&
  {Kraus}}]{Thao2020_TS_K2-25}
{Thao}, P.~C., {Mann}, A.~W., {Johnson}, M.~C., {et~al.} 2020, \aj, 159, 32

\bibitem[{{Thebault} \& {Haghighipour}(2015)}]{Thebault2015}
{Thebault}, P. \& {Haghighipour}, N. 2015, {Planet Formation in Binaries},
  309--340

\bibitem[{{Trifonov}(2019)}]{exostriker}
{Trifonov}, T. 2019, {The Exo-Striker: Transit and radial velocity interactive
  fitting tool for orbital analysis and N-body simulations}

\bibitem[{{Trifonov} {et~al.}(2021){Trifonov}, {Caballero}, {Morales},
  {Seifahrt}, {Ribas}, {Reiners}, {Bean}, {Luque}, {Parviainen}, {Pall{\'e}},
  {Stock}, {Zechmeister}, {Amado}, {Anglada-Escud{\'e}}, {Azzaro}, {Barclay},
  {B{\'e}jar}, {Bluhm}, {Casasayas-Barris}, {Cifuentes}, {Collins}, {Collins},
  {Cort{\'e}s-Contreras}, {de Leon}, {Dreizler}, {Dressing}, {Esparza-Borges},
  {Espinoza}, {Fausnaugh}, {Fukui}, {Hatzes}, {Hellier}, {Henning}, {Henze},
  {Herrero}, {Jeffers}, {Jenkins}, {Jensen}, {Kaminski}, {Kasper},
  {Kossakowski}, {K{\"u}rster}, {Lafarga}, {Latham}, {Mann}, {Molaverdikhani},
  {Montes}, {Montet}, {Murgas}, {Narita}, {Oshagh}, {Passegger}, {Pollacco},
  {Quinn}, {Quirrenbach}, {Ricker}, {Rodr{\'\i}guez L{\'o}pez}, {Sanz-Forcada},
  {Schwarz}, {Schweitzer}, {Seager}, {Shporer}, {Stangret}, {St{\"u}rmer},
  {Tan}, {Tenenbaum}, {Twicken}, {Vanderspek}, \& {Winn}}]{Trifonov2021}
{Trifonov}, T., {Caballero}, J.~A., {Morales}, J.~C., {et~al.} 2021, Science,
  371, 1038

\bibitem[{{Trifonov} {et~al.}(2018){Trifonov}, {K{\"u}rster}, {Zechmeister},
  {Tal-Or}, {Caballero}, {Quirrenbach}, {Amado}, {Ribas}, {Reiners}, {Reffert},
  {Dreizler}, {Hatzes}, {Kaminski}, {Launhardt}, {Henning}, {Montes},
  {B{\'e}jar}, {Mundt}, {Pavlov}, {Schmitt}, {Seifert}, {Morales}, {Nowak},
  {Jeffers}, {Rodr{\'\i}guez-L{\'o}pez}, {del Burgo}, {Anglada-Escud{\'e}},
  {L{\'o}pez-Santiago}, {Mathar}, {Ammler-von Eiff}, {Guenther}, {Barrado},
  {Gonz{\'a}lez Hern{\'a}ndez}, {Mancini}, {St{\"u}rmer}, {Abril}, {Aceituno},
  {Alonso-Floriano}, {Antona}, {Anwand-Heerwart}, {Arroyo-Torres}, {Azzaro},
  {Baroch}, {Bauer}, {Becerril}, {Ben{\'\i}tez}, {Berdi{\~n}as}, {Bergond},
  {Bl{\"u}mcke}, {Brinkm{\"o}ller}, {Cano}, {C{\'a}rdenas V{\'a}zquez},
  {Casal}, {Cifuentes}, {Claret}, {Colom{\'e}}, {Cort{\'e}s-Contreras},
  {Czesla}, {D{\'\i}ez-Alonso}, {Feiz}, {Fern{\'a}ndez}, {Ferro},
  {Fuhrmeister}, {Galad{\'\i}-Enr{\'\i}quez}, {Garcia-Piquer}, {Garc{\'\i}a
  Vargas}, {Gesa}, {G{\'o}mez Galera}, {Gonz{\'a}lez-Peinado}, {Gr{\"o}zinger},
  {Grohnert}, {Gu{\`a}rdia}, {Guijarro}, {de Guindos}, {Guti{\'e}rrez-Soto},
  {Hagen}, {Hauschildt}, {Hedrosa}, {Helmling}, {Hermelo}, {Hern{\'a}ndez
  Arab{\'\i}}, {Hern{\'a}ndez Casta{\~n}o}, {Hern{\'a}ndez Hernando},
  {Herrero}, {Huber}, {Huke}, {Johnson}, {de Juan}, {Kim}, {Klein},
  {Kl{\"u}ter}, {Klutsch}, {Lafarga}, {Lamp{\'o}n}, {Lara}, {Laun}, {Lemke},
  {Lenzen}, {L{\'o}pez del Fresno}, {L{\'o}pez-Gonz{\'a}lez},
  {L{\'o}pez-Puertas}, {L{\'o}pez Salas}, {Luque}, {Mag{\'a}n Madinabeitia},
  {Mall}, {Mandel}, {Marfil}, {Mar{\'\i}n Molina}, {Maroto Fern{\'a}ndez},
  {Mart{\'\i}n}, {Mart{\'\i}n-Ruiz}, {Marvin}, {Mirabet}, {Moya},
  {Moreno-Raya}, {Nagel}, {Naranjo}, {Nortmann}, {Ofir}, {Oreiro}, {Pall{\'e}},
  {Panduro}, {Pascual}, {Passegger}, {Pedraz}, {P{\'e}rez-Calpena}, {P{\'e}rez
  Medialdea}, {Perger}, {Perryman}, {Pluto}, {Rabaza}, {Ram{\'o}n}, {Rebolo},
  {Redondo}, {Reinhardt}, {Rhode}, {Rix}, {Rodler}, {Rodr{\'\i}guez},
  {Rodr{\'\i}guez Trinidad}, {Rohloff}, {Rosich}, {Sadegi},
  {S{\'a}nchez-Blanco}, {S{\'a}nchez Carrasco}, {S{\'a}nchez-L{\'o}pez},
  {Sanz-Forcada}, {Sarkis}, {Sarmiento}, {Sch{\"a}fer}, {Schiller},
  {Sch{\"o}fer}, {Schweitzer}, {Solano}, {Stahl}, {Strachan}, {Su{\'a}rez},
  {Tabernero}, {Tala}, {Tulloch}, {Veredas}, {Vico Linares}, {Vilardell},
  {Wagner}, {Winkler}, {Wolthoff}, {Xu}, {Yan}, \& {Zapatero
  Osorio}}]{Trifonov2018_carmenespaper}
{Trifonov}, T., {K{\"u}rster}, M., {Zechmeister}, M., {et~al.} 2018, \aap, 609,
  A117

\bibitem[{{Trifonov} {et~al.}(2020){Trifonov}, {Tal-Or}, {Zechmeister},
  {Kaminski}, {Zucker}, \& {Mazeh}}]{Trifonov2020}
{Trifonov}, T., {Tal-Or}, L., {Zechmeister}, M., {et~al.} 2020, \aap, 636, A74

\bibitem[{{Trotta}(2008)}]{Trotta2008}
{Trotta}, R. 2008, Contemporary Physics, 49, 71

\bibitem[{{Van Eylen} {et~al.}(2018){Van Eylen}, {Agentoft}, {Lundkvist},
  {Kjeldsen}, {Owen}, {Fulton}, {Petigura}, \& {Snellen}}]{VanEylen2018}
{Van Eylen}, V., {Agentoft}, C., {Lundkvist}, M.~S., {et~al.} 2018, \mnras,
  479, 4786

\bibitem[{{Van Eylen} {et~al.}(2021){Van Eylen}, {Astudillo-Defru}, {Bonfils},
  {Livingston}, {Hirano}, {Luque}, {Lam}, {Justesen}, {Winn}, {Gandolfi},
  {Nowak}, {Palle}, {Albrecht}, {Dai}, {Campos Estrada}, {Owen},
  {Foreman-Mackey}, {Fridlund}, {Korth}, {Mathur}, {Forveille}, {Mikal-Evans},
  {Osborne}, {Ho}, {Almenara}, {Artigau}, {Barrag{\'a}n}, {Barros}, {Bouchy},
  {Cabrera}, {Caldwell}, {Charbonneau}, {Chaturvedi}, {Cochran}, {Csizmadia},
  {Damasso}, {Delfosse}, {De Medeiros}, {D{\'\i}az}, {Doyon}, {Esposito},
  {F{\H{u}}r{\'e}sz}, {Figueira}, {Georgieva}, {Goffo}, {Grziwa}, {Guenther},
  {Hatzes}, {Jenkins}, {Kabath}, {Knudstrup}, {Latham}, {Lavie}, {Lovis},
  {Mennickent}, {Mullally}, {Murgas}, {Narita}, {Pepe}, {Persson}, {Redfield},
  {Ricker}, {Santos}, {Seager}, {Serrano}, {Smith}, {Mascare{\~n}o}, {Subjak},
  {Twicken}, {Udry}, {Vanderspek}, \& {Zapatero Osorio}}]{VanEylen2021}
{Van Eylen}, V., {Astudillo-Defru}, N., {Bonfils}, X., {et~al.} 2021, \mnras,
  507, 2154

\bibitem[{{Venturini} {et~al.}(2020){Venturini}, {Guilera}, {Haldemann},
  {Ronco}, \& {Mordasini}}]{Venturini2020}
{Venturini}, J., {Guilera}, O.~M., {Haldemann}, J., {Ronco}, M.~P., \&
  {Mordasini}, C. 2020, \aap, 643, L1

\bibitem[{{Wang} {et~al.}(2014){Wang}, {Fischer}, {Xie}, \&
  {Ciardi}}]{Wang2014}
{Wang}, J., {Fischer}, D.~A., {Xie}, J.-W., \& {Ciardi}, D.~R. 2014, \apj, 791,
  111

\bibitem[{Wilson {et~al.}(2021)Wilson, Gibson, Lothringer, Sing, Mikal-Evans,
  de~Mooij, Nikolov, \& Watson}]{Wilson2021}
Wilson, J., Gibson, N.~P., Lothringer, J.~D., {et~al.} 2021, MNRAS, 503, 4787

\bibitem[{{Winn} {et~al.}(2010){Winn}, {Fabrycky}, {Albrecht}, \&
  {Johnson}}]{Winn2010_rm}
{Winn}, J.~N., {Fabrycky}, D., {Albrecht}, S., \& {Johnson}, J.~A. 2010, \apjl,
  718, L145

\bibitem[{{Winters} {et~al.}(2019{\natexlab{a}}){Winters}, {Henry}, {Jao},
  {Subasavage}, {Chatelain}, {Slatten}, {Riedel}, {Silverstein}, \&
  {Payne}}]{Winters2019}
{Winters}, J.~G., {Henry}, T.~J., {Jao}, W.-C., {et~al.} 2019{\natexlab{a}},
  \aj, 157, 216

\bibitem[{{Winters} {et~al.}(2015){Winters}, {Henry}, {Lurie}, {Hambly}, {Jao},
  {Bartlett}, {Boyd}, {Dieterich}, {Finch}, {Hosey}, {Ianna}, {Riedel},
  {Slatten}, \& {Subasavage}}]{Winters2015}
{Winters}, J.~G., {Henry}, T.~J., {Lurie}, J.~C., {et~al.} 2015, \aj, 149, 5

\bibitem[{{Winters} {et~al.}(2019{\natexlab{b}}){Winters}, {Medina}, {Irwin},
  {Charbonneau}, {Astudillo-Defru}, {Horch}, {Eastman}, {Vrijmoet}, {Henry},
  {Diamond-Lowe}, {Winston}, {Barclay}, {Bonfils}, {Ricker}, {Vanderspek},
  {Latham}, {Seager}, {Winn}, {Jenkins}, {Udry}, {Twicken}, {Teske},
  {Tenenbaum}, {Pepe}, {Murgas}, {Muirhead}, {Mink}, {Lovis}, {Levine},
  {L{\'e}pine}, {Jao}, {Henze}, {Fur{\'e}sz}, {Forveille}, {Figueira},
  {Esquerdo}, {Dressing}, {D{\'\i}az}, {Delfosse}, {Burke}, {Bouchy},
  {Berlind}, \& {Almenara}}]{Winters2019_ltt1445}
{Winters}, J.~G., {Medina}, A.~A., {Irwin}, J.~M., {et~al.} 2019{\natexlab{b}},
  \aj, 158, 152

\bibitem[{{Zacharias} {et~al.}(2012){Zacharias}, {Finch}, {Girard}, {Henden},
  {Bartlett}, {Monet}, \& {Zacharias}}]{Zacharias2013}
{Zacharias}, N., {Finch}, C.~T., {Girard}, T.~M., {et~al.} 2012, VizieR Online
  Data Catalog, I/322A

\bibitem[{{Zechmeister} \& {K{\"u}rster}(2009)}]{Zechmeister2009_GLS}
{Zechmeister}, M. \& {K{\"u}rster}, M. 2009, \aap, 496, 577

\bibitem[{{Zechmeister} {et~al.}(2018){Zechmeister}, {Reiners}, {Amado},
  {Azzaro}, {Bauer}, {B{\'e}jar}, {Caballero}, {Guenther}, {Hagen}, \&
  {Jeffers}}]{serval}
{Zechmeister}, M., {Reiners}, A., {Amado}, P.~J., {et~al.} 2018, \aap, 609, A12

\bibitem[{{Zeng} {et~al.}(2019){Zeng}, {Jacobsen}, {Sasselov}, {Petaev},
  {Vanderburg}, {Lopez-Morales}, {Perez-Mercader}, {Mattsson}, {Li}, {Heising},
  {Bonomo}, {Damasso}, {Berger}, {Cao}, {Levi}, \& {Wordsworth}}]{Zeng2019}
{Zeng}, L., {Jacobsen}, S.~B., {Sasselov}, D.~D., {et~al.} 2019, PNAS, 116,
  9723

\end{thebibliography}

\newpage
\begin{appendix} %First appendix

\section{Further RV-only investigation}\label{appendix:gpwide}
\begin{figure}[!h]
    \centering
    \includegraphics[width=1\linewidth]{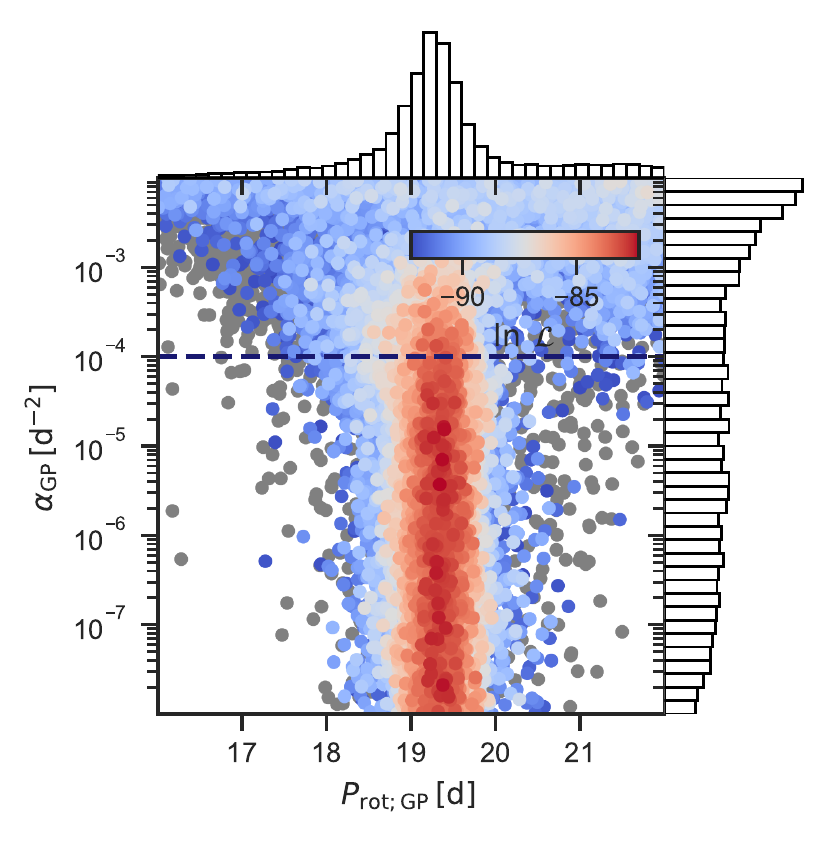}
    \caption[]{Posterior distribution of the GP parameters for $\alpha$ and $P_\textnormal{rot}$ for an RV-only run, including a QP-GP centered on the rotational period $\sim$19\,d and using an unconstrained prior on $\alpha$. Samples are color-coded in accordance to their log-likelihood values, where those with a $\Delta \ln L > 10$ compared to the best solution are colored gray. The horizontal dashed line represents the maximum value at which we constrained the $\alpha$ parameter in order to exclude the plateau-like behavior. The plot is inspired by \cite{Stock2020_threesuperearths}.
    }
    \label{fig:alphavsprot}
\end{figure}

For the main RV-only analysis (Sect.~\ref{sec:rvonlymodeling}), we decided to use the dSHO-GP kernel (Eq.~\ref{eq:dsho}) for its computationally faster capabilities in comparison to the QP-GP kernel, as presented in \texttt{george},

\begin{equation*}
k_{i,j}(\tau)=\sigma^2_\textnormal{GP}\exp\left(-\alpha\tau^2-\Gamma\sin^2\left(\frac{\pi\tau}{P_\textnormal{rot}}\right)\right),
\end{equation*}

\noindent where $\tau = |t_{i} - t_{j}|$ is the temporal distance between two points, $\sigma_\textnormal{GP}$ is the amplitude of the GP modulation, $\alpha$ is the inverse length-scale of the GP exponential component, $\Gamma$ is the amplitude of the GP sine-squared component, and $P_\textnormal{rot}$ is the rotational modulation of the GP quasi-periodic component. We followed the approach as discussed by \cite{Stock2020_threesuperearths} in order to set up the priors and thus constrained $\alpha$ to refrain from including the samples that exhibit a ``plateau-like'' behavior \citep[see Fig.~6 in][]{Stock2020_threesuperearths}. Upon visual inspection of the $\alpha$-versus-$P_\textnormal{rot}$ correlation plot when keeping $\alpha$ unconstrained (Fig.~\ref{fig:alphavsprot}), we found that the $\alpha$ prior values for this analysis should be log-uniform from $10^{-8}$ to $10^{-4}$, corresponding to timescales\footnote{$\alpha = 1/l^2$, where $l$ is the timescale for the GP variations. 
In the original \juliet\ paper \citep{juliet}, $\alpha$ was defined as $\alpha = 1/2l^2$. However, this has since been corrected.} of 100 to 10\,000 days and we set $\Gamma$ to be log-uniform from $10^{-2}$ to $10^{1}$. These constraints ensured that truly (decaying) periodic-like signals were picked up by the GP, rather than allowing the exponentially decaying component to be the dominant term, and thus acting as a white-noise filter.
As already stated in Sect.~\ref{sec:rvonlymodeling}, using either of the two kernels resulted in indistinguishable differences in their Bayesian log evidence. 

We further experimented with the two GP kernels by applying a wider prior for $P_\textnormal{rot}$, $\mathcal{U}(10\,\textnormal{d},150\,\textnormal{d})$. We denote this with GP$_\textnormal{wide}$. The motivation for the wider period prior is to evaluate whether the GP is flexible enough and can account for both the stellar activity and the longer-term signal at the same time, in other words, how the statistical metric that we use (i.e., the Bayesian log evidence) compares to that of the sinusoidal variation. The GP$_\textnormal{wide}$ often picked up signals at 80\,d, 35\,d (the alias of the stellar rotation period), and also 102\,d, but the 19\,d signal attributed to the true stellar rotation period seemed to be dominated by the others. In fact, the 1 circular Kep (2.5\,d) + GP$_\textnormal{wide}$ model had a slightly favorable log evidence ($\Delta\ln{\mathcal{Z}} \sim 2$) when compared to our final model choice of 1 circular Kep (2.5\,d) + 1 Sin ($\sim$102\,d) + dSHO-GP$_\textnormal{19d}$. This is intriguing, because based on the log evidence alone, the statistically better model is the one including the GP$_\textnormal{wide}$ component, which is indeed able to account for both signals. One reason for this behavior can be that we do not have enough of time baseline coverage to truly constrain the long-term trend. 
Hence, this could be used as a word of caution for when performing a ``blind search'' for signals by applying a wide period prior for the  GP because it could be that it does statistically have an improved value and its flexibility can account for the signals. 
Though, this model is not physically motivated considering that we know the rotation period of the star and this should be taken into account. One should thus take care when implementing a GP into the model and should ask for what purpose it is trying to achieve.

\clearpage
\section{Priors and posteriors} \label{sec:appendixpriorsposteriors} 

%%%% RV ONLY PRIORS
\begin{table*}[!h]
\small
\centering
\caption[Priors for the RV-only fits for TOI-1201]{Priors for the RV-only fits for TOI-1201 with \juliet\ in Sect.~\ref{sec:rvonlymodeling}.}
\label{tab:priors_rvonly}
\begin{tabular}{lccr}
\hline
\hline
\noalign{\smallskip}
Parameter & Prior & Unit & Description \\
\noalign{\smallskip}
\hline
\noalign{\smallskip}
\multicolumn{4}{c}{\textit{Parameters for transiting planet b}} \\ 
\noalign{\smallskip}
~~~$P_{b}$  & $\mathcal{F}(2.4919858227 ) $ & d & Period\\
~~~$t_{0,b}$ (BJD)  & $\mathcal{F}(2459169.2321864809) $ & d & Time of transit center\\
~~~$K_{b}$  & $\mathcal{U}(0.0,20) $ & \ms & RV semi-amplitude\\
~~~$S_{1,b} = \sqrt{e_c}\sin \omega_c$  & $\mathcal{F}{(0.0)}$ (circular) & $\dots$ & Parametrization for $e$ and $\omega$\\
 &  $\mathcal{U}{(-1,1)}$ (eccentric) & $\dots$ & Parametrization for $e$ and $\omega$\\
~~~$S_{2,b} = \sqrt{e_b}\cos \omega_b$  & $\mathcal{F}{(0.0)}$ (circular) & $\dots$ & Parametrization for $e$ and $\omega$\\
 &  $\mathcal{U}{(-1,1)}$ (eccentric) & $\dots$ & Parametrization for $e$ and $\omega$\\
\noalign{\smallskip}
\multicolumn{4}{c}{\textit{Parameters for long-term signal in RVs}} \\ 
\noalign{\smallskip}
~~~$P_{[c]}$  & $\mathcal{U}(60.0,150.0) $ & d & Period\\
~~~$t_{0,[c]}$  & $\mathcal{U}(2458730.0,2458840.0) $ & d & Time of transit center\\
~~~$K_{[c]}$  & $\mathcal{U}(0.0,20.0) $ & \ms & RV semi-amplitude\\
~~~$S_{1,[c]} = \sqrt{e_{[c]}}\sin \omega_{[c]}$  & $\mathcal{F}{(0.0)}$ (circular) & $\dots$ & Parametrization for $e$ and $\omega$\\
 &  $\mathcal{U}{(-1,1)}$ (eccentric) & $\dots$ & Parametrization for $e$ and $\omega$\\
~~~$S_{2,[c]} = \sqrt{e_{[c]}}\cos \omega_{[c]}$  & $\mathcal{F}{(0.0)}$ (circular) & $\dots$ & Parametrization for $e$ and $\omega$\\
 &  $\mathcal{U}{(-1,1)}$ (eccentric) & $\dots$ & Parametrization for $e$ and $\omega$\\
\noalign{\smallskip}
\multicolumn{4}{c}{\textit{RV instrumental parameters}} \\
\noalign{\smallskip}
~~~$\gamma_{\textnormal{CARMENES-VIS}}$  & $\mathcal{U}(-10.0,10.0) $ & \ms & Systemic velocity for CARMENES-VIS\\
~~~$\sigma_{\textnormal{CARMENES-VIS}}$  & $\mathcal{J}(0.01,100.0) $ & \ms & Extra jitter term for CARMENES-VIS\\
\noalign{\smallskip}
\multicolumn{4}{c}{\textit{dSHO-GP parameters}} \\
\noalign{\smallskip}
~~~$\sigma_\textnormal{GP,\ CARMENES-VIS}$   & $\mathcal{U}(0.0,10.0)$ & \ms & Amplitude of the dSHO-GP\\
~~~$Q_{0\ \textnormal{GP,\ CARMENES-VIS}}$  & $\mathcal{J}(1.0,100000.0)$ & $\dots$ & Quality factor for the secondary oscillation of the dSHO-GP\\
~~~$f_\textnormal{GP,\ CARMENES-VIS}$  & $\mathcal{U}(0.1,1.0)$ & $\dots$ & Fractional amplitude of the secondary oscillation of the dSHO-GP\\
% Fractional amplitude of the secondary oscillation to the primary of the dSHO-GP
~~~$\delta Q_\textnormal{GP,\ CARMENES-VIS}$  & $\mathcal{J}(1.0,100000.0)$ & $\dots$ & Quality factor difference between the first \\
& & & and second oscillations of the dSHO-GP\\
% Difference between the quality factors of the first and second oscillations of the dSHO-GP
~~~$P_\textnormal{rot,\ GP,\ CARMENES-VIS}$  & $\mathcal{U}(15.0,25.0)$ & d & Primary period of the dSHO-GP\\
  & $\mathcal{U}(10.0,150.0)$ (wide) & d & Primary period of the dSHO-GP\\
\noalign{\smallskip}
\multicolumn{4}{c}{\textit{QP-GP parameters}} \\
\noalign{\smallskip}
~~~$\sigma_\textnormal{GP,\ CARMENES-VIS}$  & $\mathcal{U}(0.0,10.0)$ & \ms & Amplitude of the QP-GP\\
~~~$\Gamma_\textnormal{GP,\ CARMENES-VIS}$  & $\mathcal{J}(0.01,10.0)$ & $\dots$ & Amplitude of the sine-squared component of the QP-GP\\
~~~$\alpha_\textnormal{GP,\ CARMENES-VIS}$  & $\mathcal{J}(10^{-8},0.001)$ & d$^{-2}$ & Inverse length-scale of the exponential component of the QP-GP\\
~~~$P_\textnormal{rot,\ GP,\ CARMENES-VIS}$  & $\mathcal{U}(15.0,25.0)$ & d & Rotational period of the quasi-periodic component of the QP-GP\\
  & $\mathcal{U}(10.0,150.0)$ (wide) & d & Rotational period of the quasi-periodic component of the QP-GP\\
\noalign{\smallskip}
\hline
\end{tabular}
\end{table*}

%%%% JOINT FIT PRIORS
\begin{table*}[!h]
\small
\centering
\caption[Priors for the final joint transit and RV fit for TOI-1201]{Priors for the final joint transit and RV fit for TOI-1201 with \juliet\ in Sect.~\ref{sec:jointfit}.}
\label{tab:priors_joint}
\begin{tabular}{lccr}
\hline
\hline
\noalign{\smallskip}
Parameter & Prior & Unit & Description \\
\noalign{\smallskip}
\hline
\noalign{\smallskip}
\multicolumn{4}{c}{\textit{Stellar parameters}} \\ 
\noalign{\smallskip}
~~~$\rho_{*}$  & $\mathcal{N}(5.40,.60) $ & g\,cm$^{-3}$ & Stellar density\\
\noalign{\smallskip}
\multicolumn{4}{c}{\textit{Parameters for transiting planet b}} \\ 
\noalign{\smallskip}
~~~$P_{b}$  & $\mathcal{N}(2.4919858227,0.0001) $ & d & Period\\
~~~$t_{0,b}$ (BJD)  & $\mathcal{N}(2459169.232186481,0.05) $ & d & Time of transit center.\\
~~~$r_{1,b}$  & $\mathcal{U}(0.0,1.0) $ & $\dots$ & Parametrization for p and b\\
~~~$r_{2,b}$  & $\mathcal{U}(0.0,1.0) $ & $\dots$ & Parametrization for p and b\\
~~~$K_{b}$  & $\mathcal{U}(0.0,20) $ & \ms & RV semi-amplitude\\
~~~$S_{1,b} = \sqrt{e_b}\sin \omega_b$  & 0.0 (fixed) & $\dots$ & Parametrization for $e$ and $\omega$\\
~~~$S_{2,b} = \sqrt{e_b}\cos \omega_b$  & 0.0 (fixed) & $\dots$ & Parametrization for $e$ and $\omega$\\
\noalign{\smallskip}
\multicolumn{4}{c}{\textit{Parameters for long-term RV signal}} \\ 
\noalign{\smallskip}
~~~$P_{[c]}$  & $\mathcal{U}(60.0,150.0) $ & d & Period\\
~~~$t_{0,[c]}$  & $\mathcal{U}(2458730.0,2458840.0) $ & d & Time of transit center\\
~~~$K_{[c]}$  & $\mathcal{U}(0.0,20.0) $ & \ms & RV semi-amplitude\\
~~~$S_{1,[c]} = \sqrt{e_{[c]}}\sin \omega_{[c]}$  & 0.0 (fixed) & $\dots$ & Parametrization for $e$ and $\omega$\\
~~~$S_{2,[c]} = \sqrt{e_{[c]}}\cos \omega_{[c]}$  & 0.0 (fixed) & $\dots$ & Parametrization for $e$ and $\omega$\\
\noalign{\smallskip}
\multicolumn{4}{c}{\textit{RV instrumental parameters}} \\
\noalign{\smallskip}
~~~$\gamma_{\textnormal{CARMENES-VIS}}$  & $\mathcal{U}(-10.0,10.0) $ & \ms & Systemic velocity for CARMENES-VIS\\
~~~$\sigma_{\textnormal{CARMENES-VIS}}$  & $\mathcal{J}(0.01,100.0) $ & \ms & Extra jitter term for CARMENES-VIS\\[0.1 cm]
~~~$\sigma_\textnormal{GP,\ CARMENES-VIS}$  & $\mathcal{U}(0.0,10.0)$ & \ms & Amplitude of the dSHO-GP\\
~~~$Q_{0\ \textnormal{GP,\ CARMENES-VIS}}$  & $\mathcal{J}(1.0,100000.0)$ & $\dots$ & Quality factor for the secondary oscillation of the dSHO-GP\\
~~~$f_\textnormal{GP,\ CARMENES-VIS}$  & $\mathcal{U}(0.1,1.0)$ & $\dots$ & Fractional amplitude of the secondary oscillation of the dSHO-GP\\
% Fractional amplitude of the secondary oscillation to the primary of the dSHO-GP
~~~$\delta Q_\textnormal{GP,\ CARMENES-VIS}$  & $\mathcal{J}(1.0,100000.0)$ & $\dots$ & Quality factor difference between the first\\
& &  & and second oscillations of the dSHO-GP\\
% Difference between the quality factors of the first and second oscillations of the dSHO-GP
~~~$P_\textnormal{rot,\ GP,\ CARMENES-VIS}$  & $\mathcal{U}(15.0,25.0)$ & d & Primary period of the dSHO-GP\\
\noalign{\smallskip}
\multicolumn{4}{c}{\textit{Photometry instrumental parameters}} \\
\noalign{\smallskip}
~~~$D_{\textnormal{TESS4}}$  & 1.0 (fixed)  & $\dots$ & Dilution factor for TESS4\\
~~~$M_{\textnormal{TESS4}}$  & $\mathcal{U}(-0.1,0.1)$ & $10^6$ ppm & Relative flux offset for TESS4\\
~~~$\sigma_{\textnormal{TESS4}}$  & $\mathcal{J}(10^{-7},100000.0)$ & ppm & Extra jitter term for TESS4\\[0.1 cm]
~~~$\sigma_\textnormal{GP,\ TESS4}$  & $\mathcal{J}(10^{-5},100.0)$ & ppm & Amplitude of the Matern-3/2-GP for TESS4\\
~~~$\rho_\textnormal{GP,\ TESS4}$  & $\mathcal{J}(0.001,100.0)$ & d & Length scale of the Matern-3/2-GP for TESS4\\
\noalign{\medskip}
~~~$D_{\textnormal{TESS31}}$  & 1.0 (fixed)  & $\dots$ & Dilution factor for TESS31\\
~~~$M_{\textnormal{TESS31}}$  & $\mathcal{U}(-0.1,0.1)$ & $10^6$ ppm & Relative flux offset for TESS31\\
~~~$\sigma_{\textnormal{TESS31}}$  & $\mathcal{J}(10^{-7},100000.0)$ & ppm & Extra jitter term for TESS31\\[0.1 cm]
~~~$\sigma_\textnormal{GP,\ TESS31}$  & $\mathcal{J}(10^{-8},0.0001)$ & ppm & Amplitude of the exp-GP for TESS31\\
~~~$T_{\textnormal{GP},\textnormal{TESS31}}$  & $\mathcal{J}(10^{-5},100.0)$ & d & Characteristic timescale of the exp-GP for TESS31\\
\noalign{\medskip}
~~~$q_{1,\textnormal{TESS4+TESS31}}$  & $\mathcal{U}(0.0,1.0)$ & $\dots$ & Quadratic limb-darkening parametrization for TESS4 and TESS31\\
~~~$q_{2,\textnormal{TESS4+TESS31}}$  & $\mathcal{U}(0.0,1.0)$ & $\dots$ & Quadratic limb-darkening parametrization for TESS4 and TESS31\\
\noalign{\medskip}
~~~$D_{\textnormal{OAA}}$  & 1.0 (fixed) & $\dots$ & Dilution factor for OAA\\
~~~$M_{\textnormal{OAA}}$  & $\mathcal{U}(-0.1,0.1) $ & $10^6$ ppm & Relative flux offset for OAA\\
~~~$\sigma_{\textnormal{OAA}}$  & $\mathcal{J}(10^{-7},100000.0) $ & ppm & Extra jitter term for OAA\\
~~~$q_{1,\textnormal{OAA}}$  & $\mathcal{U}(0.0,1.0) $ & $\dots$ & Linear limb-darkening parametrization for OAA\\
\noalign{\medskip}
~~~$D_{\textnormal{LCO-SSO}}$  & 1.0 (fixed) & $\dots$ & Dilution factor for LCO-SSO\\
~~~$M_{\textnormal{LCO-SSO}}$  & $\mathcal{U}(-0.1,0.1) $ & $10^6$ ppm & Relative flux offset for LCO-SSO\\
~~~$\sigma_{\textnormal{LCO-SSO}}$  & $\mathcal{J}(10^{-7},100000.0) $ & ppm & Extra jitter term for LCO-SSO\\
~~~$q_{1,\textnormal{LCO-SSO}}$  & $\mathcal{U}(0.0,1.0) $ & $\dots$ & Linear limb-darkening parametrization for LCO-SSO\\
~~~$\theta_{0,\textnormal{LCO-SSO}}$  & $\mathcal{U}(-0.1,0.1) $ &  & Linear term applied to the airmass for LCO-SSO\\
\noalign{\medskip}
~~~$D_{\textnormal{LCO-SAAO}}$  & 1.0 (fixed) & $\dots$ & Dilution factor for LCO-SAAO\\
~~~$M_{\textnormal{LCO-SAAO}}$  & $\mathcal{U}(-0.1,0.1) $ & $10^6$ ppm & Relative flux offset for LCO-SAAO\\
~~~$\sigma_{\textnormal{LCO-SAAO}}$  & $\mathcal{J}(10^{-7},100000.0) $ & ppm & Extra jitter term for LCO-SAAO\\
~~~$q_{1,\textnormal{LCO-SAAO}}$  & $\mathcal{U}(0.0,1.0) $ & $\dots$ & Linear limb-darkening parametrization for LCO-SAAO\\
~~~$\theta_{0,\textnormal{LCO-SAAO}}$  & $\mathcal{U}(-0.1,0.1) $ &  & Linear term applied to the airmass for LCO-SAAO\\
\noalign{\smallskip}
\hline
\end{tabular}
\end{table*}

%%%% ALL POSTERIORS
\begin{table}[!h]
\centering
\caption[Full set of posterior parameters used in the model for TOI-1201]{Full set of posterior parameters used in the model for TOI-1201 and described in Sect.~\ref{sec:jointfit}}
\label{tab:posteriors_full}
\begin{tabular}{lc}
\hline
\hline
\noalign{\smallskip}
Parameter & Posterior  \\
\noalign{\smallskip}
\hline
\noalign{\smallskip}
\multicolumn{2}{c}{\textit{Stellar parameters}}\\
\noalign{\smallskip}
~~~$\rho_{*}$ (g\,cm$^{-3}$) & $5.57^{+.51}_{-.48}$\\[0.1 cm]
\noalign{\smallskip}
\multicolumn{2}{c}{\textit{Posterior parameters for transiting planet b}} \\ 
\noalign{\smallskip}
~~~$P_{b}$ (d) & $2.4919863^{+0.0000030}_{-0.0000031}$\\[0.1 cm]
~~~$t_{0,b}$ (d) & $2459169.23222^{+0.00052}_{-0.00054}$\\[0.1 cm]
~~~$r_{1,b}$  & $0.603^{+0.048}_{-0.055}$\\[0.1 cm]
~~~$r_{2,b}$  & $0.04383^{+0.00096}_{-0.00110}$\\[0.1 cm]
~~~$K_{b}$ (\ms) & $4.65^{+0.60}_{-0.64}$\\[0.1 cm]
\noalign{\smallskip}
\multicolumn{2}{c}{\textit{Parameters for long-term signal in RVs}} \\
\noalign{\smallskip}
~~~$P_{[c]}$ (d)  & $102^{+21}_{-15}$\\[0.1 cm]
~~~$t_{0,[c]}$ (d)  & $2458772^{+12}_{-16}$\\[0.1 cm]
~~~$K_{[c]}$ (\ms) & $5.84^{+0.91}_{-0.87}$\\[0.1 cm]
\noalign{\smallskip}
\multicolumn{2}{c}{\textit{RV instrumental parameters}}\\
\noalign{\smallskip}
~~~$\gamma_{\textnormal{CARMENES-VIS}}$ (m\,s$^{-1}$)  & $-2.91^{+0.74}_{-0.97}$\\[0.1 cm]
~~~$\sigma_{\textnormal{CARMENES-VIS}}$ (m\,s$^{-1}$)  & $0.36^{+0.78}_{-0.29}$\\[0.1 cm]
~~~$\sigma_\textnormal{GP,\ CARMENES-VIS}$ (m\,s$^{-1}$) & $3.7^{+1.7}_{-1.1}$\\[0.1 cm]
~~~$Q_{0\ \textnormal{GP,\ CARMENES-VIS}}$ & $5.6^{+41.0}_{-3.8}$\\[0.1 cm]
~~~$f_\textnormal{GP,\ CARMENES-VIS}$  & $0.57^{+0.25}_{-0.28}$\\[0.1 cm]
~~~ $\delta Q_\textnormal{GP,\ CARMENES-VIS}$ & $179^{+5600}_{-170}$\\[0.1 cm]
~~~$P_\textnormal{rot,\ GP,\ CARMENES-VIS}$ (d) & $19.62^{+1.10}_{-0.81}$\\[0.1 cm]
\noalign{\smallskip}
\multicolumn{2}{c}{\textit{Photometry instrumental parameters}}\\
\noalign{\smallskip}
~~~$M_{\textnormal{TESS4}}$ ($10^6$ ppm)  & $-0.00003^{+0.00020}_{-0.00021}$\\[0.1 cm]
~~~$\sigma_{\textnormal{TESS4}}$ (ppm)  & $810^{+44}_{-45}$\\[0.1 cm]
\noalign{\medskip}
~~~$\sigma_\textnormal{GP,\ TESS4}$ (ppm)  & $0.001455^{+0.000110}_{-0.000098}$\\[0.1 cm]
~~~$\rho_\textnormal{GP,\ TESS4}$ (d) & $0.255^{+0.027}_{-0.022}$\\[0.1 cm]
\noalign{\medskip}
~~~$M_{\textnormal{TESS31}}$ ($10^6$ ppm)  & $-0.0003^{+0.0011}_{-0.0018}$\\[0.1 cm]
~~~$\sigma_{\textnormal{TESS31}}$ (ppm)  & $0.0075^{+3.7000}_{-0.0074}$\\[0.1 cm]
\noalign{\medskip}
~~~$\sigma_\textnormal{GP,\ TESS31}$ (ppm)  & $0.0000023^{+0.0000170}_{-0.0000019}$\\[0.1 cm]
~~~$T_\textnormal{GP,\ TESS31}$  (d) & $0.027^{+0.130}_{-0.024}$\\[0.1 cm]
\noalign{\medskip}
~~~$q_{1,\textnormal{TESS4+TESS31}}$  & $0.47^{+0.25}_{-0.22}$\\[0.1 cm]
~~~$q_{2,\textnormal{TESS4+TESS31}}$  & $0.33^{+0.28}_{-0.20}$\\[0.1 cm]
\noalign{\medskip}
~~~$M_{\textnormal{OAA}}$ ($10^6$ ppm)  & $-0.00002^{+0.00019}_{-0.00019}$\\[0.1 cm]
~~~$\sigma_{\textnormal{OAA}}$ (ppm)  & $0.072^{+32.000}_{-0.072}$\\[0.1 cm]
~~~$q_{1,\textnormal{OAA}}$  & $0.46^{+0.30}_{-0.27}$\\[0.1 cm]
\noalign{\medskip}
~~~$M_{\textnormal{LCO-SSO}}$ ($10^6$ ppm)  & $0.03181^{+0.00041}_{-0.00041}$\\[0.1 cm]
~~~$\sigma_{\textnormal{LCO-SSO}}$ (ppm)  & $926^{+140}_{-150}$\\[0.1 cm]
~~~$q_{1,\textnormal{LCO-SSO}}$  & $0.69^{+0.18}_{-0.23}$\\[0.1 cm]
~~~$\theta_{0,\textnormal{LCO-SSO}}$  & $0.0007^{+0.0003}_{-0.0003}$\\[0.1 cm]
\noalign{\medskip}
~~~$M_{\textnormal{LCO-SAAO}}$ ($10^6$ ppm)  & $0.0383^{+0.0030}_{-0.0031}$\\[0.1 cm]
~~~$\sigma_{\textnormal{LCO-SAAO}}$ (ppm)  & $112^{+860}_{-110}$\\[0.1 cm]
~~~$q_{1,\textnormal{LCO-SAAO}}$  & $0.47^{+0.24}_{-0.24}$\\[0.1 cm]
~~~$\theta_{0,\textnormal{LCO-SAAO}}$  & $0.0333^{+0.0026}_{-0.0026}$\\[0.1 cm]
\noalign{\smallskip}
\hline
\end{tabular}
\end{table}

\begin{figure*}[!h]
    \centering
    \includegraphics[width=1\linewidth]{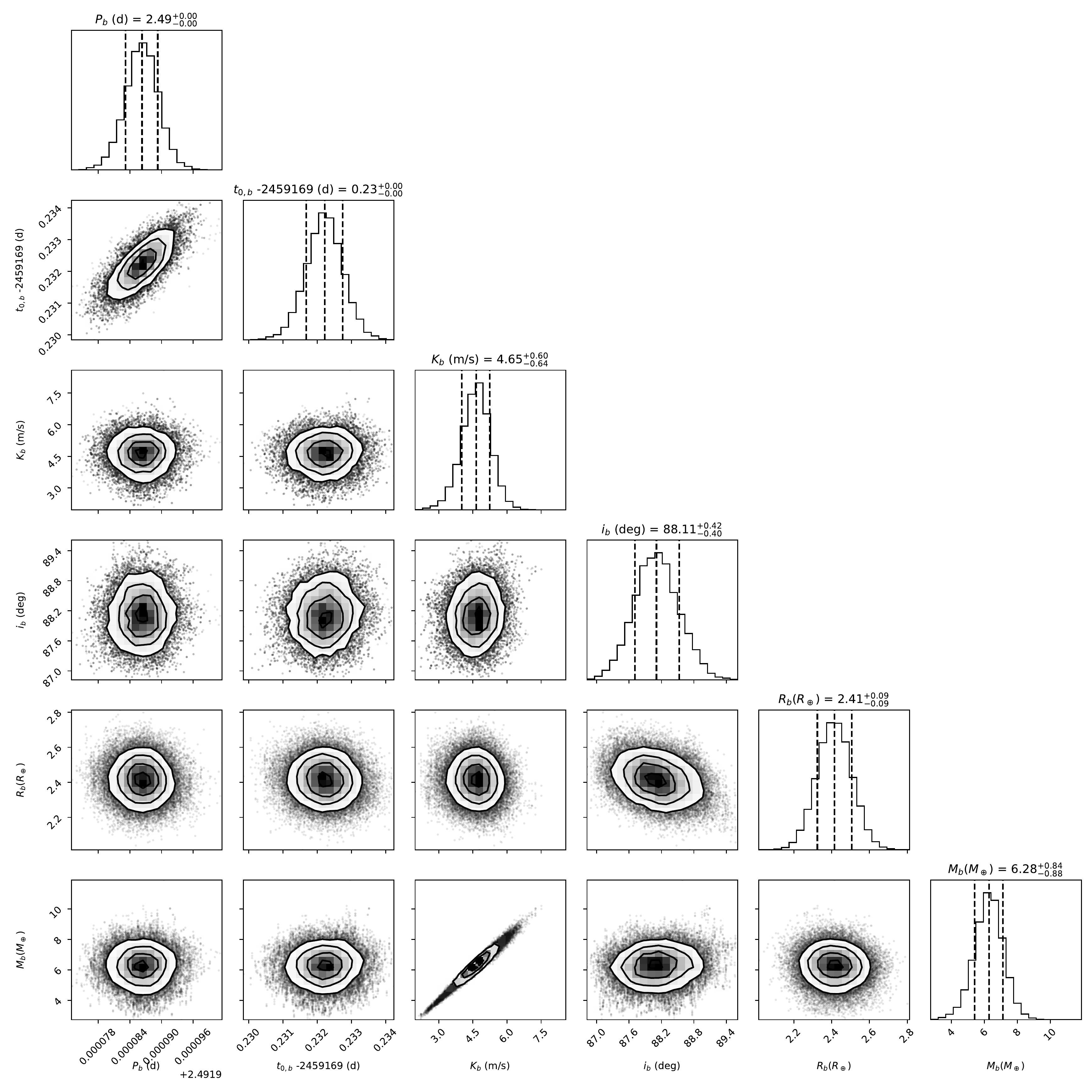}
    \caption[Posterior distributions for the transiting planet TOI-1201~b]{Posterior distributions for the transiting planet TOI-1201~b from the joint fit described in Sect.~\ref{sec:jointfit}. 
    }
    \label{fig:cornerplot_p1}
\end{figure*}

\begin{figure*}[!h]
    \centering
    \includegraphics[width=1\linewidth]{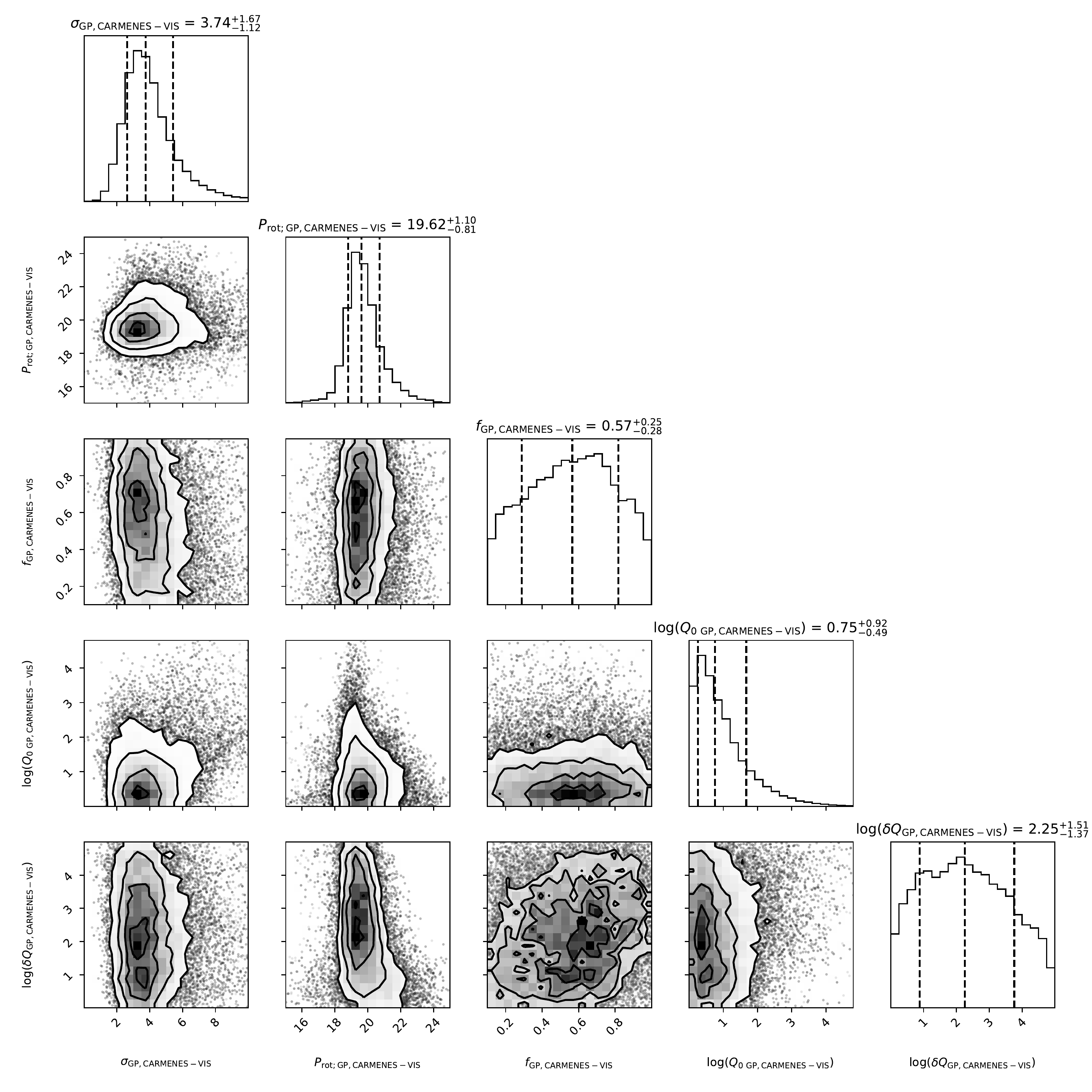}
    \caption[Posterior distributions for the dSHO-GP on the CARMENES RVs]{Posterior distributions for the dSHO-GP on the CARMENES RVs from the joint fit described in Sect.~\ref{sec:jointfit}.}
    \label{fig:cornerplot_gp}
\end{figure*}

\begin{figure*}[!h]
    \centering
    \includegraphics[width=1\linewidth]{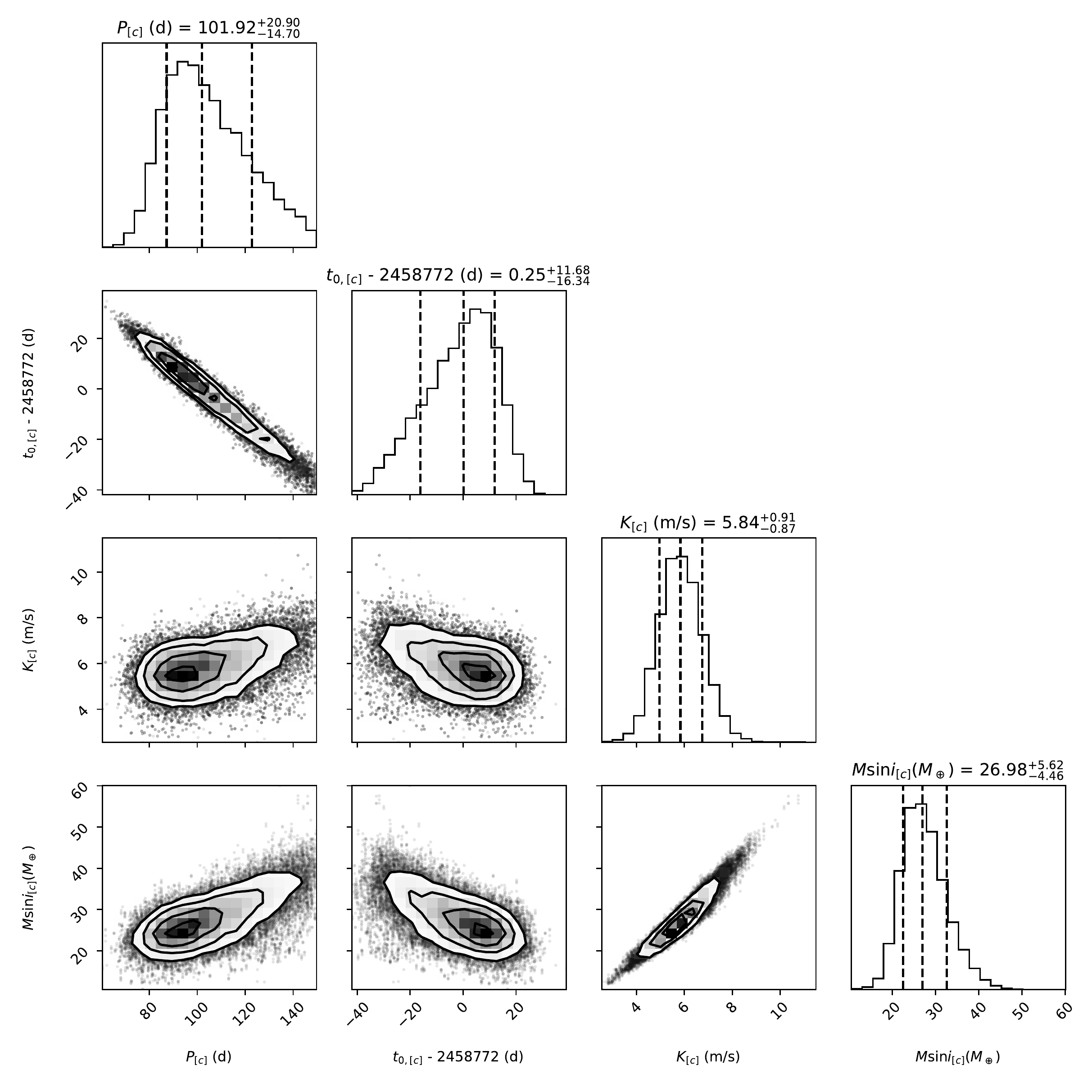}
    \caption[Posterior distributions for the long-term signal]{Posterior distributions for the long-term signal from the joint fit described in Sect.~\ref{sec:jointfit}.}
    \label{fig:cornerplot_p2}
\end{figure*}

\clearpage
\section{Two-planet model} \label{appendix:twoplanetmodel}

As discussed in Sect.~\ref{sec:rvonlymodeling}, a $\sim$102\,d signal is the strongest one found in the RVs. It is favored in the model when included as an additional sinusoidal component.
However, a GP could also model it well (see Appendix~\ref{appendix:gpwide}). If considered to be a planet, it would then have a minimum mass of $27.0^{+5.6}_{-4.5}$\,\mearth\ (Table~\ref{tab:posteriors_planet_w_p2}, Fig.~\ref{fig:cornerplot_p2}).
Thus far, we cannot immediately rule out a possible planet candidate.
Nevertheless, given that the orbital period is on the order of the time baseline, we do not have sufficient evidence to unequivocally prove the origin of the signal to be planetary. 

From a dynamical point of view, it would be an unproblematic configuration, since both planets maintain a sufficient spatial separation to preserve long-term stability.
In addition, the longer-period planet candidate, whose expected bulk mass would require a significant gaseous envelope, is far enough from the star to avoid significant atmospheric losses due to photoevaporation~\citep{Owen2012a}.
A Neptunian-mass planet at this orbit is, thus, plausible.
In fact, core accretion theory, which is considered the most promising path of planet formation, routinely predicts planets like those found in TOI-1201.
Global formation models of multi-planetary systems~\citep[e.g.,][]{Emsenhuber2020} find the tentative two-planet configuration to be a rather typical outcome~\citep[e.g.,][their Fig.~D.3]{Schlecker2020}.
New planetary population syntheses specifically tailored to M-dwarf systems also propose the regular formation of planets in the mass range of TOI-1201~b and the tentative TOI-1201~[c] \citep{Burn2021}.
Therefore, we believe that the latter is a promising candidate planetary companion. Further observations, especially with longer ground-based coverage, will be necessary to unequivocally determine the nature of the signal.

\begin{table}[!h]
\small
\centering
\caption[Same derived posterior table as Table~\ref{tab:posteriors_planet} but including the second signal]{Same derived posterior table as Table~\ref{tab:posteriors_planet} but assuming the second signal to be of planetary origin.}
\label{tab:posteriors_planet_w_p2}
\begin{tabular}{lccl}
\hline
\hline
\noalign{\smallskip}
Parameter & \multicolumn{2}{c}{Posterior\tablefootmark{(a)}} & Unit \\
 & b & [c]\tablefootmark{(b)} & \\
\noalign{\smallskip}
\hline
\noalign{\smallskip}
$P_p$ & $2.4919863^{+0.0000030}_{-0.0000031}$ & $102^{+21}_{-15}$ & d\\[0.1 cm]
$t_{0,p}$ (BJD) & $2459169.23222^{+0.00052}_{-0.00054}$ & $2458772^{+12}_{-16}$ & d\\[0.1 cm]
$r_{1,p}$ & $0.603^{+0.048}_{-0.055}$ & $\dots$ & $\dots$\\[0.1 cm]
$r_{2,p}$ & $0.04383^{+0.00096}_{-0.00110}$ & $\dots$ & $\dots$\\[0.1 cm]
$K_p$ & $4.65^{+0.60}_{-0.64}$ & $5.84^{+0.91}_{-0.87}$ & \ms\\[0.1 cm]
$e_p$ & $0.0$ (fixed) & $0.0$ (fixed) & $\dots$\\[0.1 cm]
$\omega_p$ & $90.0$ (fixed) & $90.0$ (fixed) & $\dots$\\[0.1 cm]
$p = R_{p}/R_\star$ & $0.04383^{+0.00096}_{-0.00110}$ & $\dots$ & $\dots$\\[0.1 cm]
$b = (a_{p}/R_\star) \cos{i_{p}}$ & $0.404^{+0.071}_{-0.082}$ & $\dots$ & $\dots$\\[0.1 cm]
$a_{p}/R_\star$ & $12.23^{+0.36}_{-0.36}$ & $\dots$ & $\dots$\\[0.1 cm]
\noalign{\smallskip}\noalign{\smallskip}
\hline
\noalign{\smallskip}\noalign{\smallskip}
$i_p$ & $88.11^{+0.42}_{-0.40}$ & $\dots$ & deg\\[0.1 cm]
$t_\textnormal{T}$\tablefootmark{(c)} & $1.747^{+0.096}_{-0.091}$ & $\dots$ & h\\[0.1 cm]
$M_p$ & $6.28^{+0.84}_{-0.88}$ & $>27.0^{+5.6}_{-4.5}$ & $M_\oplus$\\[0.1 cm]
$R_p$ & $2.415^{+0.091}_{-0.090}$ & $\dots$ & $R_\oplus$\\[0.1 cm]
$\rho_p$ & $2.45^{+0.48}_{-0.42}$ & $\dots$ & g cm$^{-3}$\\[0.1 cm]
$g_p$ & $10.5^{+1.8}_{-1.6}$ & $\dots$ & m s$^{-2}$\\[0.1 cm]
$a_p$ & $0.0287^{+0.0012}_{-0.0012}$ & $0.341^{+0.046}_{-0.034}$ & \au \\[0.1 cm]
$T_\textnormal{eq}$\tablefootmark{(d)} & $703^{+15}_{-14}$ & $204^{+12}_{-13}$ & K\\[0.1 cm]
$S_p$\tablefootmark{(e)} & $40.6^{+3.6}_{-3.2}$ & $0.287^{+0.072}_{-0.065}$ & $S_\oplus$\\[0.1 cm]
\noalign{\smallskip}
\hline
\end{tabular}
\tablefoot{
\tablefoottext{a}{Parameters obtained with the posterior values from Table~\ref{tab:posteriors_full}. Error bars denote the 68\,\% posterior credibility intervals.}
\tablefoottext{b}{The square brackets denote the assumption that the long-term signal is of planetary origin.}
\tablefoottext{c}{Transit duration from first contact to fourth contact.}
\tablefoottext{d}{Equilibrium temperature calculated assuming zero Bond albedo.}
\tablefoottext{e}{Insolation.}
}
\end{table}

\clearpage
\section{Additional figures}

% \tess\ long term phot phase folded
\begin{figure}[!h]
    \centering
    \includegraphics[width=0.9\linewidth]{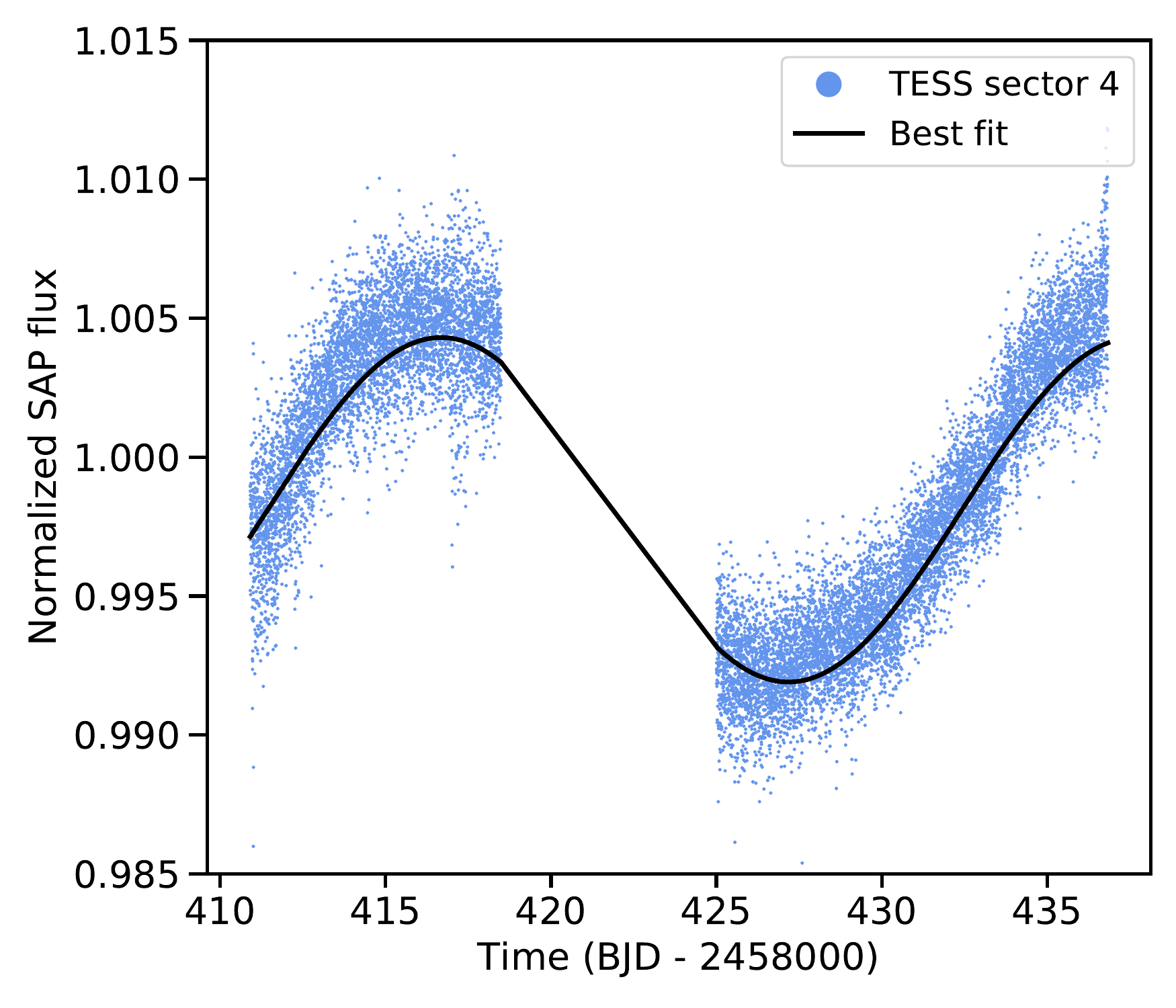}
    \caption[Phase-folded \tess\ light curve exhibiting the stellar rotation period for TOI-1201]{Phase-folded \tess\ \texttt{SAP} light curve from sector 4, exhibiting the stellar rotation period at 20.5\,d for TOI-1201. The \tess\ \texttt{SAP} light curve from sector 31 shows a different shape, probably attributed to instrumental effects or contamination by the companion.
    }
    \label{fig:longtermtess}
\end{figure}

% Boxplot for model comparison
\begin{figure*}[!h]
    \centering
    \includegraphics[width=1\linewidth]{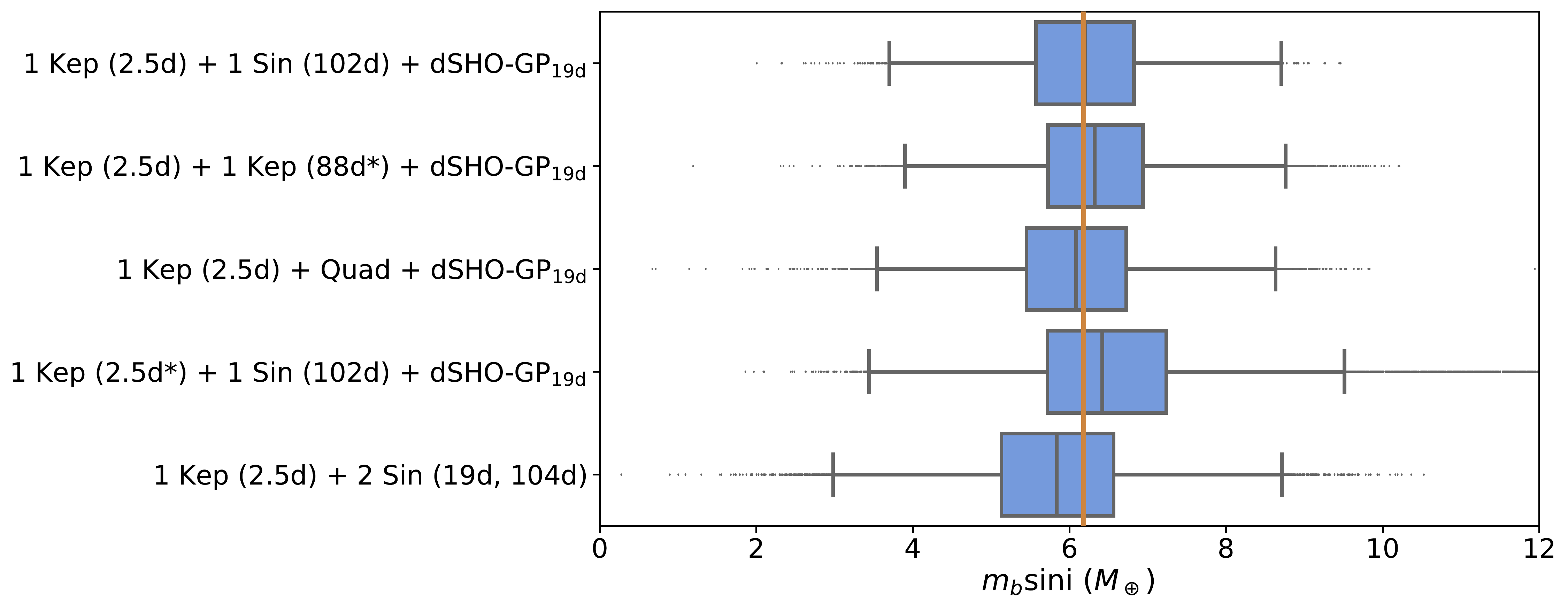}
    \caption[RV-only posterior distributions of the minimum mass of TOI-1201~b as a function of model choice]{Posteriors of the minimum mass of the 2.5\,d transiting planet around TOI-1201 depending on the model choice for the RV-only fits. The vertical orange line is the median value corresponding to the 1 Kep (2.5\,d) + 1 Sin (102\,d) + dSHO-GP$_\textnormal{19d}$ model (top). The 25\,\% and 75\,\% quartiles are represented as the blue box, the extending black lines show the rest of the distribution, and the dots are considered ``outliers.'' The model names correspond to those in Table~\ref{tab:modelcomparison}.}
    \label{fig:boxplot_rvmodels}
\end{figure*}

% GLS periodograms of the RVs and activity indicators for the companion 
\begin{figure*}[!h]
    \centering
    \includegraphics[width=1\linewidth]{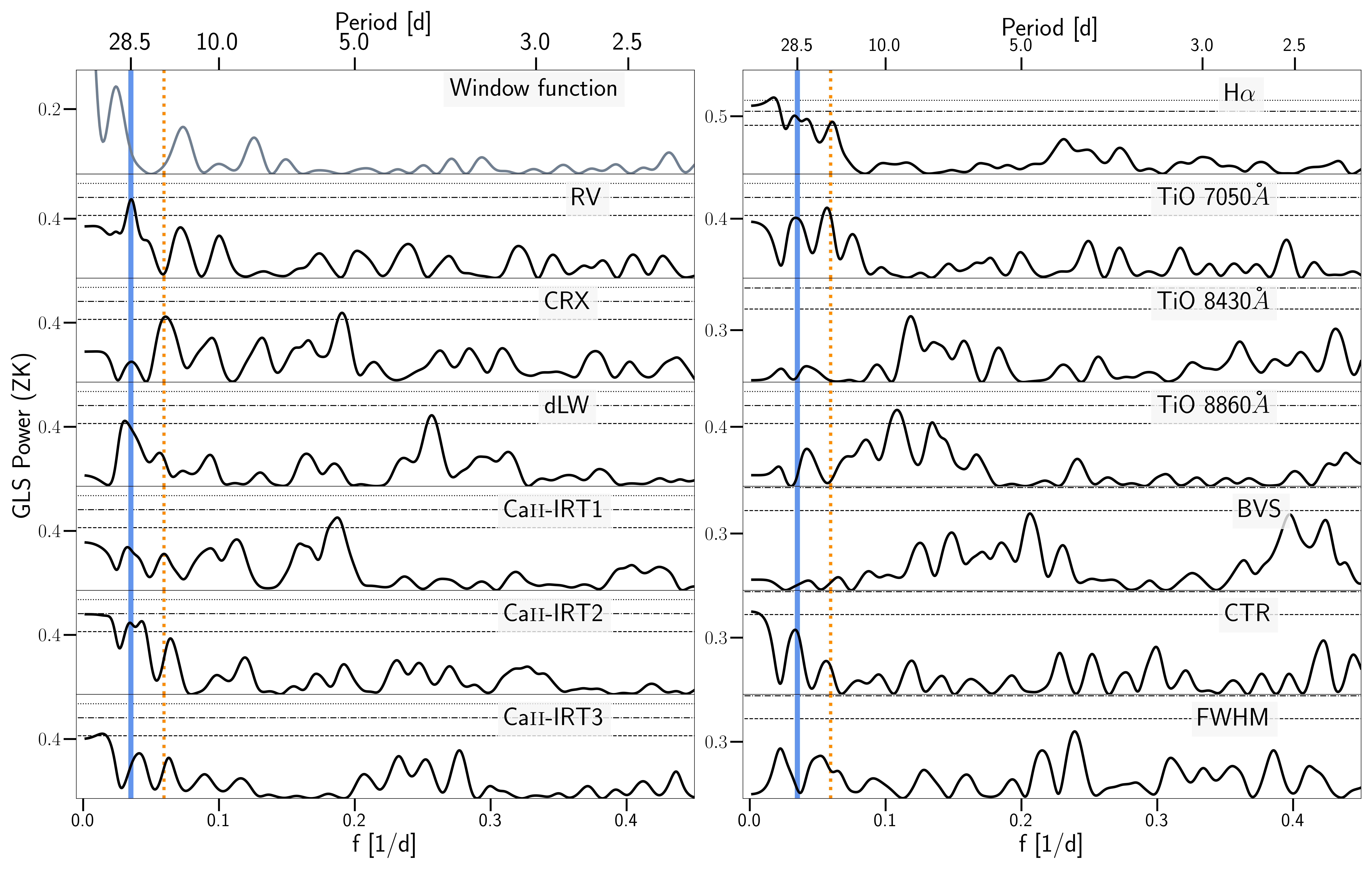}
    \caption[GLS periodograms of the RVs and various stellar activity indicators from the CARMENES spectroscopic data for the companion of TOI-1201]{GLS periodograms of the RVs and various stellar activity indicators from the CARMENES spectroscopic data for PM~J02489--1432E. The vertical solid blue line corresponds to the main signal found in the RVs (28.5\,d). The vertical dotted orange line corresponds to the 41\,d alias ($\sim$16.8\,d). The horizontal dotted, dot-dashed, and dashed lines represent the 10\,\%, 1\,\%, and 0.1\,\% FAP levels, respectively. 
    }
    \label{fig:compgls}
\end{figure*}

% RV timeseries for the companion of TOI-1201
\begin{figure*}[!h]
\centering
\begin{minipage}{0.92\textwidth}
  \centering
  \includegraphics[width=1\linewidth]{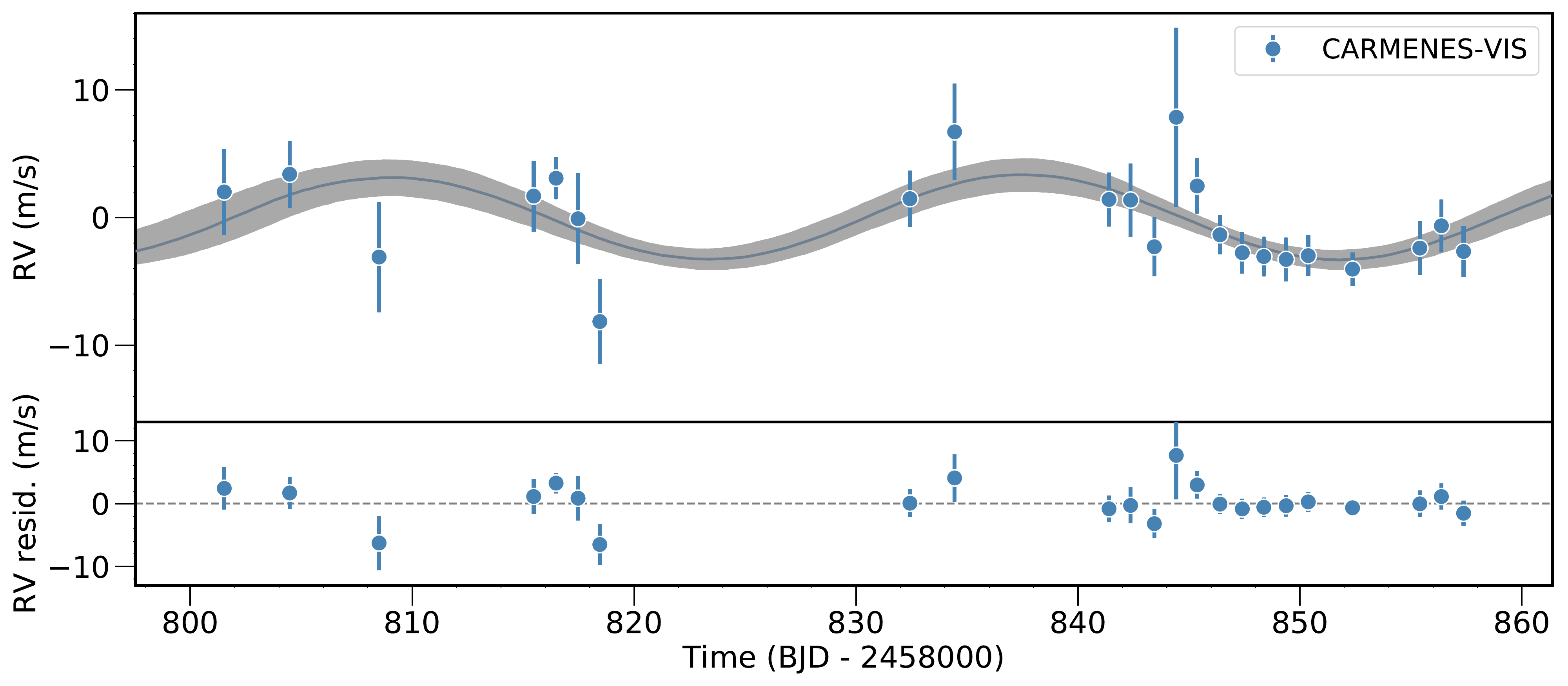}
\end{minipage}
\begin{minipage}{.45\textwidth}
  \centering
  \includegraphics[width=1\linewidth]{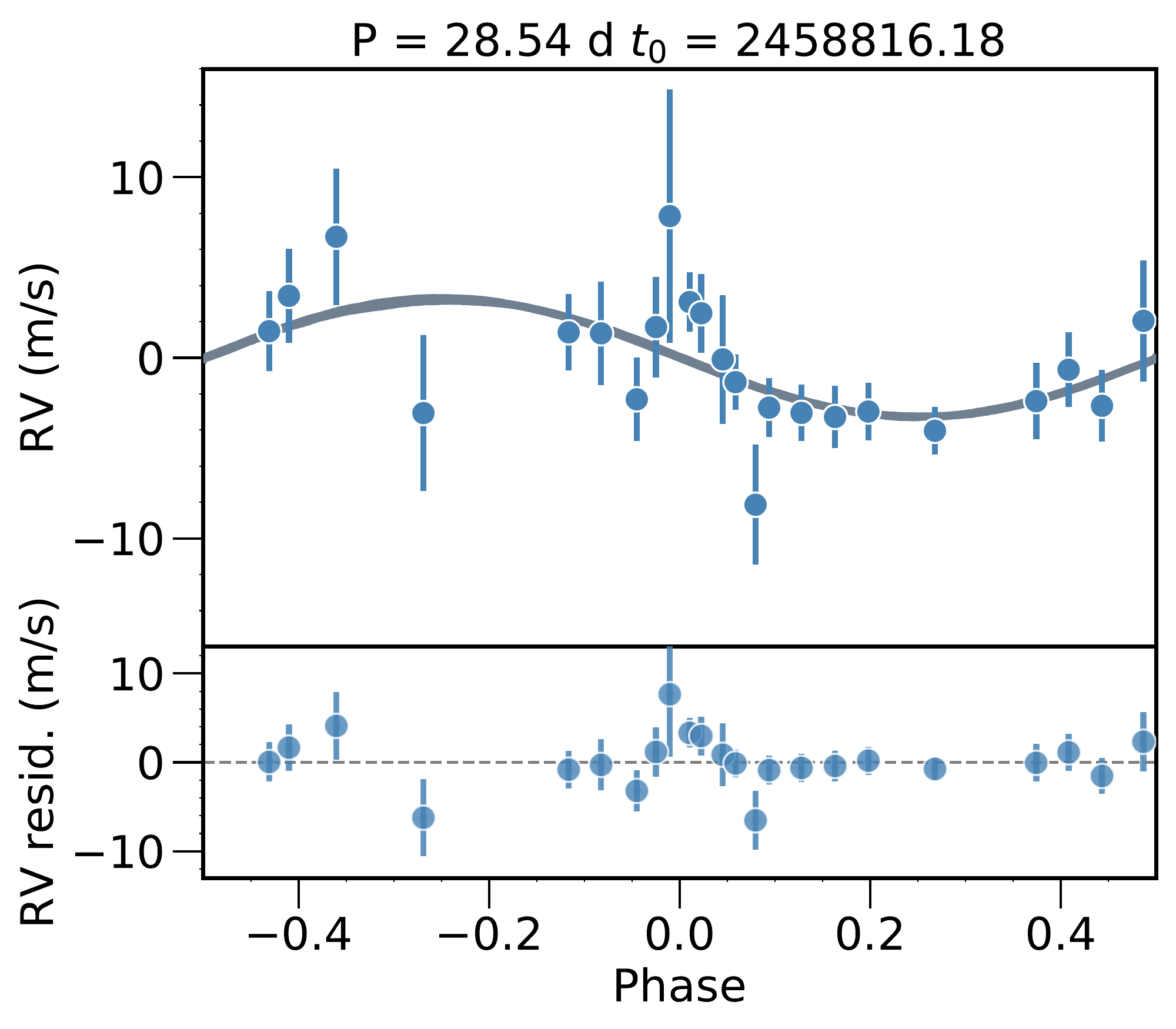}
\end{minipage}
\caption[RV time series and phase-folded RVs for PM~J02489--1432E]{CARMENES RV data for PM~J02489--1432E (TOI-393) with the best-fit model from the RV-only fit overplotted (dark gray), described in Sect.~\ref{sec:companionrvs}. \textit{Top:} RVs time series. The light gray band represents the 68\,\% credibility interval. \textit{Bottom:} RVs phase-folded to the period of the significant signal at 28.5\,d. The bottom panel of each plot shows the residuals after the model is subtracted.}
\label{fig:rvs_comp}
\end{figure*}

% Figure of the detection limits
\begin{figure}[!h]
\centering
\includegraphics[width=1\linewidth]{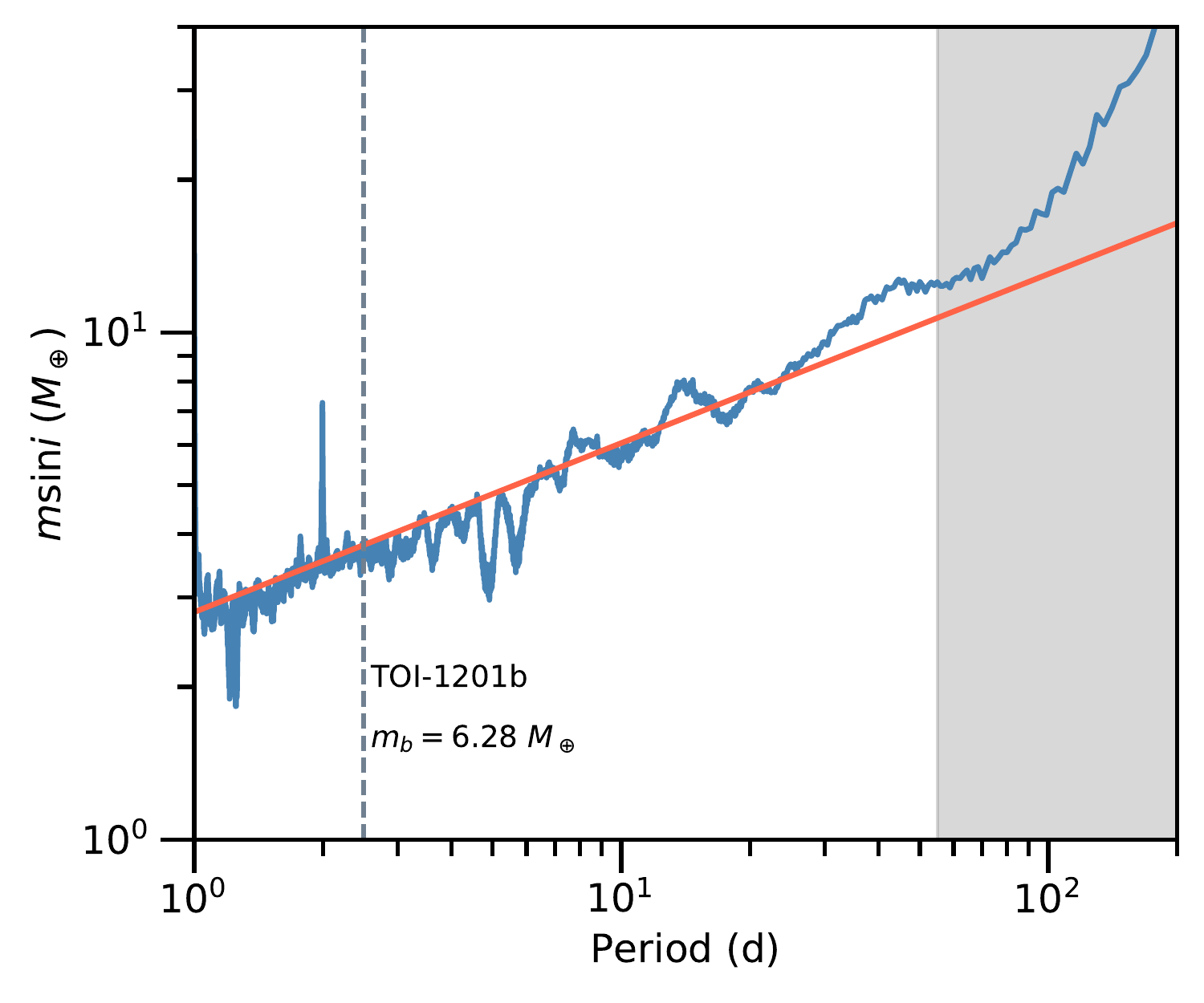}
\caption[]{Detection limits on $m\sin i$ following \cite{Bonfils2013} for the RV residuals of PM~J02489--1432E after subtracting out the most prominent 28.5\,d signal (Sect.~\ref{sec:companionrvs}). Any planet above the blue line is excluded with 99.9\% confidence. The orange line is the mass corresponding to a semi-amplitude of 3\,\ms. TOI-1201\,b is indicated by a dashed gray line. Planets with long-period values greater than the time baseline are marked with a gray shaded region.}
\label{fig:detectionlimits}
\end{figure}

\clearpage
\section{Data tables} \label{sec:appendixdata}

%% TOI-1201 RV DATA
\begin{table}[!h]
\centering
\caption{CARMENES VIS RV data of TOI-1201.}
\label{tab:rvdata}
\begin{tabular}{c
                S[table-format=2]
                S[table-format=2]
                }
\hline
\hline
\noalign{\smallskip}
BJD (TDB\tablefootmark{*}) & \multicolumn{1}{c}{RV (m\,s$^{-1}$)} & \multicolumn{1}{c}{$\sigma_\textnormal{RV}$ (m\,s$^{-1}$)}  \\
\noalign{\smallskip}
\hline
\noalign{\smallskip}
2458796.50268 & -11.12 & 2.86 \\[0.1 cm]
2458801.48909 & -10.09 & 3.92 \\[0.1 cm]
2458804.45542 & 0.33 & 3.20 \\[0.1 cm]
2458808.52929 & -1.10 & 4.24 \\[0.1 cm]
2458814.47394 & -4.51 & 3.76 \\[0.1 cm]
2458815.44971 & -13.25 & 3.33 \\[0.1 cm]
2458816.45605 & -11.81 & 2.17 \\[0.1 cm]
2458817.43864 & -6.04 & 1.91 \\[0.1 cm]
2458818.43170 & -10.98 & 3.23 \\[0.1 cm]
2458829.47849 & -3.24 & 3.02 \\[0.1 cm]
2458832.40663 & 5.73 & 2.22 \\[0.1 cm]
2458834.41799 & 6.48 & 3.89 \\[0.1 cm]
2458841.37576 & 3.47 & 1.97 \\[0.1 cm]
2458842.38984 & 9.88 & 2.48 \\[0.1 cm]
2458843.41633 & 4.96 & 2.18 \\[0.1 cm]
2458844.39932 & 14.76 & 4.22 \\[0.1 cm]
2458845.34605 & 3.36 & 2.05 \\[0.1 cm]
2458846.37295 & 2.57 & 1.59 \\[0.1 cm]
2458847.37900 & 7.87 & 1.72 \\[0.1 cm]
2458848.34964 & -5.31 & 1.77 \\[0.1 cm]
2458849.35491 & 4.27 & 1.52 \\[0.1 cm]
2458850.35018 & -1.90 & 1.89 \\[0.1 cm]
2458855.42501 & -3.53 & 1.74 \\[0.1 cm]
2458856.35167 & -0.20 & 2.26 \\[0.1 cm]
2458857.35123 & 6.76 & 1.73 \\[0.1 cm]
2458860.37498 & 2.18 & 3.63 \\[0.1 cm]
2458881.32010 & -5.56 & 2.18 \\[0.1 cm]
2458882.31533 & 0.88 & 1.81 \\[0.1 cm]
2458890.30180 & -11.40 & 3.28 \\[0.1 cm]
2458891.29664 & -8.28 & 1.95 \\[0.1 cm]
2458894.29514 & -10.75 & 1.80 \\[0.1 cm]
2458895.31304 & -9.93 & 2.99 \\[0.1 cm]
2458897.30005 & -6.50 & 2.31 \\[0.1 cm]
\noalign{\smallskip}
\hline
\end{tabular}
\tablefoot{\tablefoottext{*}{Barycentric dynamical time.}}
\end{table}

%% TOI-1201 COMPANION RV DATA
\begin{table}[!h]
\centering
\caption{CARMENES VIS RV data of PM~J02489--1432E (TOI-393).}
\label{tab:rvdatacomp}
\begin{tabular}{c
                S[table-format=2]
                S[table-format=2]
                }
\hline
\hline
\noalign{\smallskip}
BJD (TDB\tablefootmark{*}) & \multicolumn{1}{c}{RV (m\,s$^{-1}$)} & \multicolumn{1}{c}{$\sigma_\textnormal{RV}$ (m\,s$^{-1}$)} \\
\noalign{\smallskip}
\hline
\noalign{\smallskip}
2458801.53021 & 3.20 & 3.36 \\[0.1 cm]
2458804.47688 & 4.59 & 2.61 \\[0.1 cm]
2458808.50608 & -1.90 & 4.33 \\[0.1 cm]
2458815.47257 & 2.88 & 2.78 \\[0.1 cm]
2458816.48146 & 4.28 & 1.66 \\[0.1 cm]
2458817.47079 & 1.10 & 3.57 \\[0.1 cm]
2458818.45424 & -6.95 & 3.32 \\[0.1 cm]
2458832.42705 & 2.67 & 2.22 \\[0.1 cm]
2458834.44005 & 7.91 & 3.78 \\[0.1 cm]
2458841.39830 & 2.61 & 2.11 \\[0.1 cm]
2458842.36846 & 2.56 & 2.87 \\[0.1 cm]
2458843.43878 & -1.09 & 2.31 \\[0.1 cm]
2458844.42653 & 9.05 & 7.01 \\[0.1 cm]
2458845.36897 & 3.67 & 2.18 \\[0.1 cm]
2458846.39647 & -0.16 & 1.54 \\[0.1 cm]
2458847.40170 & -1.57 & 1.63 \\[0.1 cm]
2458848.37188 & -1.86 & 1.56 \\[0.1 cm]
2458849.37805 & -2.09 & 1.73 \\[0.1 cm]
2458850.37392 & -1.79 & 1.59 \\[0.1 cm]
2458852.36953 & -2.85 & 1.31 \\[0.1 cm]
2458855.40269 & -1.19 & 2.12 \\[0.1 cm]
2458856.37286 & 0.55 & 2.07 \\[0.1 cm]
2458857.37366 & -1.46 & 1.99 \\[0.1 cm]
\noalign{\smallskip}
\hline
\end{tabular}
\tablefoot{\tablefoottext{*}{Barycentric dynamical time.}}
\end{table}

\begin{table}[!h]
\centering
\caption{Astrometric data of the binary WDS~J02490--1432 (KPP~2871).}
\label{tab:binary}
\begin{tabular}{cccl}
\hline
\hline
\noalign{\smallskip}
Epoch & $\rho$ & $\theta$ & Origin \\
~ & (arcsec) & (deg) & \\
\noalign{\smallskip}
\hline
\noalign{\smallskip}
1953.927 & 7.17 $\pm$ 0.30 & 100.7 & POSS-I Red \\ % XE772 (0FHS) SuperCOSMOS, MJD = 34716.2
1986.935 & 7.76 $\pm$ 0.20 & 97.7 & UKST Infrared \\ % IS615 (A38Q) SuperCOSMOS, MJD = 46772.5
1988.838 & 7.90 $\pm$ 0.20 & 96.4 & UKST Blue \\ % (...) SuperCOSMOS, MJD = 47467.6
1989.742 & 7.81 $\pm$ 0.20 & 100.0 & UKST Red \\ % ER615 (A2IU) SuperCOSMOS, MJD = 47797.7
1998.582 & 8.19 $\pm$ 0.13 & 98.3 & 2MASS$^a$ \\ %, MJD = 51026.4
1999.782 & 8.184 & 99.0 $\pm$ 0.2 & UCAC4$^b$ \\
2000.775 & 8.23 $\pm$ 0.10 & 98.7 & DENIS \\ %, MJD = 51827.5
2012.110 & 8.39542 & 98.788 & KPP$^c$ \\ % 2019JDSO...15...21K
2015.000 & 8.4 $\pm$ 1.0 & 98.96 & \gaia\ DR1$^d$ \\ %
2015.500 & 8.391 $\pm$ 0.061 & 98.96 & \gaia\ DR2$^e$ \\ %, MJD = 57174.0
2016.000 & 8.408 $\pm$ 0.034 & 98.94 & \gaia\ EDR3 \\ %
\noalign{\smallskip}
\hline
\end{tabular}
\tablefoot{
\tablefoottext{a}{WDS tabulates $\rho$ = 8.15\,arcsec and $\theta$ = 98.3\,deg (Thurgood Marshall High School, priv. comm.).}
\tablefoottext{b}{From \citet{Zacharias2013}.}
\tablefoottext{c}{From \citet{KnappNanson2019}.}
\tablefoottext{d}{WDS tabulates $\rho$ = 8.34363\,arcsec and $\theta$ = 99.001\,deg \citep{KnappNanson2019}.}
\tablefoottext{e}{WDS tabulates $\rho$ = 8.34896$\pm$0.00004\,arcsec and $\theta$ = 98.999\,deg \citep{El-BradyRix2018}.}}
\end{table}

\end{appendix}

\end{document}